\documentclass[a4paper,11pt]{article}
\pdfoutput=1 

\usepackage{jheppub2} 
\usepackage{bbm,bm,graphicx,mathtools,xcolor,slashed}
\usepackage{amssymb,amsmath}

\usepackage{booktabs}

\usepackage{hyperref}
\usepackage{orcidlink}

\usepackage{tabularx}
\usepackage{makecell}

\newcommand{\diff}{\mathrm{d}}
\newcommand{\Diff}{{\mathcal{D}}}
\newcommand{\tr}{\operatorname{tr}}

\newcommand{\im}{\mathrm{i}}

\newcommand{\calP}{\mathcal{P}}

\newcommand{\calZ}{\mathcal{Z}}

\newcommand{\rme}{\mathrm{e}}

\newcommand{\ev}[1]{\left\langle{#1}\right\rangle}
\newcommand{\Ns}{N_\mathrm{s}}
\newcommand{\Nt}{N_\mathrm{t}}

\DeclareMathOperator{\real}{Re}
\DeclareMathOperator{\imag}{Im}

\newcommand{\calN}{\mathcal{N}}
\newcommand{\vsub}[2]{#1_{\mathrm{#2}}}
\newcommand{\vsup}[2]{#1^{\mathrm{#2}}}

\newcommand{\SW}{\vsub{S}{W}}
\newcommand{\Qtop}{\vsub{Q}{top}}
\newcommand{\qtop}{\vsub{q}{top}}
\newcommand{\tflow}{\vsub{t}{flow}}

\newcommand{\rev}[1]{\textcolor{black}{#1}}

\preprint{YITP-26-11, RIKEN-iTHEMS-Report-26}

\title{\boldmath Numerical Hints for Dyon Condensation at $\theta=2\pi$ via Wilson--'t~Hooft Loops in $SU(2)$ Yang--Mills Theory}

\author[a,b]{Hiromasa Watanabe \orcidlink{0000-0003-1515-1277},}
\affiliation[a]{Yukawa Institute for Theoretical Physics, Kyoto University,
Kitashirakawa Oiwakecho, Sakyo-ku, Kyoto 606-8502, Japan}
\affiliation[b]{Department of Physics, and Research and Education Center for Natural Sciences, Keio University,
4-1-1 Hiyoshi, Yokohama, Kanagawa 223-8521, Japan}
\emailAdd{hiromasa.watanabe@yukawa.kyoto-u.ac.jp}

\author[c]{Issaku Kanamori \orcidlink{0000-0003-4467-1052},}
\affiliation[c]{RIKEN Center for Computational Science,
Kobe 650-0047, Japan}

\author[d]{Okuto Morikawa \orcidlink{0000-0002-0044-4491},}
\affiliation[d]{Center for Interdisciplinary Theoretical and Mathematical Sciences
(iTHEMS), RIKEN, Wako 351-0198, Japan}

\author[e,f]{Yuki Nagai \orcidlink{0000-0001-5098-5440},}
\affiliation[e]{Information Technology Center, The University of Tokyo, Kashiwa, Chiba 277–0882, Japan}
\affiliation[f]{Department of Advanced Materials Science, 
The University of Tokyo, Kashiwa, Chiba, 277-8561, Japan}

\author[a]{Yuya Tanizaki \orcidlink{0000-0003-1283-1808},}

\author[g,h,c]{Akio Tomiya \orcidlink{0000-0001-9374-3716}}
\affiliation[g]{Department of Information and Mathematical Sciences,
Tokyo Woman's Christian  University, Tokyo 167-8585, Japan}
\affiliation[h]{Department of Physics, Kyoto University, Kyoto 606-8502, Japan}

\abstract{%
Yang--Mills theories at $\theta$ and $\theta+2\pi$ are unitarily equivalent, but their $2\pi$ periodicity has a nontrivial realization. 
\rev{
Recent developments in generalized symmetries rigorously prove that confinement vacua at $\theta=0$ and $2\pi$ should belong to different symmetry-protected topological (SPT) states with the $1$-form center symmetry. 
}
For its examination, we measure the Wilson--'t~Hooft loop operators at $\theta=2\pi$ for the $SU(2)$ Wilson lattice gauge action and discuss their long-distance behaviors. 
This requires us to identify the gauge topological charge in the presence of defects, and we employ the $1$-form covariant DBW2 gradient flow to smear lattice gauge fields. 
\rev{
We find a clear perimeter-law signal for the dyonic Wilson--'t Hooft loop at $\theta=2\pi$, providing numerical evidence in the pure $SU(2)$ Yang--Mills theory for the theoretically expected dyon condensation at $\theta=2\pi$. 
}
}

\begin{document}
\maketitle
\flushbottom

\section{Introduction}
\label{sec:introduction}

In $4$d Yang--Mills (YM) theory, colored degrees of freedom are expected to be confined due to the formation of the electric flux tube~\cite{Wilson:1974sk}, which explains why only color-singlet hadrons appear as asymptotic states. 
It is an important open problem in particle/hadron physics to understand its microscopic mechanism and to develop computational techniques for a reliable prediction of confinement physics.

The $\theta$ parameter of the YM theory turns out to provide valuable information about confinement. 
Let us give an intuitive demonstration of its relevance using one of the scenarios, called dual superconductivity~\cite{Nambu:1974zg, Mandelstam:1974pi, Polyakov:1975rs, tHooft:1977nqb}, where one explains the formation of the color-electric flux tube using the dual Meissner effect caused by the condensation of color-magnetic monopoles. 
Introducing the $\theta$ parameter to this scenario, the magnetic monopole with the charge $m$ acquires the fractional effective electric charge $\vsub{e}{eff}=\frac{\theta}{2\pi}m$ due to the Witten effect~\cite{Witten:1979ey}. 
Thus, the $\theta$ parameter directly affects the condensing objects in the confinement vacuum, and its effect in the confinement phase is much more dramatic compared with the ones in weakly-coupled Higgs-type phases.

Especially in the large-$N$ limit of the $SU(N)$ YM theory, it has been known that there exists a 1st-order phase transition at $\theta=\pi$ that separates the confinement vacua at $\theta=0$ and $\theta=2\pi$~\cite{Witten:1980sp, DiVecchia:1980yfw, Witten:1998uka}. 
In the dual-superconductivity scenario, the condensing particles at $\theta=0$ and $\theta=2\pi$ are monopoles and dyons, respectively. Thus, at least one phase transition would be necessary to exchange the condensation modes when the $\theta$ parameter is varied from $0$ to $2\pi$~\cite{tHooft:1981bkw}, 
\rev{and its explicit demonstration is given for a lattice Abelian model~\cite{Cardy:1981qy, Cardy:1981fd}.}
While this is just a model-dependent explanation, the presence of the monopole- and dyon-condensation vacua can be explicitly confirmed for the softly-broken $\calN=2$ supersymmetric YM theory~\cite{Seiberg:1994rs, Seiberg:1994aj}. 
\rev{
Complementing this dynamical picture based on the Abelianization ansatz, the recent development of generalized symmetries provides a gauge-invariant topological characterization of the distinction between the confinement vacua at $\theta=0$ and $2\pi$: they carry different symmetry-protected topological (SPT) labels for the $\mathbb{Z}_{N}$ $1$-form center symmetry~\cite{Gaiotto:2017yup}.\footnote{While the $1$-form symmetry characterization does not assume any Abelization of the dynamics, one can find the consistency betwen this abstract and the conventional dynamical arguments when the Abelianization actually happens. For the Cardy--Rabinovici model~\cite{Cardy:1981qy, Cardy:1981fd}, this is explicitly demonstrated in \cite{Honda:2020txe, Hayashi:2022fkw, Katayama:2025pmz}.}
}

This formal discussion based on the $1$-form symmetry has an impact not only in the context of the pure YM theory but also for the case of Quantum Chromo-Dynamics (QCD). 
For QCD, dynamical quarks in the fundamental representation explicitly violate the $\mathbb{Z}_N$ $1$-form symmetry. 
Despite this fact, the data of the $1$-form symmetry in the pure YM theory is encoded in the $U(1)$ baryon-number symmetry~\cite{Tanizaki:2018wtg} (see also Refs.~\cite{Gaiotto:2017tne, Tanizaki:2017mtm, Yonekura:2019vyz, Anber:2019nze}). 
As a result, the above distinction of confinement vacua for pure YM case affects the global structure of the $\eta'$ field for QCD associated with the anomalous $U(1)$ axial (or $\vsub{U(1)}{A}$) symmetry~\cite{Tanizaki:2022ngt, Tanizaki:2022plm, Hayashi:2023wwi, Hayashi:2024gxv, Hayashi:2024qkm}, and the periodicity of $\eta'$ and the SPT label for the YM confinement vacua are intertwined~\cite{Hayashi:2024qkm}: 
The minimal-winding $\eta'$ vortex should live on the boundary of the sheet of the fractional quantum Hall liquid~\cite{Komargodski:2018odf}, which never occurs if the $\vsub{U(1)}{{A}}$ violation comes solely from the small-instanton Kobayashi--Maskawa--'t~Hooft vertex~\cite{Kobayashi:1970ji, Kobayashi:1971qz, Maskawa:1974vs, tHooft:1976rip}. 
As these recent developments clarify, the realization of the $\vsub{U(1)}{{A}}$ anomaly in the QCD vacuum requires sophisticated understandings of the confinement mechanism at general values of $\theta$ for pure YM theory. 
This strongly motivates us to examine, in particular, the dyon condensation at $\theta=2\pi$ using the numerical lattice Monte Carlo simulation.

For this goal, we need a gauge-invariant criterion to diagnose the monopole versus dyon condensation, and the Wilson--'t~Hooft classification~\cite{tHooft:1977nqb, tHooft:1979rtg} plays the pivotal role there.\footnote{The dual superconductivity scenario assumes effective Abelianization of the dynamics in the non-Abelian gauge theory, and the notion of the monopole or dyon does not generically have a fully non-Abelian gauge-invariant definition. Here, we explain the fully gauge-invariant diagnostics to distinguish phases. } 
It requires not only the Wilson loop $W(C)$, describing the worldline of weight-charge color-electric test particles, but also the 't~Hooft loop $H(C,\Sigma)$, describing the worldline of weight-charge color-magnetic test particles (with the attachment of the $1$-form symmetry generator on the open surface $\Sigma$ satisfying  $\partial\Sigma=C$)\footnote{Weight-charge electric and magnetic particle do not satisfy the mutual-locality condition of Refs.~\cite{Dirac:1931kp, Schwinger:1966nj, Zwanziger:1968rs, Goddard:1976qe}, and thus they cannot be genuine particles at the same time. 
Note that all the dynamical excitations in the YM theory with adjoint matter have the electric/magnetic charges belonging to the root lattice~\cite{Goddard:1976qe}, so they are mutually local and genuine particles.  
For the $SU(N)$ gauge group, the weight-charge electric particles are chosen to be genuine ones, and the weight-charge magnetic particles are attached with the physical Dirac string, whose worldsheet gives the $1$-form symmetry generator~\cite{Aharony:2013hda, Kapustin:2014gua, Gaiotto:2014kfa}.}. 
In the dual-superconductivity picture, the ’t~Hooft loop $H(C,\Sigma)$ obeys the perimeter law in the monopole-condensed phase due to the Debye screening, while both $W(C)$ and $HW(C,\Sigma)$ obey the area law. In the dyon-condensed phase, however, it is the dyonic line $HW(C,\Sigma)$ that obeys the perimeter law, while $W(C)$ and $H(C,\Sigma)$ obey the area law. 
This suggests the $N$-ality of the electric charge attached to the perimeter-law dyonic line specifies the SPT level, which is shown in \cite{tHooft:1979rtg, Nguyen:2023fun, Maeda:2025ycr, Hayashi:2026ijm} (see also Refs.~\cite{Gukov:2013zka, Kapustin:2013qsa, Kapustin:2013uxa}). 
For $\theta=0$, the lattice Monte Carlo simulation shows the perimeter law of the 't~Hooft loop~\cite{deForcrand:2000fi}, which is consistent with the monopole-condensation picture.  

\rev{
In this paper, we study the long-distance behaviors of the Wilson--'t~Hooft loop operators for the $SU(2)$ Wilson lattice gauge theory at $\theta=2\pi$ 
and test the theoretically expected dyon condensation
}
(see Table~\ref{tab:W-tH_classification_SU2}). 
We should point out that there are many previous lattice numerical studies that investigate the $\theta$-dependence of the free energy (see, \textit{e.g.}, Refs.~\cite{DElia:2012pvq, DElia:2013uaf, Bonati:2015sqt, Bonati:2016tvi, Bonati:2019kmf, Bonanno:2023hhp, Kitano:2020mfk, Kitano:2021jho, Yamada:2024vsk, Yamada:2024pjy, Hirasawa:2024fjt}) and they suggest some evidence for the spontaneous $CP$ breaking at $\theta=\pi$ for low temperatures, which is consistent with the level-crossing scenario between the monopole- and dyon-condensation vacua. 
\rev{
While these works share the same spirit with our study, let us emphasize that we take a complementary strategy, which detects the dyon condensation by computing the Wilson--'t~Hooft loops at $\theta=2\pi$.
Our clearest direct numerical result is the perimeter-law behavior of the dyonic Wilson--'t Hooft loop.
The direct 't Hooft-loop data are also consistent with the expected area-law scaling, although the available scale-separation window and statistics do not permit a conclusive asymptotic determination.
Together with the Wilson--'t~Hooft classification~\cite{tHooft:1977nqb, tHooft:1979rtg}, these results provide numerical evidence that the confinement vacuum at $\theta=2\pi$ realizes the dyon condensation phase and carries a nontrivial SPT label.
}

\begin{table}
    \centering
    \begin{tabular}{c|ccc|c}
        \hline
         Condensation&  $\langle W(C)\rangle$&  $\langle H(C,\Sigma)\rangle$ & $\langle HW(C,\Sigma)\rangle$ & $\mathbb{Z}_2^{[1]}$ sym.\\
         \hline\hline
         Electric & Perimeter & Area & Area & Broken\\
         Monopole   & Area & Perimeter & Area & Unbroken; Trivial gap\\
         Dyon & Area & Area$^\star$ & Perimeter$^\star$ & Unbroken; SPT\\
         \hline
    \end{tabular}
    \caption{Wilson--'t~Hooft classification for gapped phases of $SU(2)$ gauge theories (with adjoint matters). The names of the phases (or condensation modes) in the leftmost column are heuristic. 
    In this work, we examine whether the confinement vacuum at $\theta=2\pi$ corresponds to dyon condensation by testing the starred `$^\star$' entries.
    }
    \label{tab:W-tH_classification_SU2}
\end{table}

The organization of this paper is as follows. 
In Sec.~\ref{sec:theory}, we first explain the theoretical background for the topological aspects of the $SU(N)$ lattice gauge theory, and then we develop the reweighting formula to compute Wilson--'t~Hooft loops at $\theta=2\pi$.  
In Sec.~\ref{sec:setup}, we explain the lattice setup for the numerical computations and also the detailed strategy for computing our observables. 
In Sec.~\ref{sec:results}, the numerical results for the 't~Hooft and dyonic loops at $\theta=2\pi$ are presented, and Sec.~\ref{sec:discussion} is devoted to the summary. 
In Appendices~\ref{app:Qtop_R=12_12x8} and \ref{app:more_numerical_results}, we show some of the additional numerical results, which include the results at $\theta=4\pi$.

\section{Theoretical Backgrounds}
\label{sec:theory}

In this section, we describe the theoretical backgrounds relevant to the numerical study of this paper. 

\subsection{Wilson lattice action, 1-form symmetry, and the \texorpdfstring{$\theta$}{theta}-periodicity anomaly}

Let us start with the Wilson lattice action on the hypercubic lattice-regularized $4$-torus, 
\begin{equation}
    \SW[U_\ell, B_p]=-\frac{\beta}{2N} \sum_{p:\,\mathrm{plaquette}}\left[\rme^{-\frac{2\pi\im}{N}B_p}\tr(U_p)+\rme^{\frac{2\pi\im}{N}B_p}\tr(U_p^\dagger)\right], 
    \label{eq:WilsonAction}
\end{equation}
where $U_\ell$ are $SU(N)$ link variables, $U_p=\calP\prod_{\ell\in \partial p} U_\ell$ is the path-ordered product of $U_\ell$'s along the plaquette $p$, and $B_p\in \{0,1,\ldots, N-1\}$ are the $\mathbb{Z}_N$-valued plaquette variables. 
Here, we treat $U_\ell$'s as dynamical variables and $B_p$'s as background fields. 
We note that, in Refs.~\cite{Halliday:1981te,Halliday:1981tm}, $B_p$'s are treated as dynamical fields to study $SO(3)$ gauge theory with dynamical $\mathbb{Z}_2$ monopoles, but we always treat such $\mathbb{Z}_2$ monopoles as the probe monopoles for the $SU(2)$ Yang--Mills theory as we shall discuss more detail later for the 't~Hooft loop operator. 

This action~\eqref{eq:WilsonAction} is $SU(N)$ gauge invariant under $U_\ell\mapsto g_{x_1}^\dagger U_\ell g_{x_2}$ for the link $\ell$ connecting from the site $x_1$ to the site $x_2$ with $g_x\in SU(N)$, but it also has the background $\mathbb{Z}_N$ $1$-form gauge invariance,
\begin{equation}
    B_p \mapsto B_p + (\diff \lambda)_p, \qquad U_\ell \mapsto \rme^{\frac{2\pi\im}{N}\lambda_\ell}U_\ell, 
    \label{eq:1FormGaugeTransformation}
\end{equation}
with $\lambda_\ell\in \{0,1,\ldots, N-1\}$. 
Here, $(\diff \lambda)_p=\sum_{\ell\in \partial p}(-1)^{\mathrm{sign}_p(\ell)} \lambda_\ell$, where $(-1)^{\mathrm{sign}_p(\ell)}$ is $+1$ if $\ell$ has the positive relative orientation with $p$ and $-1$ if it is opposite. 
Therefore, the plaquette variable $B_p$ corresponds to the background $2$-form gauge field for the $\mathbb{Z}_N$ $1$-form symmetry of the pure $SU(N)$ YM theory~\cite{Gaiotto:2014kfa}. 

When the background gauge field $B_p$ is flat, \textit{i.e.},
\begin{equation}
    (\diff B)_c=0 \bmod N \quad\text{for any cube $c$},
\end{equation}
the activation of $B_p$ is equivalent to considering the YM theory with the 't~Hooft twisted boundary condition. 
When we consider the non-flat $B$-field, it introduces the monopole singularity for the cube $c$ with $(\diff B)_c\neq0$, which defines the 't~Hooft loop $H(\tilde{C},\tilde{\Sigma})$. 
Here, $\tilde{C}$ refers the loop on the dual lattice, which penetrates the cubes with $(\diff B)_c$, and $\tilde{\Sigma}$ is the open surface on the dual lattice, which is transverse to the plaquette with $B_p\neq0$ and satisfies $\partial\tilde{\Sigma}=\tilde{C}$: That is, we set $B$ as the delta-function $2$-form of the open surface $\tilde{\Sigma}$, $B=\delta[\tilde{\Sigma}]$.\footnote{The delta-function $p$-form $\delta[M]$ of the $p$-dimensional submanifold is defined by $\int \omega\wedge \delta[M]=\int_M \omega$ for all the differential $p$-forms $\omega$. } 
For the 't~Hooft loop, more details shall be discussed later in Sec.~\ref{sec:tHooftLoop}. 

When $B_p$ is flat, one may impose L\"{u}scher's admissibility condition~\cite{Luscher:1981zq, Abe:2023ncy} in the continuum limit $\beta\to \infty$ to make the gauge topology well-defined; 
\begin{equation}
    \sup_{p}\| \rme^{-\frac{2\pi\im}{N}B_p}U_p - \bm{1}\| <\varepsilon,
    \label{eq:Admissibility}
\end{equation}
where $\varepsilon$ ($\ll 1$) is a fixed parameter independent of $\beta$.\footnote{When $B_p$ is not flat, no configuration satisfies the admissibility condition, so we need to remove the plaquettes in the vicinity of the dislocation $\diff B\neq 0$ in taking ``$\sup_p$'' for the admissibility condition. This is natural as the Coulomb magnetic field $\propto r^{-2}$ around the monopole singularity, which gives an $O(1)$ value for the field strength in the lattice scale. Removal of the admissibility constraint in the vicinity of $\diff B\neq0$ should be regarded as a part of the definition for the 't~Hooft loop with the lattice $\theta$ angle. As its toy model, the vortex operator in the $2$d lattice compact boson has been studied with the admissibility constraint in Ref.~\cite{Abe:2023uan}. 
The recently developed Villain-type formulation~\cite{Chen:2024ddr, Zhang:2024sgm} would give a more straightforward lattice definition of the 't~Hooft operator with the $\theta$ angle, but we will not explore it here. 
For the Abelian models, the definition of 't~Hooft loops is straightforward, and see, \textit{e.g.},  \cite{Sulejmanpasic:2019ytl, Anosova:2022cjm, Katayama:2025pmz}. }
We define the space of link variables satisfying the admissibility condition as $\mathfrak{A}_\varepsilon[B_p]=\{\{U_\ell\}_{\ell}\mid \sup_p \| \rme^{-\frac{2\pi\im}{N}B_p}U_p - \bm{1}\| <\varepsilon\}$. 
Under the admissibility constraint, we can unambiguously assign the topological charge $\Qtop[U_\ell,B_p]$ for the lattice gauge fields by using L\"uscher's geometric construction~\cite{Luscher:1981zq}, which is manifestly local and gauge-invariant. 
With the presence of the flat $B_p$ field, it satisfies the quantization condition~\cite{vanBaal:1982ag}, 
\begin{equation}
    \vsup{\Qtop}{(geom)}[U_\ell,B_p]=\underbrace{-\frac{1}{N}\int_{T^4} \frac{1}{2}B\cup B}_{\text{$\frac{1}{N}$-quantized}} + \mathbb{Z}. 
    \label{eq:GeometricTopologicalCharge}
\end{equation}
Let us define the lattice YM partition function at finite $\theta$ with the admissibility constraint, 
\begin{equation}
    \calZ_{\theta}[B_p]= 
    \int_{\mathfrak{A}_\varepsilon[B_p]} \hspace{-1.5em}\Diff U_\ell\, 
    \exp\left\{-\SW[U_\ell,B_p]+\im \theta \vsup{\Qtop}{(geom)}[U_\ell,B_p]\right\}, 
\end{equation}
then we find the mild anomalous violation of the $2\pi$-periodicity of $\theta$ under the presence of the flat $B$-field background~\cite{Gaiotto:2017yup, Abe:2023ncy}:
\begin{equation}
    \calZ_{\theta+2\pi}[B_p]=\exp\left(-\frac{2\pi\im}{N}\int_{T^4}\frac{1}{2}B\cup B\right) \calZ_{\theta}[B_p]. 
    \label{eq:MixedAnomaly}
\end{equation}
This exact relation for the YM partition function can be thought of as the generalization of the 't~Hooft anomaly, which is referred to as the global inconsistency or the coupling-space anomaly~\cite{Gaiotto:2017yup, Tanizaki:2017bam, Komargodski:2017dmc, Kikuchi:2017pcp, Tanizaki:2018xto, Cordova:2019jnf, Cordova:2019uob}. 
Its anomaly matching condition gives the rigorous proof for the presence of at least one phase transition when we change $\theta$ from $0$ to $2\pi$ if the $1$-form symmetry is unbroken: The SPT level for the unbroken $\mathbb{Z}_N$ $1$-form symmetry have to be shifted by $1$ when we change $\theta$ by $2\pi$, but it is a discrete $\mathbb{Z}_N$ label and requires the quantum phase transition for its jump.

\subsection{DBW2 flow: Lattice gradient flow stabilizing topological sectors}

While the admissibility constraint~\eqref{eq:Admissibility} provides the theoretically beautiful lattice setup for the gauge-field topology, it makes the system in the vicinity of the continuum limit and requires a quite huge lattice box to study the confinement property, which is impractical for actual numerical simulations. 
Usually, instead of imposing the admissibility condition, we apply certain smearing procedures for the lattice gauge field to remove its lattice scale fluctuations and then calculate its topology by integrating the lattice version of the topological charge density, $\frac{\varepsilon_{\mu\nu\rho\sigma}}{32\pi^2}\tr(F_{\mu\nu}F_{\rho\sigma})$, or by counting the index of lattice Dirac operator.

One of the commonly used smearing algorithms is the gradient flow~\cite{Luscher:2010iy}: We introduce the flow action $\vsub{S}{flow}[U_\ell, B_p]$ as the cost function for the smearing, and we apply the gradient flow formally written as 
\begin{equation}
    \frac{\diff}{\diff t}U_\ell(t) = -\frac{\delta}{\delta U_\ell}\vsub{S}{flow}[U_\ell(t), B_p], 
\end{equation}
with the initial condition $U_\ell(t=0)=U_\ell$ with respect to the fictitious time referred to as flow time. 
For the precise meaning of the link-differential operation $\frac{\delta}{\delta U_\ell}$, see the Appendix of Ref.~\cite{Luscher:2010iy}. 
Along the flow, $\vsub{S}{flow}[U_\ell(t),B_p]$ monotonically decreases unless it is on the critical point, which solves the classical equation of motion of $\vsub{S}{flow}$. 
We require that the gradient-flow equation has the gauge covariance under $SU(N)$ and also $\mathbb{Z}_N$ $1$-form transformations, which is equivalent to require that $\vsub{S}{flow}[U_\ell, B_p]$ satisfies the $SU(N)$ gauge invariance and the $\mathbb{Z}_N$ $1$-form background gauge invariance. 
As a result, we shall manifestly respect the $\mathbb{Z}_N$ $1$-form global symmetry.  

There is some freedom in the choice of the flow action, and we do not have to match it with the Boltzmann weight for generating the gauge fields. Different choices of the flow action provide the different composite-operator renormalization schemes, and they should lead to equivalent results~\cite{Bonati:2014tqa, Alexandrou:2017hqw}. 
Here, let us consider the following $1$-parameter family of lattice action constructed from the $1\times 1$ and $2\times 1$ plaquettes; 
\begin{equation}
    \vsub{S}{flow}
    =-\sum_{x,\mu\nu}\real\left\{\tr[(1-8c_1)\rme^{-\frac{2\pi\im}{N}B_{x,\mu\nu}} U^{(1\times 1)}_{x,\mu\nu}
    +2c_1\, \rme^{-\frac{2\pi\im}{N}(B_{x,\mu\nu}+B_{x+\hat{\mu},\mu\nu})}U^{(2\times 1)}_{x,\mu\nu}]\right\}, 
    \label{eq:1Parameter_FlowAction}
\end{equation}
where $U^{(1\times 1)}_{x,\mu\nu}=U_{x,\mu}U_{x+\hat{\mu},\nu}U^\dagger_{x+\hat{\nu},\mu}U^\dagger_{x,\nu}$ and $U^{(2\times 1)}_{x,\mu\nu}=U_{x,\mu}U_{x+\hat{\mu},\mu}U_{x+2\hat{\mu},\mu}U^{\dagger}_{x+\hat{\mu}+\hat{\nu},\mu}U^\dagger_{x+\hat{\nu},\mu}U^\dagger_{x,\nu}$ are the $1\times 1$ and $2\times 1$ plaquettes along the $\mu$-$\nu$ direction starting and ending at the site~$x$. 
The coefficients for the first and second terms are chosen so that the leading term of $\vsub{S}{flow}$ in the classical continuum approximation becomes the classical Yang--Mills action and independent of $c_1$~\cite{Luscher:1984xn}:
\begin{equation}
    \vsub{S}{flow}\approx \frac{a^4}{2}\sum_{x,\mu\nu}\tr(F_{\mu\nu}^2)-(1+12 c_1)\frac{a^6}{12}\sum_{x,\mu\nu}\tr[(D_\mu F_{\mu\nu})^2]+ O(a^8), 
    \label{eq:1Parameter_FlowAction_Continuum}
\end{equation}
where we introduce the lattice unit $a$ for the continuum approximation. 

The negative sign in front of the dimension-$6$ operator in \eqref{eq:1Parameter_FlowAction_Continuum} crucially affects the consideration of the topological stability. 
If we consider the one-instanton configuration in the infinitely large lattice, this suggests that the size of the instanton $\rho$ becomes smaller along the gradient flow if $1+12 c_1>0$. 
When the instanton size $\rho$ becomes comparable with the lattice unit $a$, one can no longer discriminate such an instanton from the lattice artifact and the topological charge jumps, which never occur in the continuum spacetime. 
If we choose $1+12 c_1<0$, the size of the instanton grows as a function of the flow time, and we can circumvent this instability of the topological sector at least for the one-instanton configuration~\cite{Iwasaki:1983bv, Iwasaki:1983iya, GarciaPerez:1993lic, deForcrand:1995bq}.

In a recent work~\cite{Tanizaki:2024zsu}, some of the present authors studied the (in)stability of the topological sectors in the long flow-time limit in the $SU(2)$ pure lattice gauge theory with several choices of $c_1$: Wilson action ($c_1=0$), tree-level Symanzik-improved action ($c_1=-\frac{1}{12}$), Iwasaki action ($c_1=-0.331$)~\cite{Iwasaki:1983iya}, and DBW2 action ($c_1=-1.4088$)~\cite{QCD-TARO:1999mox}. 
There, it is found that the DBW2 flow gives the fast convergence to the integer-quantized topology around $t/a^2\approx 1$ and the converged topological number never changes at least within $1\lesssim t/a^2\le 50$ for the $SU(2)$ Monte Carlo samples with $\beta=2.45$ and $L^4=12^4$. 
In Ref.~\cite{Butti:2025rnu}, qualitatively the same behavior of the DBW2 flow is reported for a wider range of parameter setups in QCD-like theories with $2$-index quarks. 
For the study of $\theta=2\pi$, fast convergence to integer-quantized topology is crucial as we shall explain in Sec.~\ref{sec:tHooftLoop}, and thus we employ the DBW2 flow for the smearing procedures in this paper following Refs.~\cite{Tanizaki:2024zsu, Butti:2025rnu}; we set $c_1=-1.4088$ in the flow action \eqref{eq:1Parameter_FlowAction}.

To define the lattice topological charge density $\qtop^{(t)}(x)$ using the DBW2-flowed configuration $U_\ell(t)$, we employ the gluonic definition,\footnote{The other standard choice for the lattice topological charge is to use the fermionic definition, which counts the zero modes of the overlap Dirac operator. 
In our situation, however, it is inappropriate to use the fermionic definition for the lattice topological charge coupled to $\theta$. 
Firstly, our study relies heavily on the existence of the $\mathbb{Z}_N$ $1$-form symmetry, so every observable should consistently couple to the background gauge field $B_p$. In the standard use of the overlap index, the Dirac operator is taken as the one for the fundamental representation, which explicitly violates the $1$-form symmetry. This problem can be resolved by using the adjoint overlap operator instead of the fundamental one, but it requires some extra work. 
Secondly, as a more serious issue, the 't~Hooft loop necessarily violates the admissibility condition at the monopole singularities, which causes the nonlocality for the overlap index and leads to the mismatch between the gluonic and fermionic definitions by the $\eta$-invariant. As a matter of principle, every extended observable in field theory should respect the locality, which is manifest for 't~Hooft loops at finite $\theta$ angles if we use the gluonic definition for the lattice topological charge. } 
which integrates the lattice version of the topological charge density $\frac{\varepsilon_{\mu\nu\rho\sigma}}{32\pi^2}\tr(F_{\mu\nu}F_{\rho \sigma})$. 
For this purpose, let us introduce the clover-leaf field strength $\vsup{C}{clov}_{x,\mu\nu}[U_\ell(t),B_p]$ and its rectangular analogue $\vsup{C}{rect}_{x,\mu\nu}[U_\ell(t), B_p]$, and both of them are required to be $1$-form gauge invariant and $SU(N)$ gauge covariant at the site $x$. 
Let us write the explicit form for the clover-leaf field strength, 
\begin{align}
    \vsup{C}{clov}_{x,\mu\nu}[U_\ell(t),B_p]&=\frac{1}{4}\imag\left[\rme^{-\frac{2\pi\im}{N}B_{x,\mu\nu}}U_{x,\mu\nu}(t)+\rme^{-\frac{2\pi\im}{N}B_{x-\hat{\mu},\mu\nu}} U_{x,\nu-\mu}(t)\right.\notag\\
    &\qquad\qquad \left. +\, \rme^{-\frac{2\pi\im}{N}B_{x-\hat{\mu}-\hat{\nu},\mu\nu}} U_{x,-\mu-\nu}(t) + \rme^{-\frac{2\pi\im}{N}B_{x-\hat{\nu},\mu\nu}} U_{x,-\nu\mu}(t) \right], 
\end{align}
and $\vsup{C}{rect}_{x,\mu\nu}[U_\ell(t), B_p]$ is its straightforward extension for the average of the $2\times 1$ and $1\times 2$ plaquette counterparts as we did for $U^{(2\times 1)}_{x,\mu\nu}$ in \eqref{eq:1Parameter_FlowAction}. We then define the flowed topological charge density,
\begin{equation}
    \qtop^{(t)}(x)=\frac{1}{32\pi^2}\sum_{\mu\nu\rho\sigma}\varepsilon_{\mu\nu\rho\sigma}\left\{(1-8b_1)\tr[\vsup{C}{clov}_{x,\mu\nu}\vsup{C}{clov}_{x,\rho\sigma}]+2b_1 \tr[\vsup{C}{rect}_{x,\mu\nu}\vsup{C}{rect}_{x,\rho\sigma}]\right\}, 
\end{equation}
and the total topological charge is defined as $\Qtop^{(t)}[U_\ell,B_p]=\sum_x \qtop^{(t)}[U_\ell, B_p]$. 
To eliminate the $O(a^6)$ term for the charge density $\qtop^{(t)}(x)$ in the classical continuum limit, we choose the tree-level Symanzik improvement, $b_1=-\frac{1}{12}$.

\subsection{'t Hooft loop and its computational strategy at \texorpdfstring{$\theta=2\pi$}{theta=2pi}}
\label{sec:tHooftLoop}

We are now ready to discuss the 't~Hooft loop operator $H(\tilde{C},\tilde{\Sigma})$ at general values of $\theta$, where $\tilde{C}$ is the closed loop in the dual lattice and $\tilde{\Sigma}$ is the open surface also in the dual lattice with $\partial\tilde{\Sigma}=\tilde{C}$. 
The background $B$-field is specified as the delta-function $2$-form of the surface, 
\begin{equation}
    B_p=\delta[\tilde{\Sigma}]_p, 
\end{equation}
and this is a non-flat background gauge field as $\diff B=\delta[\partial \tilde{\Sigma}]=\delta[\tilde{C}]$. 
We then define the 't~Hooft loop operator $H(\tilde{C},\tilde{\Sigma})$ as the defect operator, whose correlation functions with electric observables $\mathcal{O}[U_\ell,B_p]$ (such as Wilson loops) are given by
\begin{equation}
    \langle H(\tilde{C},\tilde{\Sigma}) \mathcal{O}\rangle_{\theta}
    =\frac{1}{\calZ_\theta}\int\Diff U_\ell\, \mathcal{O}[U_\ell,B_p] \exp\left\{-\SW[U_\ell, B_p]+\im \theta \Qtop^{(t)}[U_\ell, B_p]\right\}, 
    \label{eq:tHooftLoop}
\end{equation}
where 
\begin{equation}
    \calZ_\theta=\int\Diff U_\ell\,  \exp\left\{-\SW[U_\ell, 0]+\im \theta \Qtop^{(t)}[U_\ell, 0]\right\}. 
\end{equation}
This expression can also be understood naturally from the fact that the 't~Hooft loop expectation value corresponds physically to the free energy of creating magnetic probes~\cite{deForcrand:2000fi, deForcrand:2005pb}.
The dependence on the choice of $\tilde{\Sigma}$ is only topological, which forbids the local counterterm along $\tilde{\Sigma}$. 
This fact is important to ensure that the string tension for the 't~Hooft loop gives a well-defined order parameter, as it cannot be removed from the renormalization.

We note that the $2\pi$-periodicity for $\calZ_\theta$ is explicitly violated. 
The admissibility condition is not imposed in the above definition, and the topological charge $\Qtop^{(t)}[U_\ell,0]$ does not show the exact quantization to integers. 
To achieve the $2\pi$ periodicity as much as possible, we multiply an overall rescaling factor $Z\approx 1$ to the topological charge, $\Qtop^{(t)}\Rightarrow Z \Qtop^{(t)}$, in the above definition so that its distribution localizes around the integers when $B$ is absent: This can be regarded as the multiplicative renormalization factor for $\theta$ due to the choice of the regularization scheme violating the integer quantization of the topological charge (see, \textit{e.g.}, Ref.~\cite{DElia:2003zne}).

To evaluate \eqref{eq:tHooftLoop} using the Monte Carlo method, we rewrite it as follows and apply the reweighting method: 
\begin{align}
    &\langle H(\tilde{C},\tilde{\Sigma}) \mathcal{O}\rangle_{\theta}\notag\\
    &=\left(\frac{\calZ_\theta}{\calZ_0}\right)^{-1} 
    \times \frac{1}{\calZ_0}\int \Diff U_\ell\exp\left\{-\SW[U_\ell,B_p]\right\}\notag\\
    &\qquad \times \frac{\int\Diff U_\ell\, \mathcal{O}[U_\ell,B_p] \exp\left\{-\SW[U_\ell, B_p]+\im \theta Z\Qtop^{(t)}[U_\ell, B_p]\right\}}{\int \Diff U_\ell\exp\left\{-\SW[U_\ell,B_p]\right\}}\notag\\
    &=\left(\frac{\calZ_\theta}{\calZ_0}\right)^{-1} 
    \times \langle H(\tilde{C},\tilde{\Sigma})\rangle_{\theta=0} \times \frac{\int\Diff U_\ell\, \mathcal{O}[U_\ell,B_p] \exp\left\{-\SW[U_\ell, B_p]+\im \theta Z \Qtop^{(t)}[U_\ell, B_p]\right\}}{\int \Diff U_\ell\exp\left\{-\SW[U_\ell,B_p]\right\}}.  
    \label{eq:tHooftLoop_Reweighting}
\end{align}
Let us write the Monte Carlo average with the weight function $\exp\{-S[U_\ell]\}$ as $\vsup{\langle \bullet\rangle}{MC}_{S[U_\ell]}$. 
Then, the first factor in \eqref{eq:tHooftLoop_Reweighting} can be expressed as 
\begin{align}
    \frac{\calZ_\theta}{\calZ_0}
    &=\vsup{\left\langle \exp\left\{\im \theta Z \Qtop^{(t)}[U_\ell,0]\right\}\right\rangle}{MC}_{\SW[U_\ell,0]}, 
    \label{eq:ReweightingFactor}
\end{align}
and this is expected to behave as $\exp[-(E(\theta)-E(0))V]$, where $E(\theta)$ is the ground-state energy at $\theta$ and $V$ is the spacetime volume. 
Since this appears in the denominator, it typically causes the severe sign problem unless $\theta = O(1/\sqrt{V})$. 
By using the DBW2 flow, we achieve the convergence of $\Qtop^{(t)}[U_\ell,0]$ to almost integers (after multiplying the rescaling factor $Z$) within a reasonable flow time $t$, and then we shall confirm that the operator $\exp(\im \theta Z \Qtop^{(t)}[U_\ell,0])$ at $\theta=2\pi$ has a sharp distribution localized around $1$. 
Thus, the DBW2 flow enables us to successfully circumvent the sign problem potentially caused by this reweighting factor at $\theta=2\pi$.\footnote{Let us here emphasize that the rounding trick of the topological charge does not work for our purpose. When we insert the 't~Hooft loop, $B=\delta[\tilde{\Sigma}]$, $\Qtop$ no longer quantizes into integers even in the continuum limit, and thus rounding off $\Qtop$ to the nearest integers cannot be justified in the presence of the 't~Hooft loops.} 

The second factor in \eqref{eq:tHooftLoop_Reweighting} is nothing but the expectation value of the 't~Hooft loop at $\theta=0$~\cite{deForcrand:2000fi, deForcrand:2005pb}:
\begin{align}
    \langle H(\tilde{C},\tilde{\Sigma})\rangle_{\theta=0} &= \frac{\int \Diff U_\ell \exp\left\{-\SW[U_\ell,B_p]\right\}}{\int \Diff U_\ell \exp\left\{-\SW[U_\ell,0]\right\}}
    = \vsup{\left\langle \rme^{-(\SW[U_\ell, B_p]-\SW[U_\ell,0])}\right\rangle}{MC}_{\SW[U_\ell,0]}. 
\end{align}
We note that this observable $\rme^{-(\SW[U_\ell, B_p]-\SW[U_\ell,0])}$ is extended over the $2$d surface $\tilde{\Sigma}$, and thus this Monte Carlo average can suffer from the severe overlap problem when the size of the surface $\tilde{\Sigma}$ becomes large. 
In Ref.~\cite{deForcrand:2000fi} (see also \cite{Kovacs:2000sy, Hoelbling:2000su,deForcrand:2004jt}), the multi-step reweighting algorithm is introduced to circumvent this problem: We consider the monotonically increasing sequence of the dual surfaces, $\tilde{\Sigma}_0=\emptyset\subset \tilde{\Sigma}_1\subset \cdots \subset \tilde{\Sigma}_n=\tilde{\Sigma}$, and we require that $\tilde{\Sigma}_k\setminus\tilde{\Sigma}_{k-1}$ is just a single dual plaquette for each $k=1$, \dots, $n$. 
We rewrite the above Monte Carlo average as 
\begin{align}
    \langle H(\tilde{C},\tilde{\Sigma})\rangle_{\theta=0} 
    = \prod_{k=1}^{n}\vsup{\left\langle \rme^{-(\SW[U_\ell, \delta[\tilde{\Sigma}_{k}]]-\SW[U_\ell,\delta[\tilde{\Sigma}_{k-1}]])}\right\rangle}{MC}_{\SW[U_\ell,\delta[\tilde{\Sigma}_{k-1}]]}, 
    \label{eq:multi_step_reweighting}
\end{align}
and then the observable becomes local on each step $k$, so no overlap problem appears.\footnote{Let us mention one subtlety though. Since the number of steps $n$ is given by the area of the dual surface $\tilde{\Sigma}$ in the lattice unit, we need to evaluate the reweighting factor of each step precisely enough to obtain $\langle H(\tilde{C},\tilde{\Sigma})\rangle_{\theta=0}$ within reasonable errors when $\tilde{\Sigma}$ becomes larger. For the lattice sizes used in this paper, this turns out to be well under control. } 
Since it is already well-established in Ref.~\cite{deForcrand:2000fi} that this shows the perimeter law in the confined phase at $\theta=0$, we are not going to reproduce it in this paper (while we confirmed it in the test run as a sanity check of our simulation code implementing the background field).

The main target of this paper is to compute the third factor in \eqref{eq:tHooftLoop_Reweighting}: 
\begin{align}
    \frac{\int\Diff U_\ell\, \mathcal{O}[U_\ell,B_p] \exp\left\{-\SW[U_\ell, B_p]+\im \theta Z \Qtop^{(t)}[U_\ell, B_p]\right\}}{\int \Diff U_\ell \exp\left\{-\SW[U_\ell,B_p]\right\}} 
    = \vsup{\left\langle \mathcal{O}[U_\ell, B_p] \rme^{\im \theta Z \Qtop^{(t)}[U_\ell, B_p]}\right\rangle}{MC}_{\SW[U_\ell, B_p]}. 
    \label{eq:determining-factor}
\end{align}
We note that, under the presence of the non-flat $B$-field, the topological charge does not quantize to integers at all.
Since the first factor gives an overall constant and the second factor shows the perimeter law, the question of whether the 't~Hooft loop shows the area law at $\theta=2\pi$ is equivalent to asking if 
$\vsup{\left\langle \rme^{2\pi\im  Z \Qtop^{(t)}[U_\ell, B_p]}\right\rangle}{MC}_{\SW[U_\ell, B_p]}$ shows the area law in terms of the dual loop $\tilde{C}$ with $\diff B=\delta[\tilde{C}]$. 
When we want to look at the dyonic loop, we take $\mathcal{O}[U_\ell, B_p]$ as the Wilson loop $W(C)$ along some loop $C$ adjacent to the dual loop $\tilde{C}$ and compute $\vsup{\left\langle W(C)\rme^{2\pi\im  Z \Qtop^{(t)}[U_\ell, B_p]}\right\rangle}{MC}_{\SW[U_\ell, B_p]}$, and this factor determines if the dyonic loop shows the area or perimeter law at $\theta=2\pi$.

\section{Simulation Setups}
\label{sec:setup}

We numerically study the $SU(2)$ Wilson action with the lattice coupling $\beta=2.45$ at two different lattice sizes $\Ns^3\times\Nt = 12^3\times 8$ and $16^3\times 10$, where the systems show confinement at $\theta=0$~\cite{deForcrand:2000fi}. 
In the present work, the 't~Hooft loop is inserted on the lattice as a probe since the $\mathbb{Z}_2$ gauge field $B$ is not dynamical.
We fix the defect insertions $\tilde{\Sigma}$ as rectangular surfaces extended in the temporal and spatial directions, which wrap around the temporal circle.
More concretely, we take the origin of the lattice sites as the base point, namely $\vsub{x}{base} = (1,1,1,1)$, and place the rectangular surface starting from that point. 
Keeping the temporal size as $\Nt$, we vary the spatial size $R$ and measure the damping rate of the expectation value of loop operators with $R$.
The simulation parameters of the ensembles generated by the Monte Carlo simulation are listed in Table~\ref{tab:simulation_parameters}, and for each parameter setup, we store $\vsub{N}{conf}$ configurations, saved every $\vsub{N}{skip}$ updates; we performed $\vsub{N}{conf}\times \vsub{N}{skip}$ heat-bath updates for each $R$ after thermalization. 
The simulation code is built on JuliaQCD \cite{Nagai:2024yaf} in the Julia language \cite{bezanson2012juliafastdynamiclanguage}, and we validated the implementation by reproducing the benchmark results and consistency checks reported in Ref.~\cite{deForcrand:2000fi} at $\theta=0$, using the same measurement strategy and update procedure for the $12^3\times 8$ lattice: We confirmed the perimeter law of the 't~Hooft loop at $\theta=0$ via the multi-step reweighting \eqref{eq:multi_step_reweighting} following Ref.~\cite{deForcrand:2000fi}.


\begin{table}[t]
    \centering
    \begin{tabular}{cccccc}
        \toprule
        $\beta$ & $\Ns$ & $\Nt$ & $R$ & $\vsub{N}{conf}$ & $\vsub{N}{skip}$ \\
        \midrule
         2.45 & 12 & 8 & 0 & 10,000 & 10 \\
         &  &  & $1$, \dots, $12$ & 40,000 & 10 \\
         \midrule
         2.45 & 16 & 10 & 0 & 5,000 & 10 \\
         &  &  & $6$, \dots, $10$ & 80,000 & 10 \\
         \bottomrule
    \end{tabular}
    \caption{Simulation parameters. Ensembles for ``$R = 0$'' are generated without the background field.}
    \label{tab:simulation_parameters}
\end{table}

Generating ensembles without the 't~Hooft loop insertion (\textit{i.e.}, ``$R = 0$'' in Table~\ref{tab:simulation_parameters}) corresponds to the standard Monte Carlo sampling with the Wilson plaquette action, and those configurations are used to determine the rescaling factor $Z$ of the topological charge and the reweighting factor $\mathcal{Z}_{\theta=2\pi}/\mathcal{Z}_0$. 
To determine the rescaling factor $Z$, we consider the DBW2 flow of the ``$R=0$'' configurations by the flow time $\tflow$ and compute $\Qtop^{(t)}(j)$ for those configurations $j=1$, \dots, $\vsub{N}{conf}$. 
The histogram of the smeared topological charges with $B=0$ turns out to show peaks that are slightly below the exact integers.
To compensate for the discrepancy, we introduce the rescaling factor by solving the minimization problem:
\begin{equation}
    \min_Z\left[\sum_{j=1}^{\vsub{N}{conf}}\left(\mathrm{round}(Z\Qtop^{(\tflow)}(j)) - Z\Qtop^{(\tflow)}(j)\right)^2\right],
    \label{eq:eval_Z_factor}
\end{equation}
where $\mathrm{round}(x)=\lfloor x-\frac{1}{2}\rfloor +\frac{1}{2}$.  
This factor $Z$ depends on the definition of lattice topological charge, \textit{i.e.}, the choice of topological charge operator and flow time, and we compute that one by one.
Note that $Z$ is also a quantity estimated by Monte Carlo sampling and has a statistical uncertainty.
We take this into account for the later analysis by using the bootstrap method.
Once the rescaling factor is obtained for each lattice definition of the topological charge operator, we employ it for the analysis of the 't~Hooft  and dyonic loops at $\theta=2\pi$.

\begin{figure}[t]
    \centering
    \includegraphics[width=0.5\linewidth]{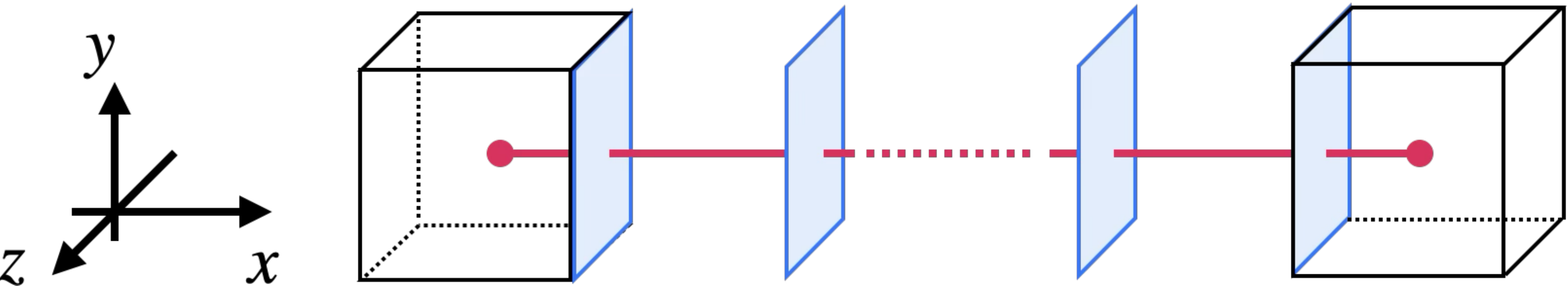}
    \caption{
    Schematic picture of monopoles on the dual lattice at a time slice.
    The background fields $B$ living on the colored plaquettes pierced by a Dirac string are turned on in the present setup.
    }
    \label{fig:monopole_dual_lattice}
\end{figure}

The ensembles with $R\neq0$ are generated by the action in which the background fields $B$ living on the plaquette inside the rectangular loop $R\times \Nt$ are turned on: This inserts the static monopole singularities at $(x,y,z)=(\frac{1}{2},\frac{1}{2},\frac{1}{2})$ and $(\frac{1}{2}+R,\frac{1}{2},\frac{1}{2})$, and the $\mathbb{Z}_2$ Dirac string is extended along $[\frac{1}{2},\frac{1}{2}+R] \times (\frac{1}{2},\frac{1}{2})\times S^1_{\Nt}$ (see Fig.~\ref{fig:monopole_dual_lattice}). 
These configurations are used to compute the dominant factors for the 't~Hooft and dyonic loops: 
For the 't~Hooft loop, we calculate
\begin{equation}
    \vsup{\left\langle \exp(2\pi\im Z\Qtop^{(\tflow)}[U_\ell, \delta[\tilde{\Sigma}]])\right\rangle}{MC}_{\SW[U_\ell, \delta[\tilde{\Sigma}]]},
    \label{eq:H_MC}
\end{equation} 
which is the dominant factor for the 't~Hooft loop at $\theta=2\pi$ in Eq.~\eqref{eq:tHooftLoop_Reweighting}. 
For the dyonic loop, we calculate\footnote{For the monopole singularity at $(x,y,z)=(\frac{1}{2},\frac{1}{2},\frac{1}{2})$, there are eight nearest neighboring sites (corners of $[0,1]^3$) for the location of the electric charge when defining the dyonic line operator. 
It would be ideal to compute the off-diagonal correlators of those eight choices, but we pick just a specific choice in this paper. } 
\begin{equation}
    \vsup{\left\langle W^{(\tflow')}(C;\vsub{x}{base})\,
    \exp(2\pi\im Z\Qtop^{(\tflow)}[U_\ell, \delta[\tilde{\Sigma}]])\right\rangle}{MC}_{\SW[U_\ell, \delta[\tilde{\Sigma}]]}, 
    \label{eq:HW_MC}
\end{equation} 
where $W^{(\tflow')}(C;\vsub{x}{base})$ is the Wilson loop of size $R\times\Nt$ on the $(x,t)$-plane starting from $\vsub{x}{base}$, which is constructed by the flowed configuration by the flow time $\tflow'$.\footnote{The other option for defining dyonic temporal lines is to introduce the Polyakov loops at $(1,1,1)$ and $(R+1,1,1)$, and we show the result for the $12^3\times 8$ lattice in Appendix~\ref{app:more_numerical_results}. 

In our definition of using the rectangular Wilson loop, the difference is the presence of the Wilson line that goes back and forth along the $x$-direction extended by $R$. One may wonder if its presence causes the problem for discussing the perimeter law of the dyonic line at $\theta=2\pi$, and we claim it does not. Let $U$ be the $x$-direction Wilson line valued in $SU(2)$, and then its contribution can be decomposed as $U\otimes U^\dagger= 1\oplus \mathrm{Adj}[U]$. The singlet contribution is identical to the insertion of two Polyakov loops (up to a numerical factor). The triplet contribution should behave as $\rme^{-\# R}$ and becomes subdominant when $R\gg 1$ in the confinement vacua. As we use the region $2d\ll R\ll \Ns-2d$ to discuss the area versus perimeter law, either definition of dyonic lines should work. 
\label{ftnt:def_dyonline}}
More concretely, the closed contour $C$ is chosen such that the probe charges located at $(x,y,z)=(1,1,1)$ and $(R+1,1,1)$ undergo the time evolution toward the periodic temporal direction.
One may independently choose $\tflow$ for the topological charge and $\tflow'$ 
for the Wilson loops,
but we will mainly work on the choice $\tflow'=\tflow$ in this paper. 

The choice of the flow time $\tflow$ has some arbitrariness as long as it satisfies the following properties:

\begin{itemize}
    \item \textbf{Lower bound of $\tflow$:} To compute the topological charge for the lattice configurations, we need to remove the short-range fluctuations via the gradient flow, which requires the diffusion length $d=\sqrt{8\tflow}$ should satisfy $d\gg 2$~\cite{Luscher:2010iy}, \textit{i.e.}
    \begin{equation}
        \tflow \gg \frac{1}{2}. 
    \end{equation}
    In addition, we have to require $\calZ_{\theta=2\pi}/\calZ_{\theta=0}$ should be almost $1$ to evade the sign problem successfully. 

    \item \textbf{Upper bound of $\tflow$:} To compute the long-range part of the interparticle potential, twice the smeared size $d$ for the (induced) electric charges should be small enough compared with their separation. Since the maximal separation of the test charges is $\vsub{R}{max}=\frac{\Ns}{2}$, we need $2d\ll \frac{\Ns}{2}$, \textit{i.e.}, 
    \begin{equation}
        \tflow\ll \frac{1}{2}\left(\frac{\Ns}{8}\right)^2. 
    \end{equation}


\end{itemize}
For $\Ns=12$, these requirements tell $0.5\ll \tflow\ll 1.1$ and it would be natural to choose $\tflow\approx 0.8$ while we have to admit that the scale separation is not achieved well. 
For $\Ns=16$, the upper bound is replaced by $\tflow\ll 2.0$, and we have the better scale separation if we could again choose $\tflow\approx 1.0$ assuming the convergence of the topological charge is fast enough. 

\section{Numerical Results}
\label{sec:results}

\subsection{Reweighting Factor \texorpdfstring{$\calZ_{\theta=2\pi}/\calZ_{\theta=0}$}{Z2pi/Z0}}

\begin{figure}[t]
    \centering
    \includegraphics[width=0.5\linewidth]{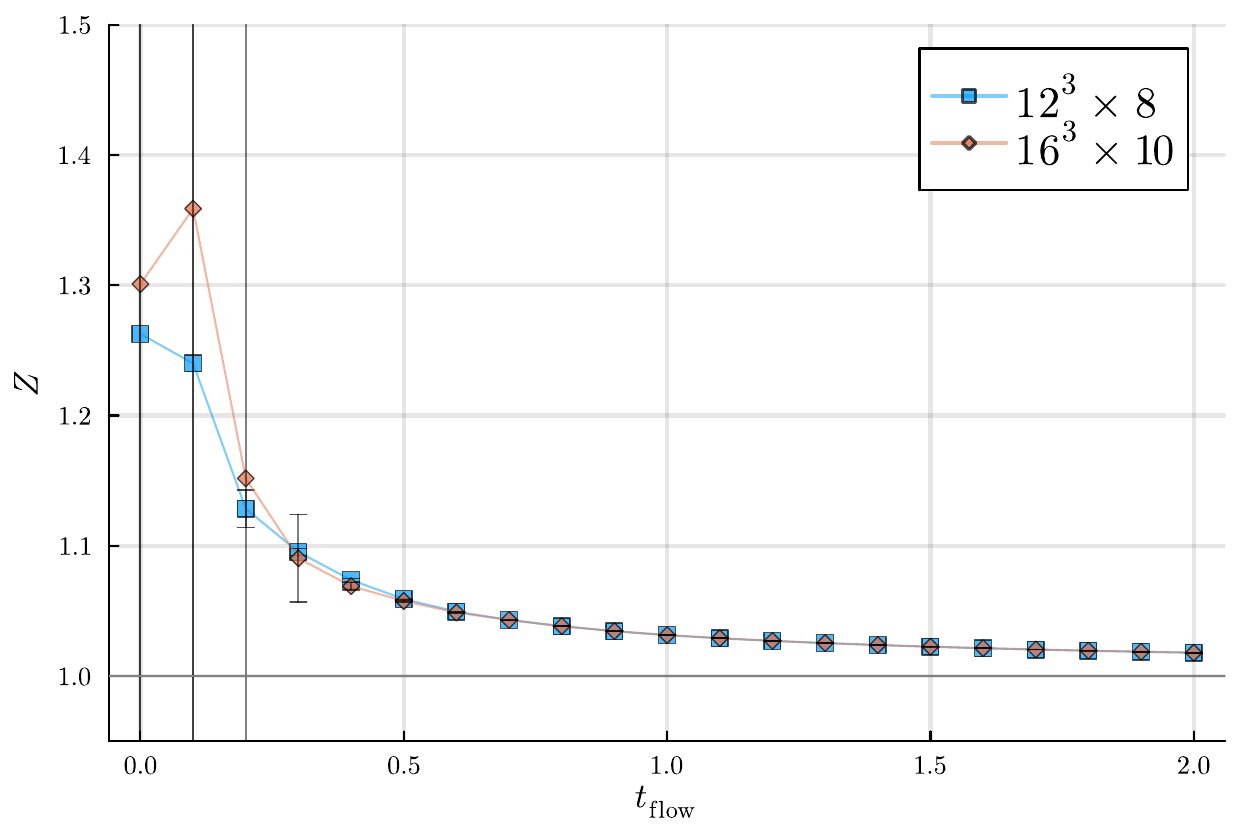}
    \caption{Flow-time dependence of the rescaling factor $Z$.}
    \label{fig:Z_factor}
\end{figure}

\begin{figure}[t]
    \centering
    \includegraphics[width=0.49\linewidth]{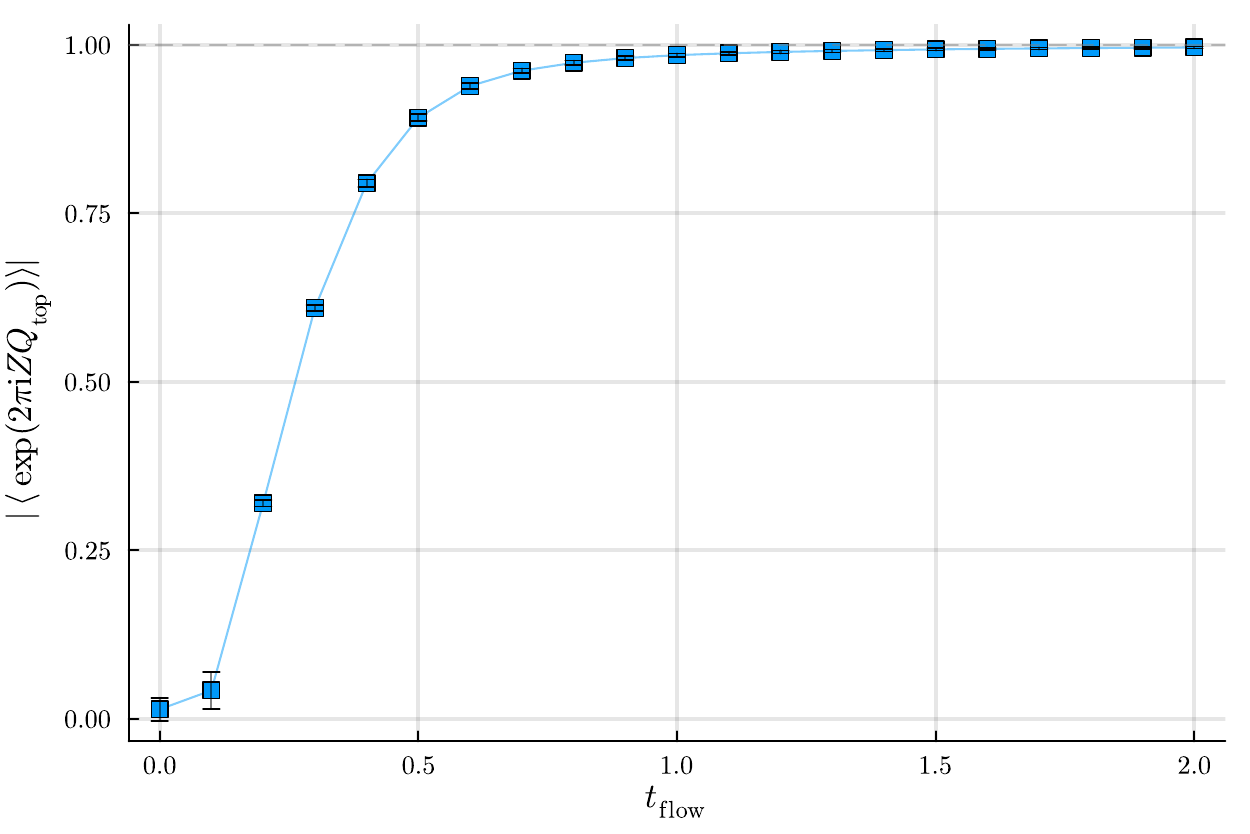}
    \includegraphics[width=0.49\linewidth]{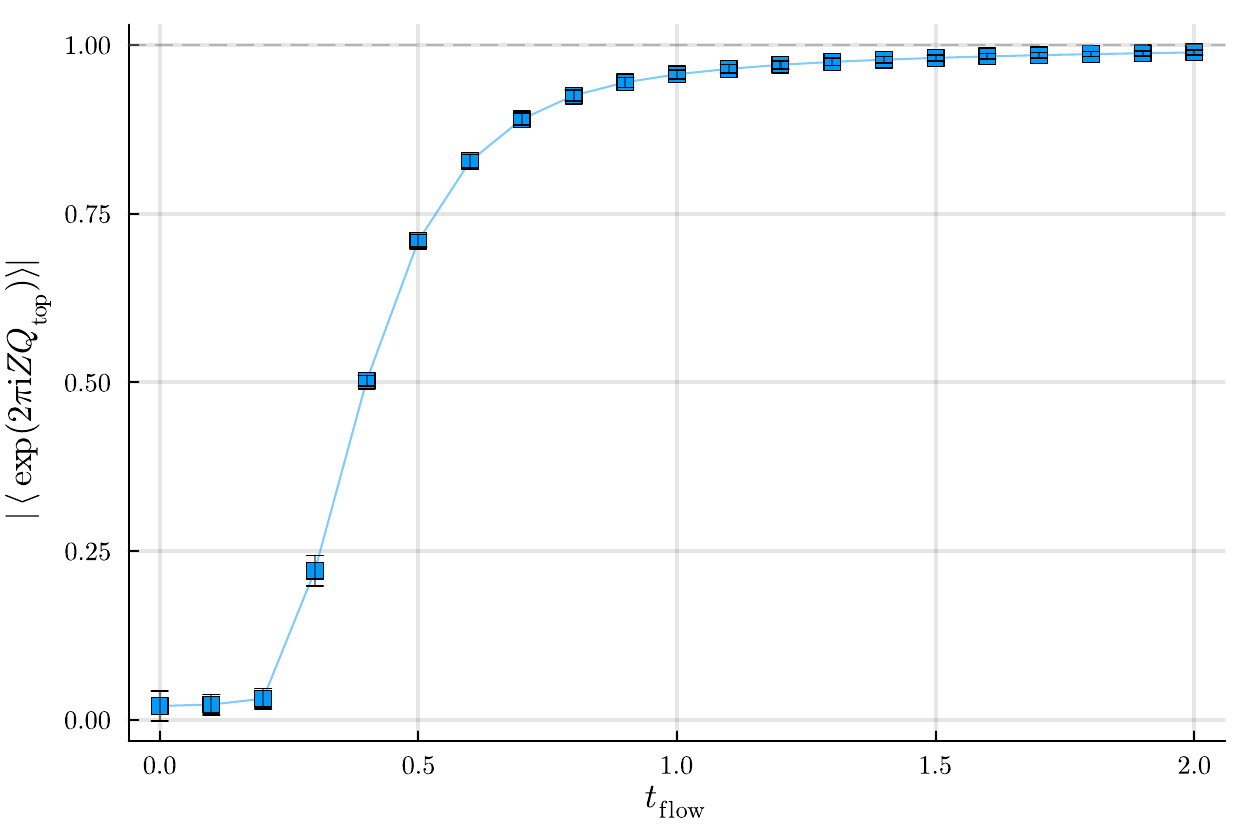}
    \caption{
    Absolute values of the reweighting factors, $\left|\mathcal{Z}_{\theta=2\pi}/\mathcal{Z}_{\theta=0}\right|$, for the $12^3 \times 8$ lattice (left) and the $16^3 \times 10$ lattice (right).
    }
    \label{fig:reweighting_factor}
\end{figure}

\begin{figure}[t]
    \centering
    \includegraphics[width=0.32\linewidth]{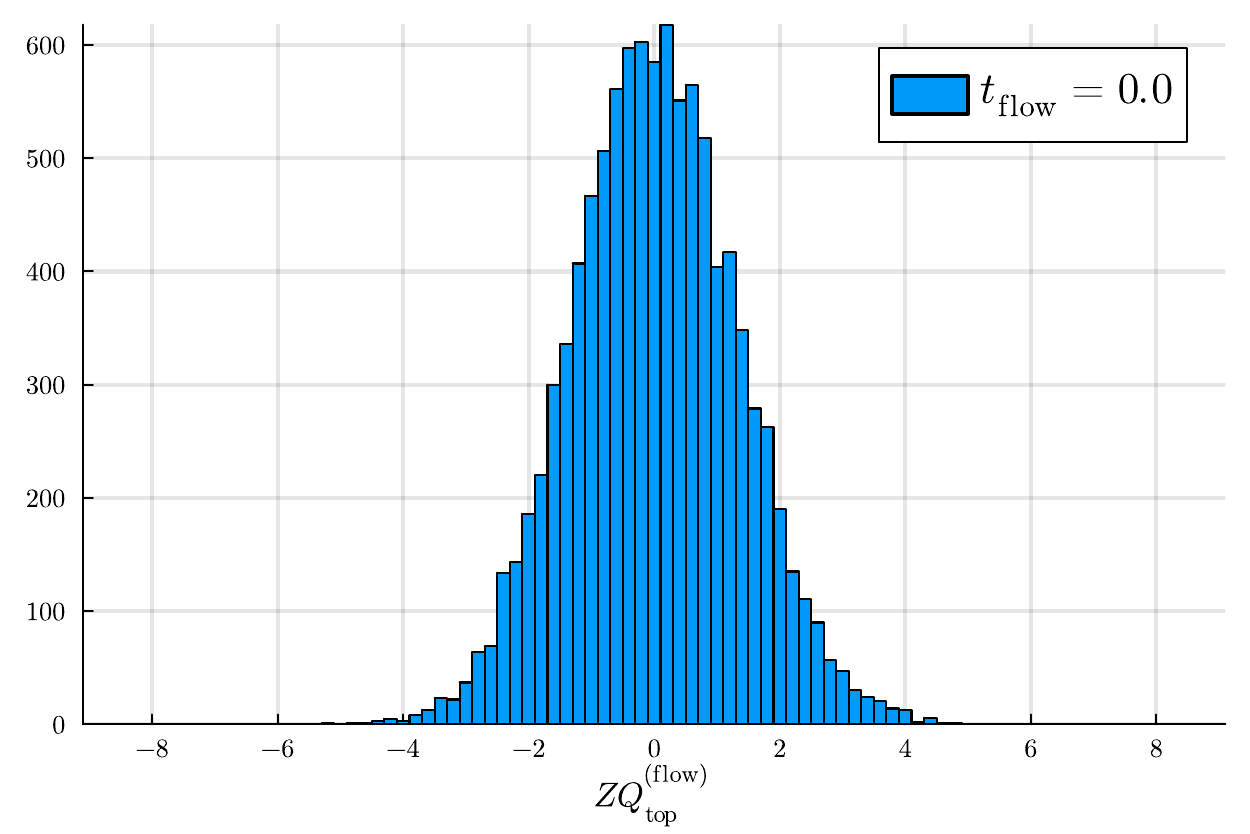}
    \includegraphics[width=0.32\linewidth]{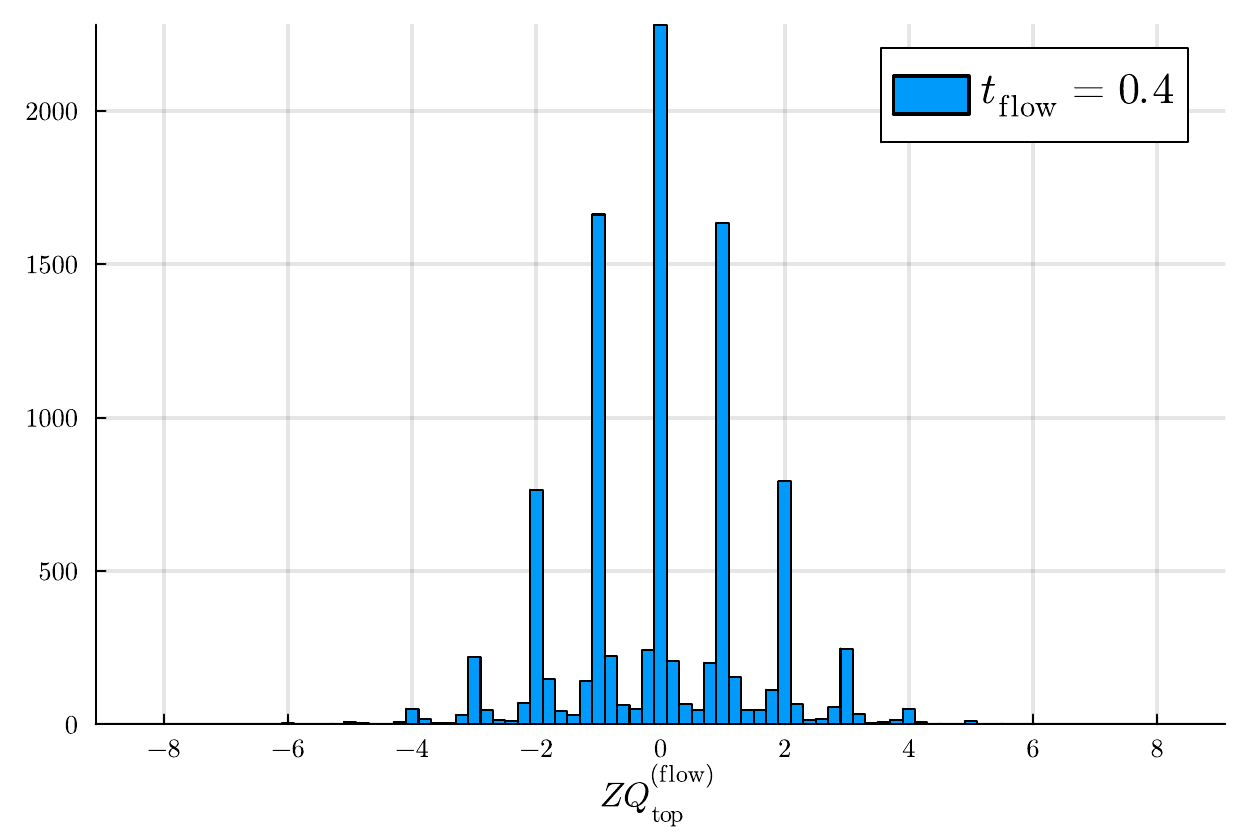}
    \includegraphics[width=0.32\linewidth]{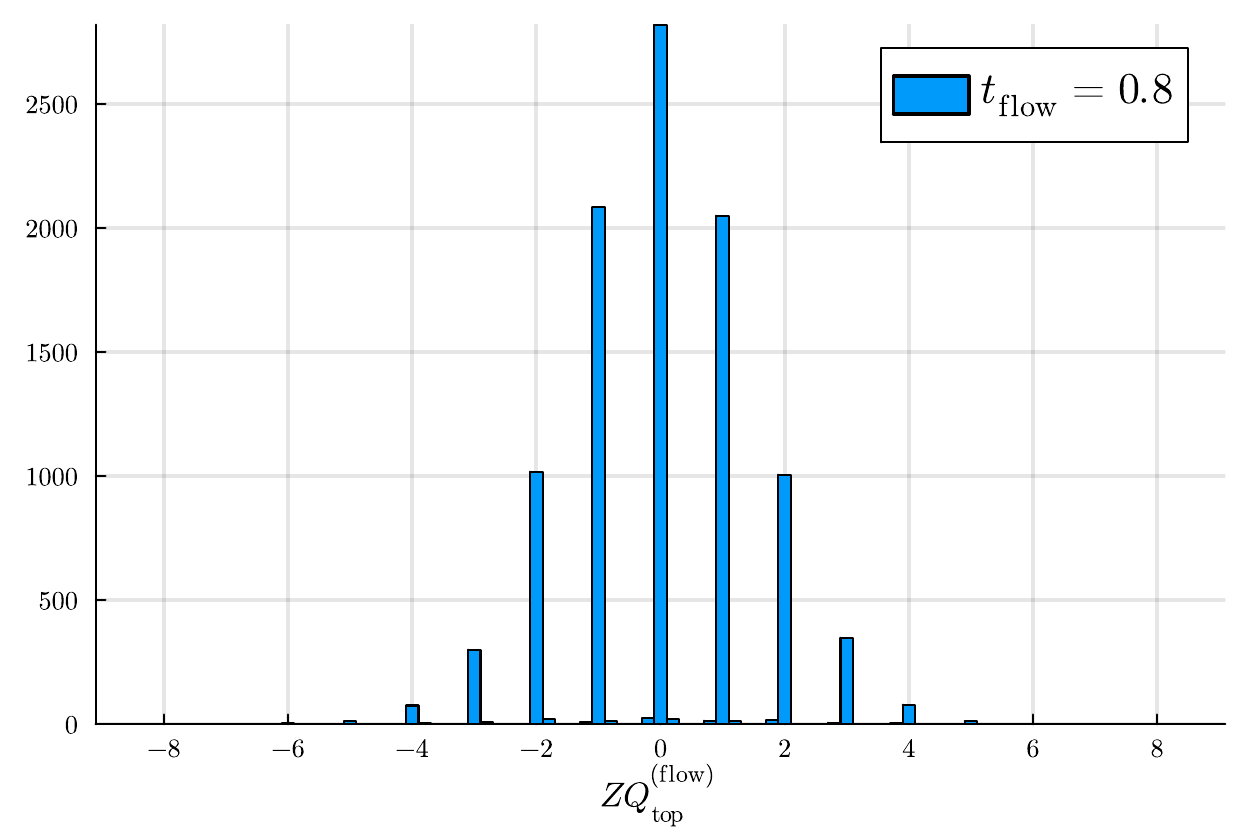}
    \caption{Histograms of $Z\Qtop^{(\tflow)}$ of the $12^3\times 8$ lattice at flow times $\tflow = 0$, $0.4$, and $0.8$. Rapid convergence toward the integer-quantized topological sectors is observed due to the DBW2 flow. }
    \label{fig:hist_ZQ_woB_12x8}
\end{figure}

\begin{figure}[t]
    \centering
    \includegraphics[width=0.32\linewidth]{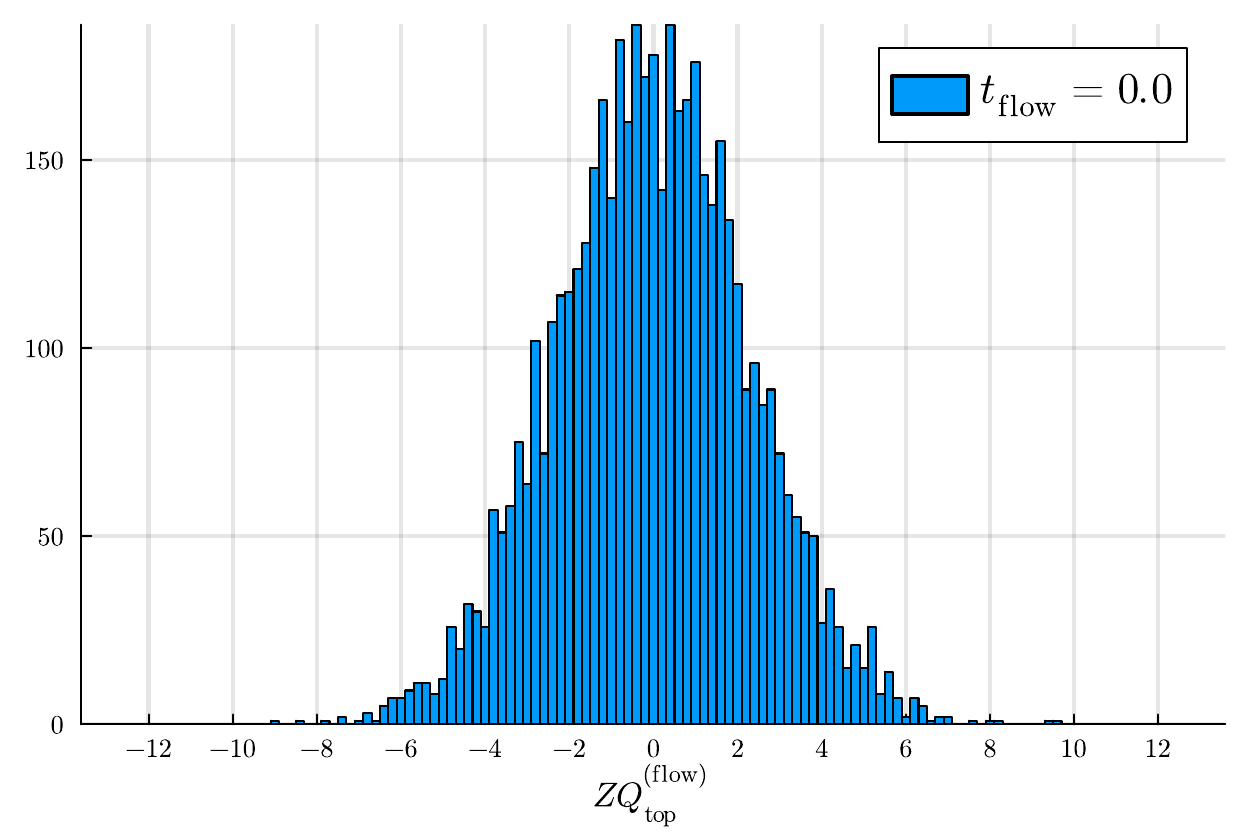}
    \includegraphics[width=0.32\linewidth]{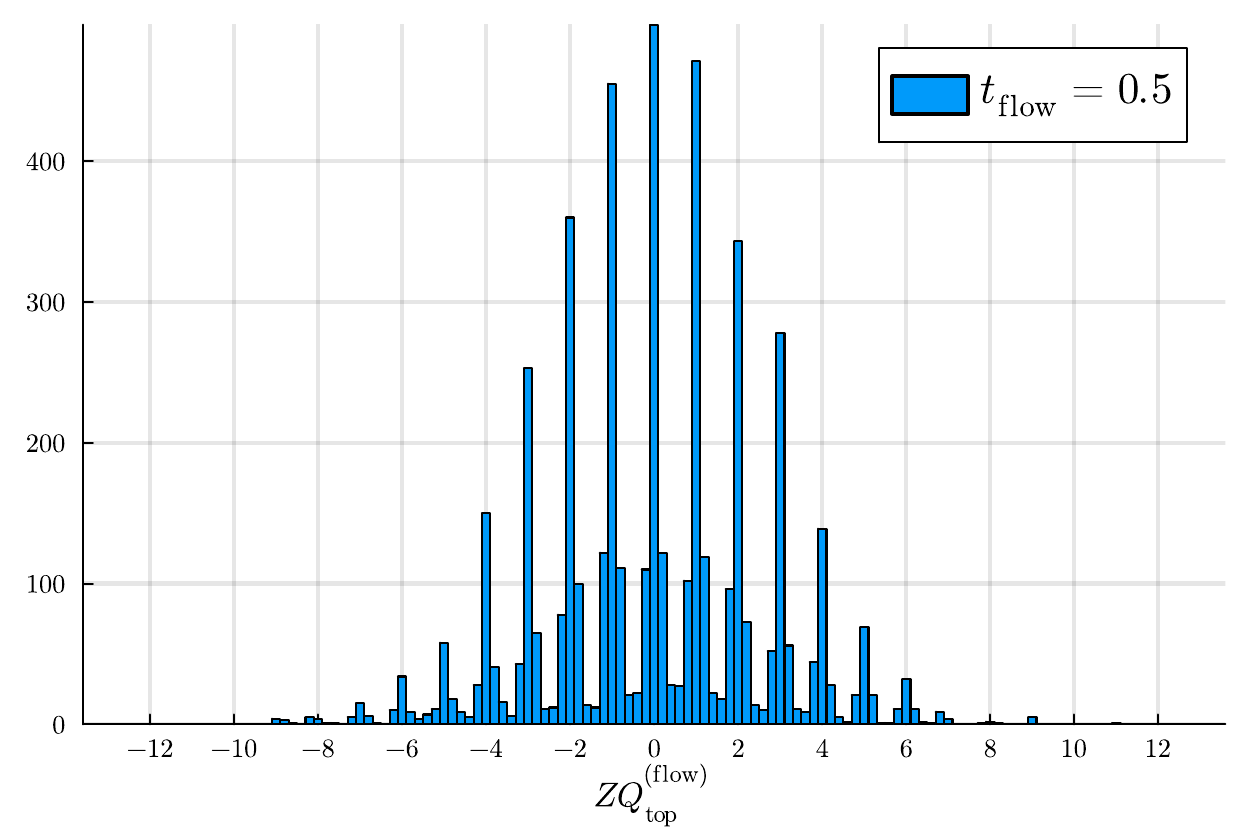}
    \includegraphics[width=0.32\linewidth]{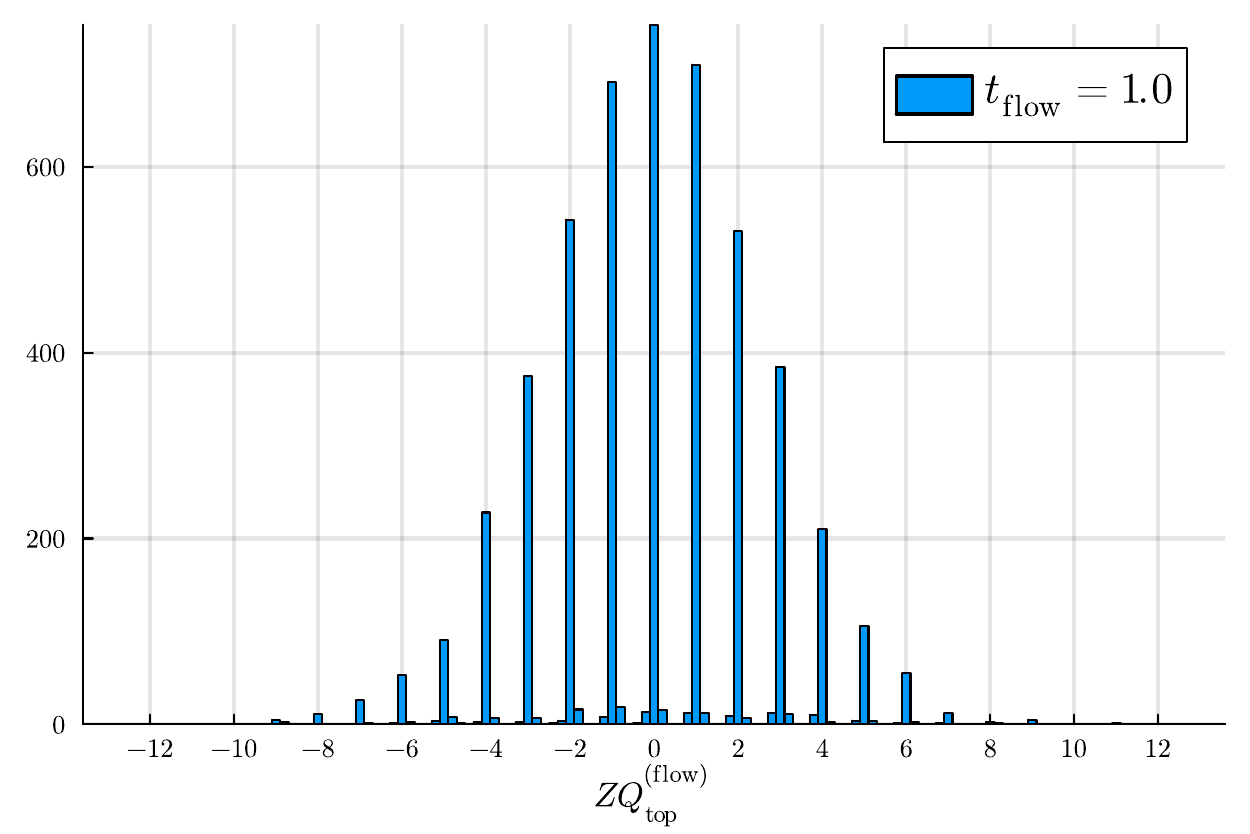}
    \caption{Histograms of $Z\Qtop^{(\tflow)}$ of the $16^3\times 10$ lattice at flow times $\tflow = 0$, $0.5$, and $1.0$.}
    \label{fig:hist_ZQ_woB_16x10}
\end{figure}

We begin with computing the rescaling factor $Z$ through \eqref{eq:eval_Z_factor}.
Figure~\ref{fig:Z_factor} represents the factors at each flow time for both lattice sizes.
The blow-up of the error in the small-flow-time region originates from the insufficient integer-quantized behavior of the topological charge, which leads to an inaccurate determination of the factor.
The factor decreases toward unity along the flow, with statistical errors becoming smaller, and the results for the $12^3\times 8$ and $16^3\times 10$ lattices turn out to be almost identical to each other for $\tflow\gtrsim 0.5$.

Figure~\ref{fig:reweighting_factor} represents the absolute value of the reweighting factor \eqref{eq:ReweightingFactor}, $|\calZ_{2\pi}/\calZ_0|$, 
and they increase and approach unity within short flow times via the DBW2 flow. 
This is a consequence of the recovery of the integer-valued nature of the topological charge after smearing: the topological charge computed by the unsmeared configurations is contaminated by UV fluctuation, and phase cancellations among the non-integer samples decrease the expectation value of the reweighting factor.
Since gradient flow removes such fluctuations gradually, many configurations acquire nearly integer-valued topological charge. 
The recovery of integer quantization under the DBW2 flow can be seen in Fig.~\ref{fig:hist_ZQ_woB_12x8} for the $12^3\times 8$ lattice and Fig.~\ref{fig:hist_ZQ_woB_16x10} for the $16^3\times 10$ lattice. 
For both cases, the DBW2 flow stabilizes the topological sector rapidly, and the topological charge on most configurations has approached an integer value at flow time $\tflow \approx 1.0$.



\subsection{'t Hooft Loop at \texorpdfstring{$\theta=2\pi$}{theta=2pi}}

\begin{figure}[t]
    \centering
    \includegraphics[width=0.32\linewidth]{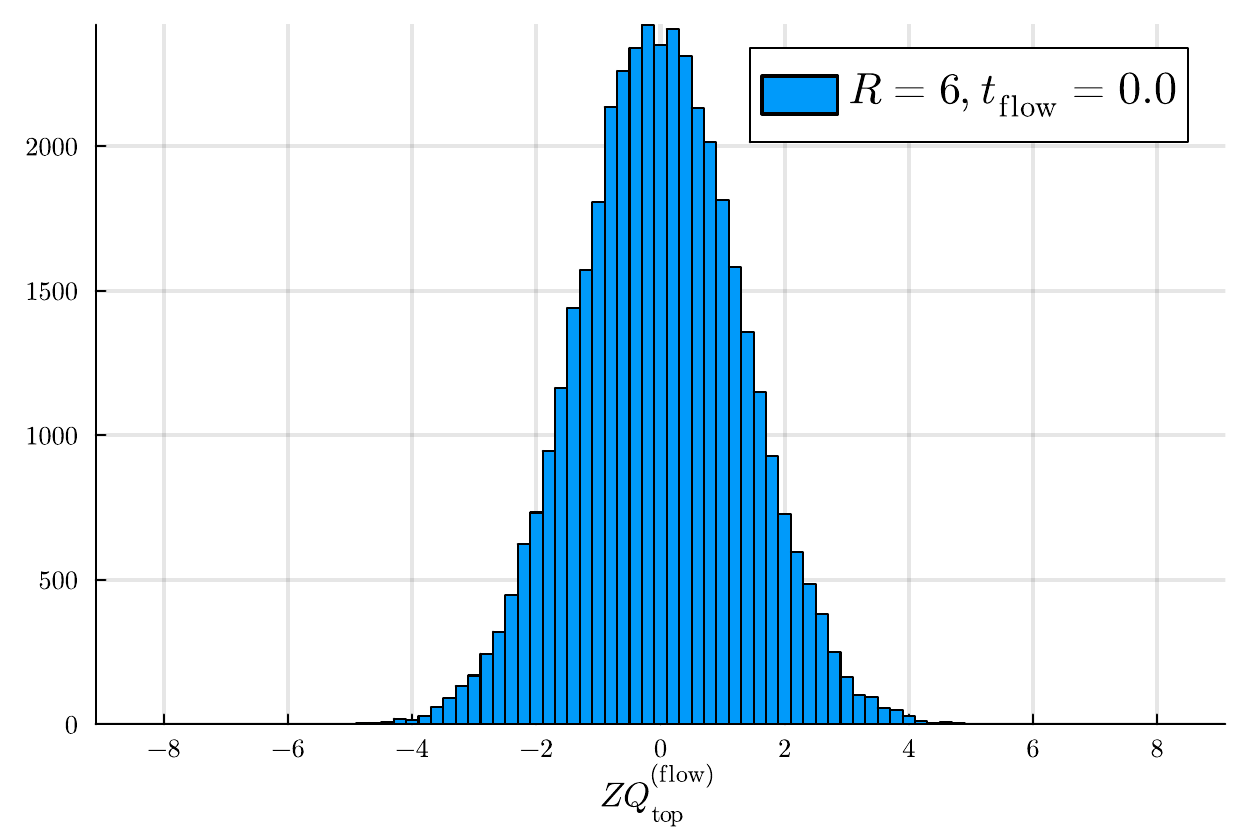}
    \includegraphics[width=0.32\linewidth]{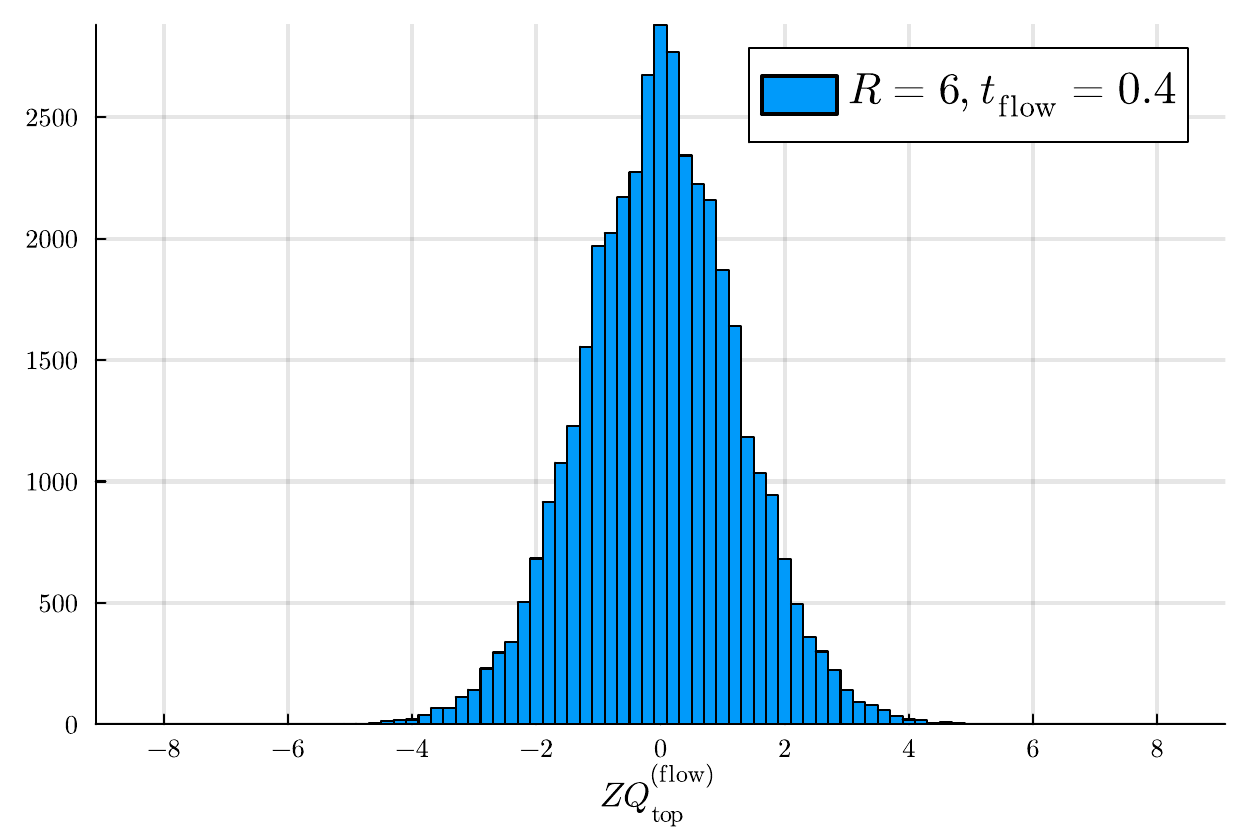}
    \includegraphics[width=0.32\linewidth]{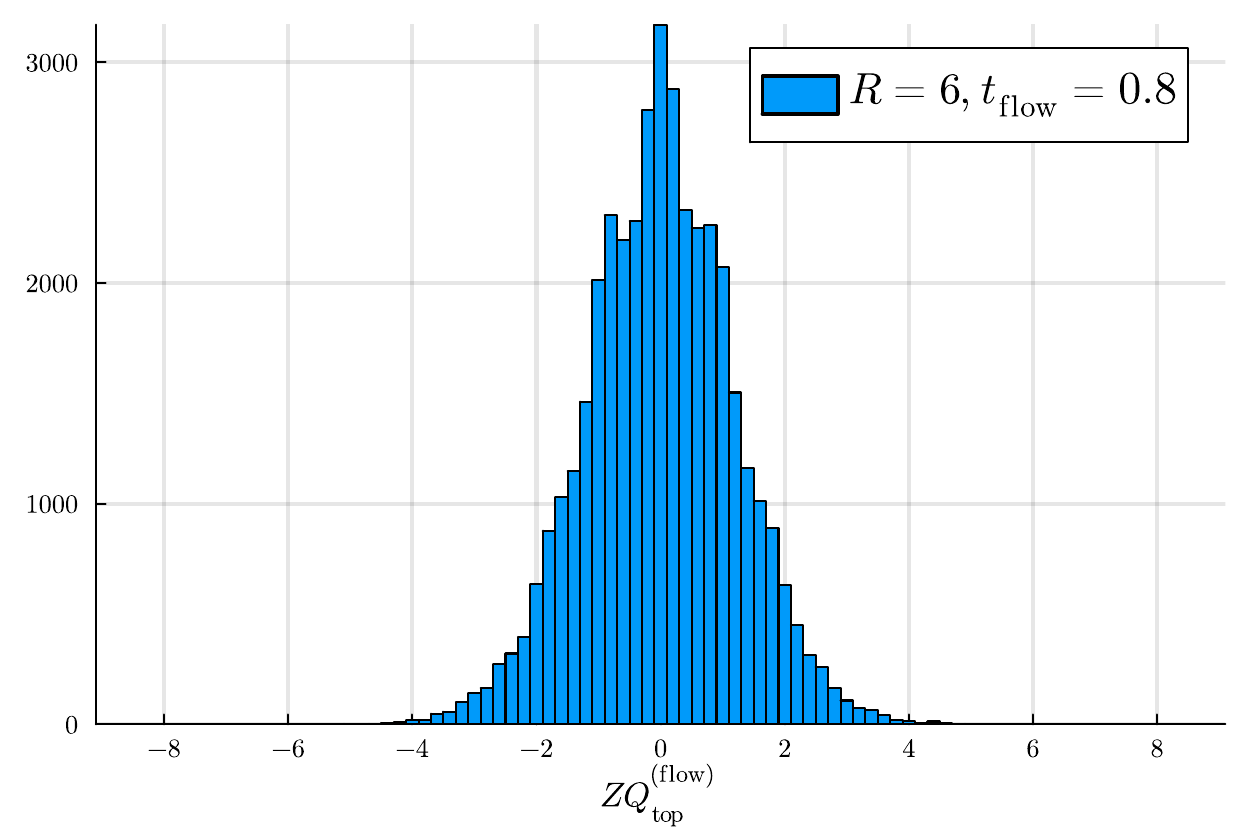}
    \caption{Histograms of $Z\Qtop^{(\tflow)}$ of the $12^3\times 8$ lattice at flow times $\tflow = 0$, $0.4$, and $0.8$.
    The spatial size of the inserted 't~Hooft loop is $R=\Ns/2=6$. 
    In the presence of 't~Hooft loops, topological charges no longer quantize to integers.}
    \label{fig:hist_ZQtop_12x8}
\end{figure}

In this section, we show the numerical results of the 't~Hooft loop expectation value at $\theta=2\pi$.
A significant modification by introducing the non-flat background field $B$ is that topological charges no longer enjoy any type of quantization:
Figure~\ref{fig:hist_ZQtop_12x8} represents the histograms of $Z\Qtop^{(\tflow)}[U_\ell,B_p]$ for the $12^3\times 8$ at several flow times with the insertion of the 't~Hooft loop with $R=\Ns/2$ as a typical example. 
One can explicitly see the absence of quantization for $Z\Qtop^{(\tflow)}$, which makes the expectation values of 't~Hooft loops at $\theta=2\pi$ much smaller than the ones at $\theta=0$. 
Our question is if it leads to the area law of the 't~Hooft loop at $\theta=2\pi$ instead of the perimeter law as expected in Table~\ref{tab:W-tH_classification_SU2}. 

As discussed in Secs.~\ref{sec:tHooftLoop} and~\ref{sec:setup}, we concentrate on evaluating the expectation value~\eqref{eq:H_MC}, 
$\vsup{\left\langle \exp(2\pi\im Z\Qtop^{(\tflow)}[U_\ell, \delta[\tilde{\Sigma}]])\right\rangle}{MC}_{\SW[U_\ell, \delta[\tilde{\Sigma}]]}$. 
For convenience, we compute the expectation values of its real and imaginary parts separately.
The $R$ dependence of the real part determines if the 't~Hooft loop at $\theta=2\pi$ obeys the area or perimeter law. 
On the other hand, the expectation value of the imaginary part should vanish since the distribution of topological charge is symmetric around the origin for our reflection-symmetric setups, which can be used as the consistency (or sanity) check of our results. 

\begin{figure}[t]
    \centering
    \includegraphics[width=0.49\linewidth]{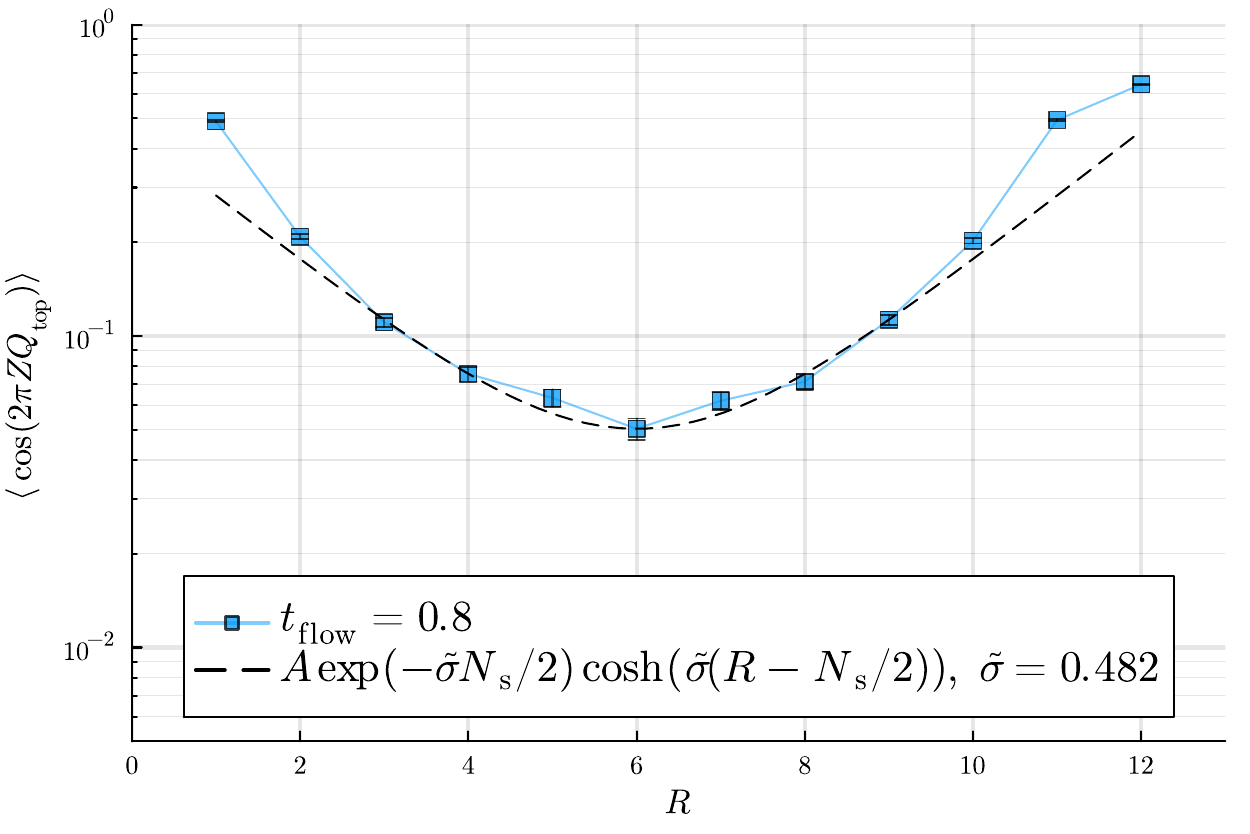}
    \includegraphics[width=0.49\linewidth]{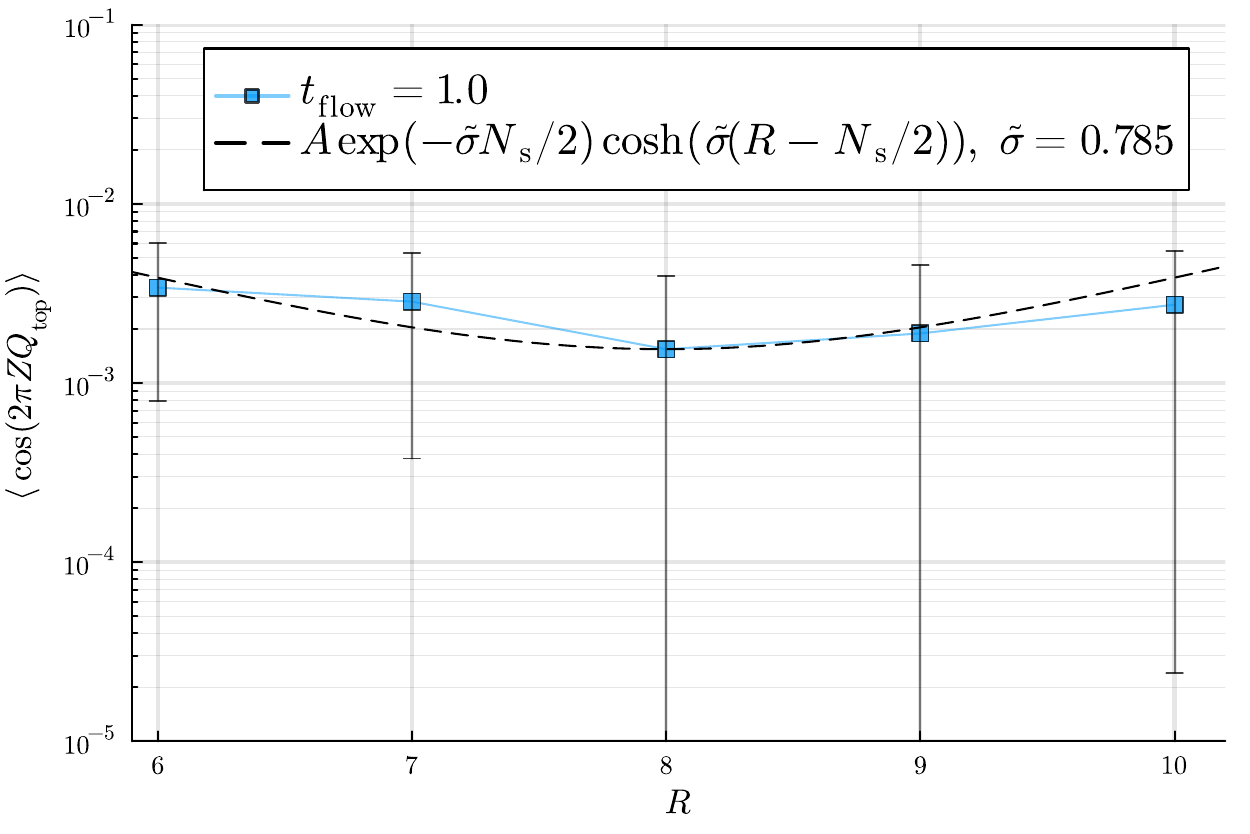}
    \caption{ 
    The real part of \eqref{eq:H_MC}, $\ev{\cos(2\pi Z\Qtop)}$, for the $12^3\times 8$ lattice (left) and the $16^3\times 10$ lattice (right). 
    The dashed curves show $A'\cosh[\tilde\sigma(R-\Ns/2)]$ and are included only as visual guides for comparison with typical area-law behavior. They are not fitting curves: $\tilde\sigma$ is determined from Polyakov-loop correlators for each lattice setup, while the overall coefficient $A'$ is fixed so that the curve matches the central value at $R=\Ns/2$.
    }
    \label{fig:H_theta=2pi}
\end{figure}

\begin{figure}[t]
    \centering
    \includegraphics[width=0.32\linewidth]{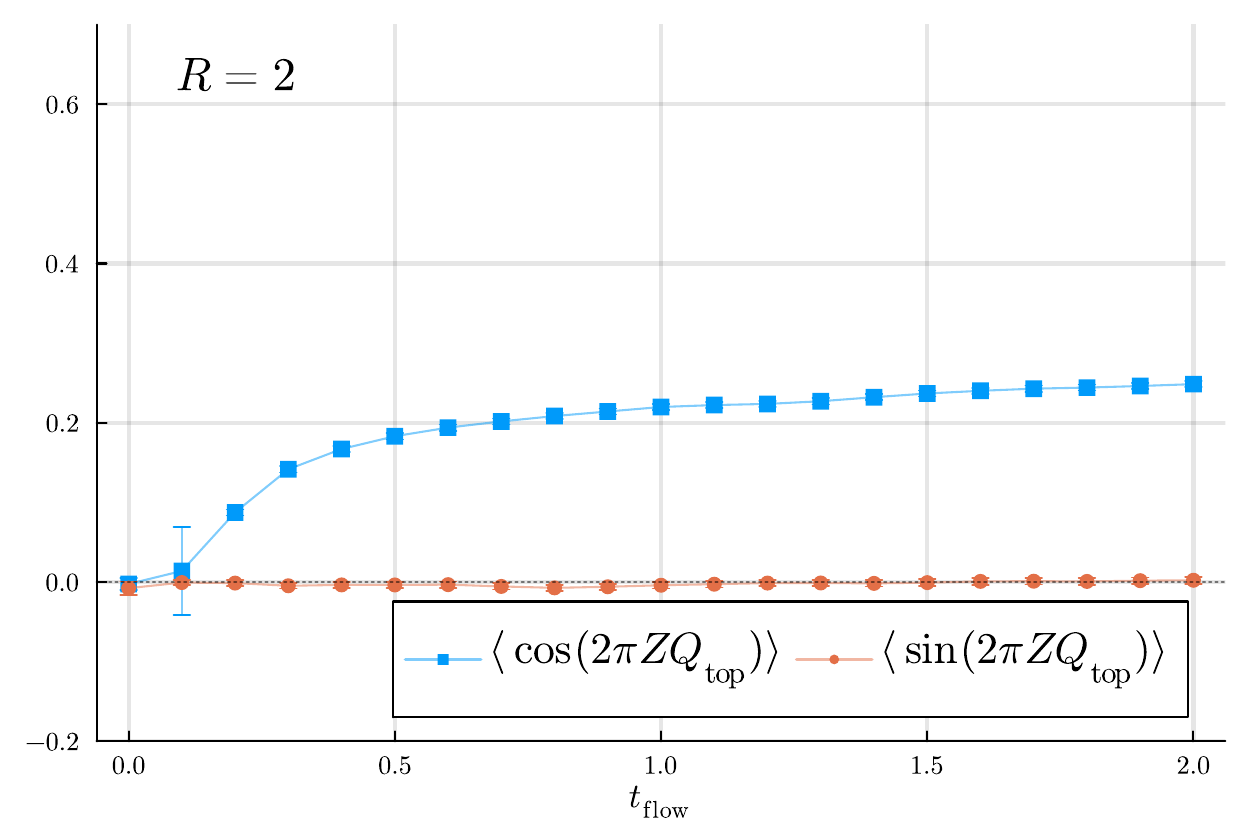}
    \includegraphics[width=0.32\linewidth]{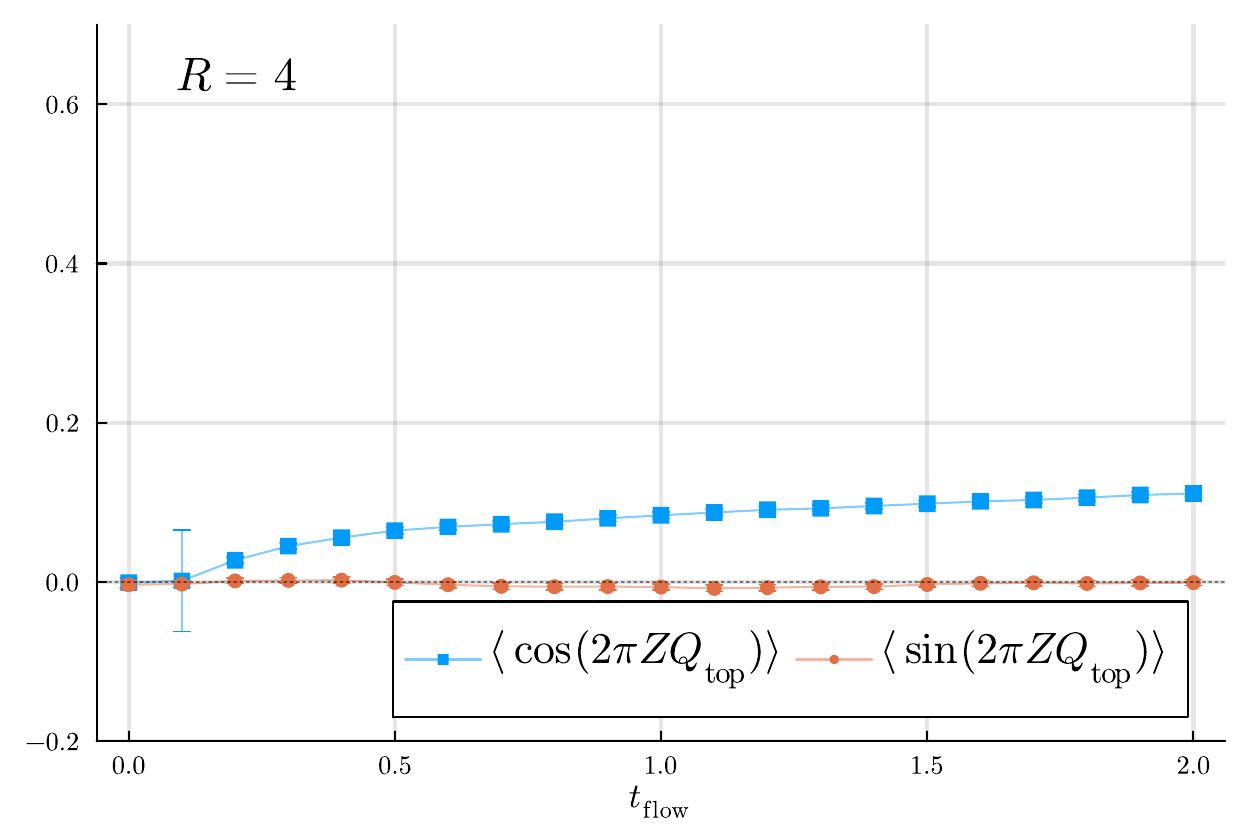}
    \includegraphics[width=0.32\linewidth]{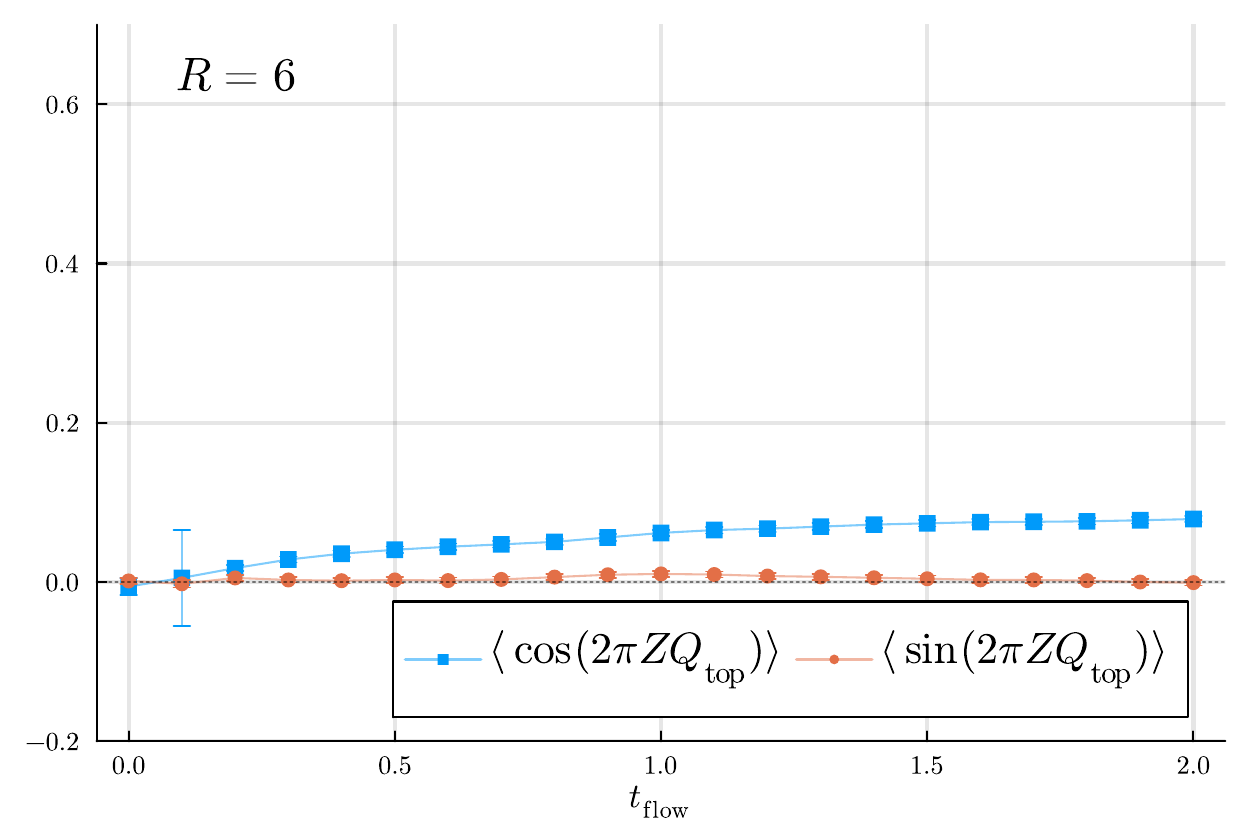}
    \\
    \includegraphics[width=0.32\linewidth]{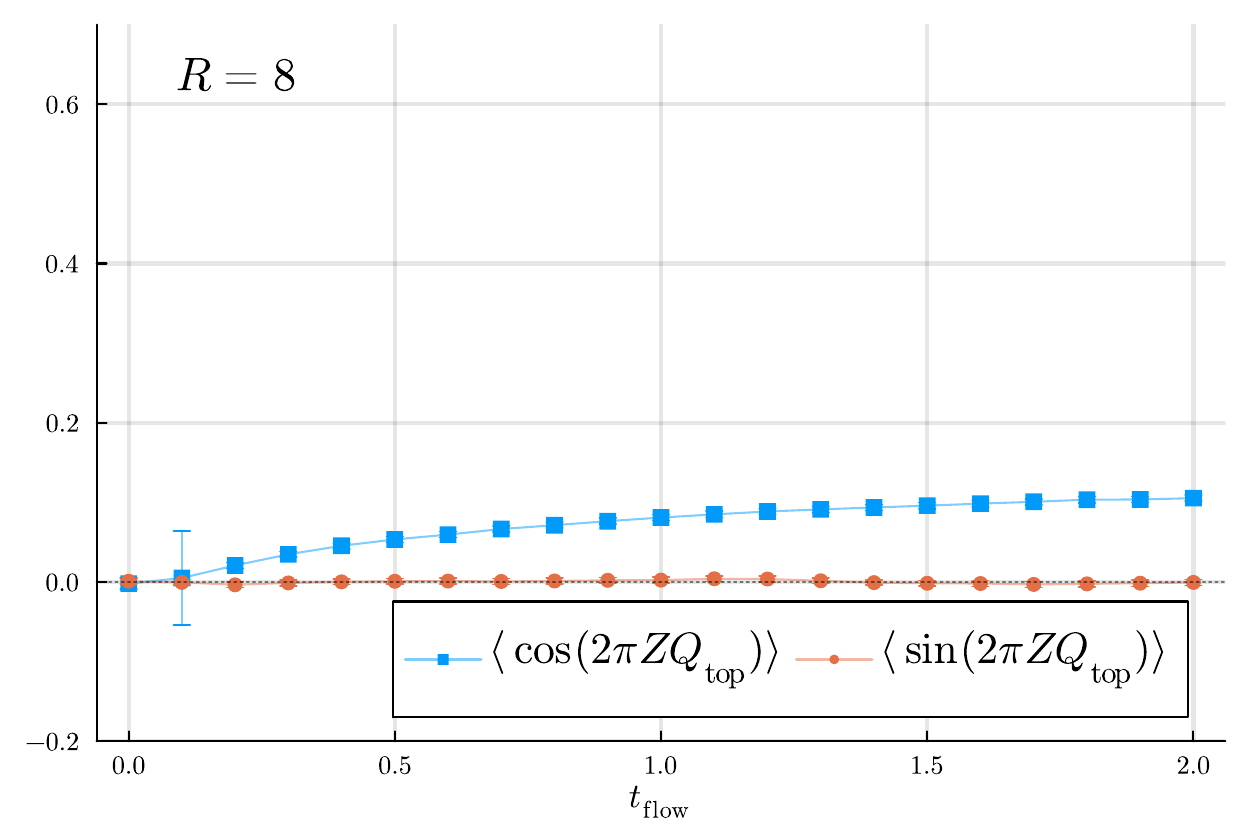}
    \includegraphics[width=0.32\linewidth]{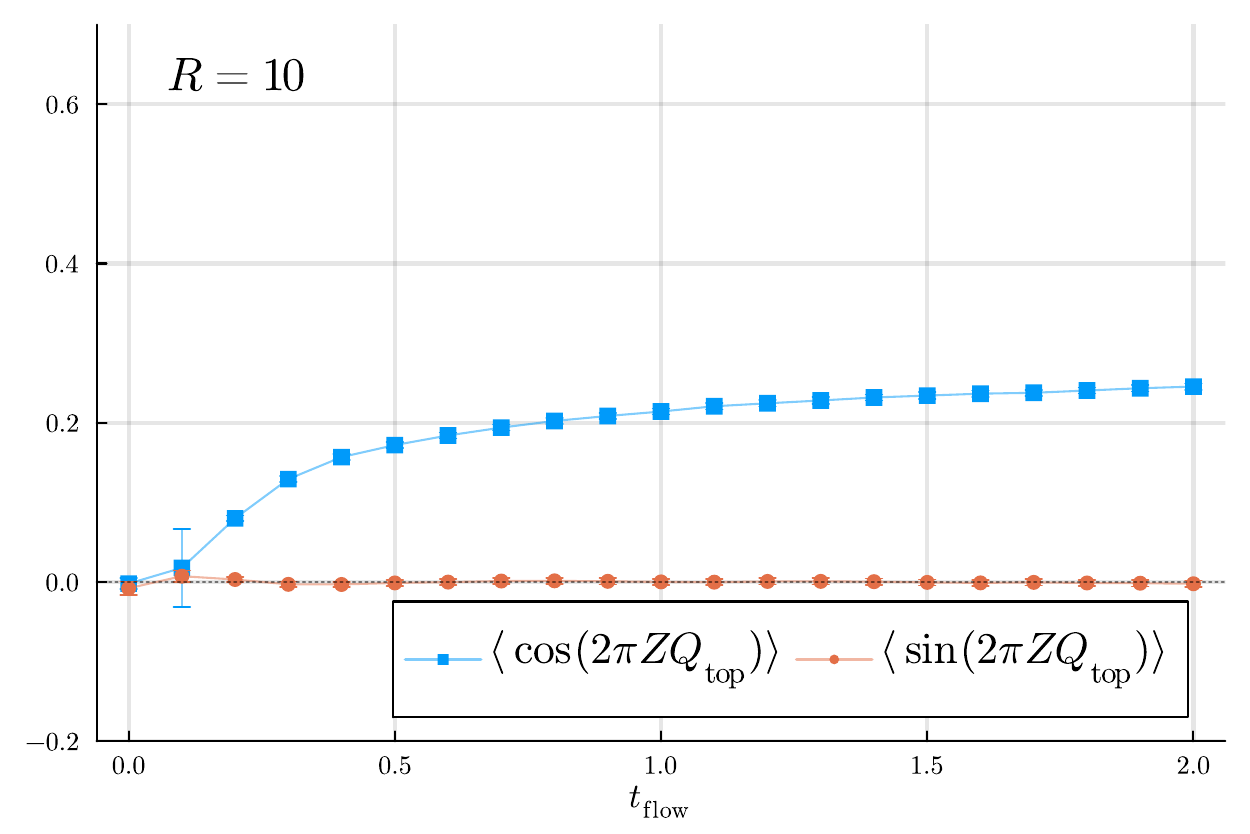}
    \includegraphics[width=0.32\linewidth]{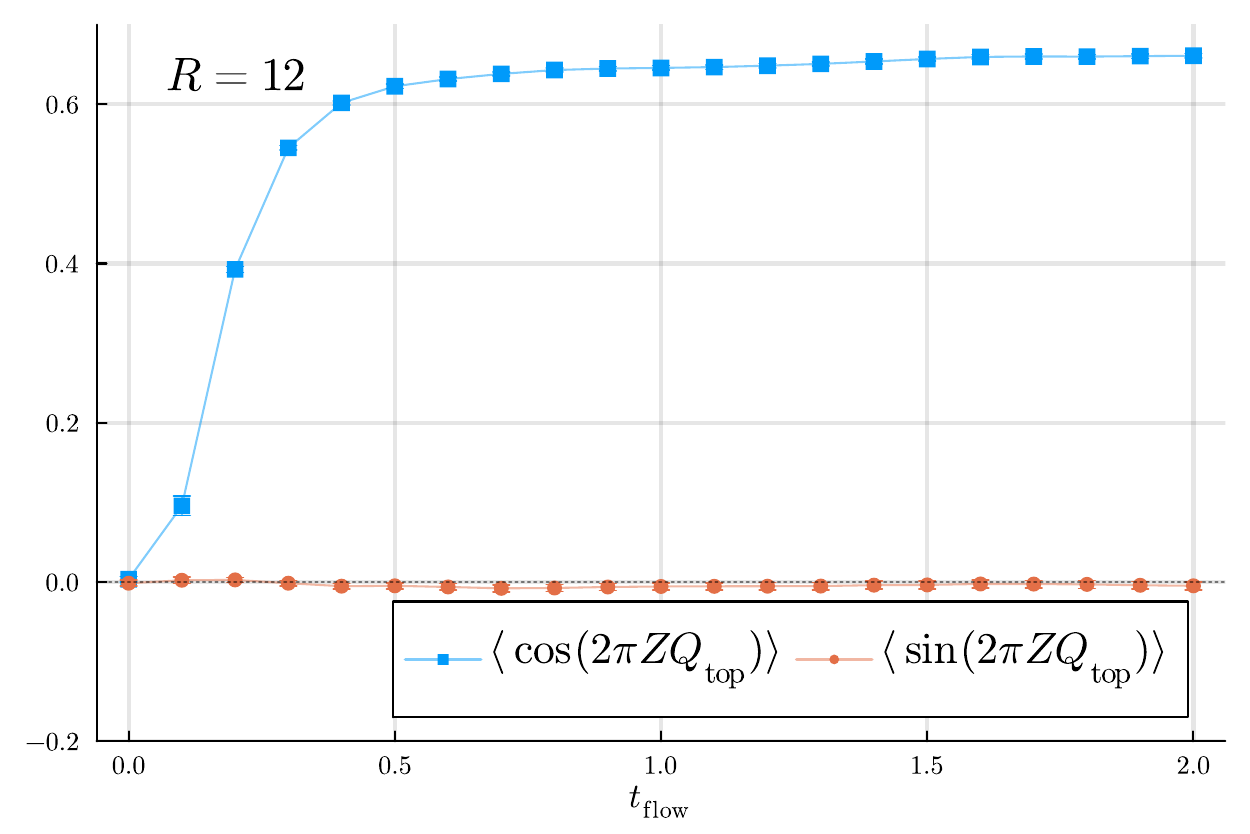}
    \caption{
    Flow-time dependence of the real and imaginary parts of \eqref{eq:H_MC} with $R=2$, $4$, \dots, $12$ on the $12^3\times 8$ lattice.
    }
    \label{fig:flowtime-dep_12x8_H_theta=2pi}
\end{figure}

\begin{figure}[t]
    \centering
    \includegraphics[width=0.32\linewidth]{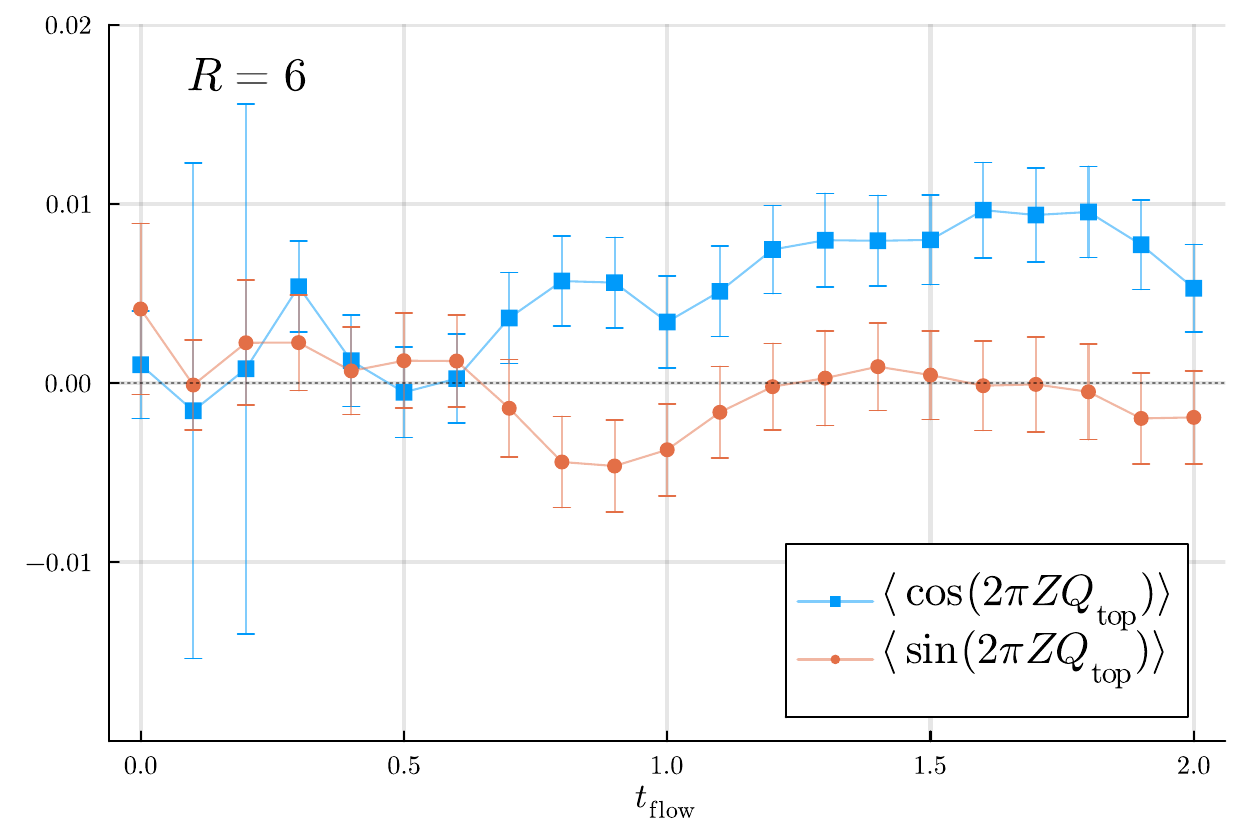}
    \includegraphics[width=0.32\linewidth]{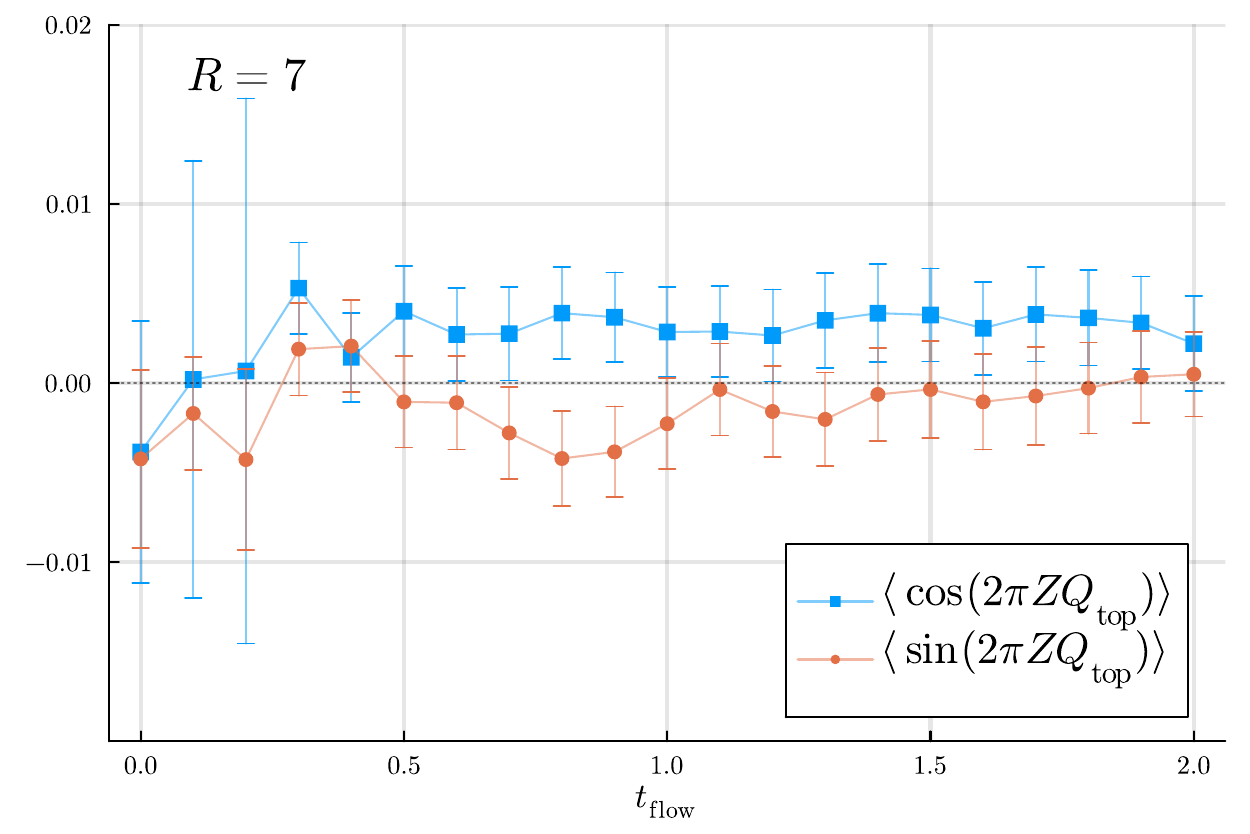}
    \includegraphics[width=0.32\linewidth]{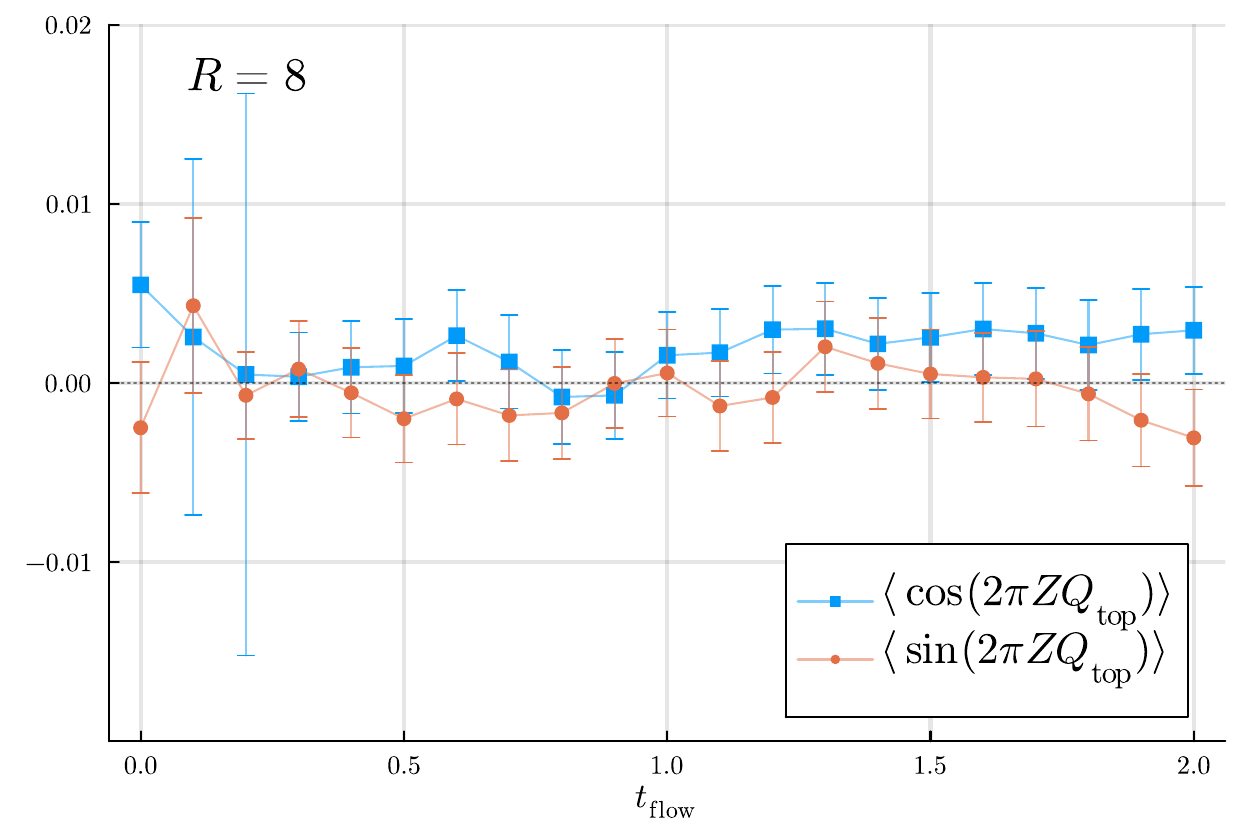}
    \\
    \includegraphics[width=0.32\linewidth]{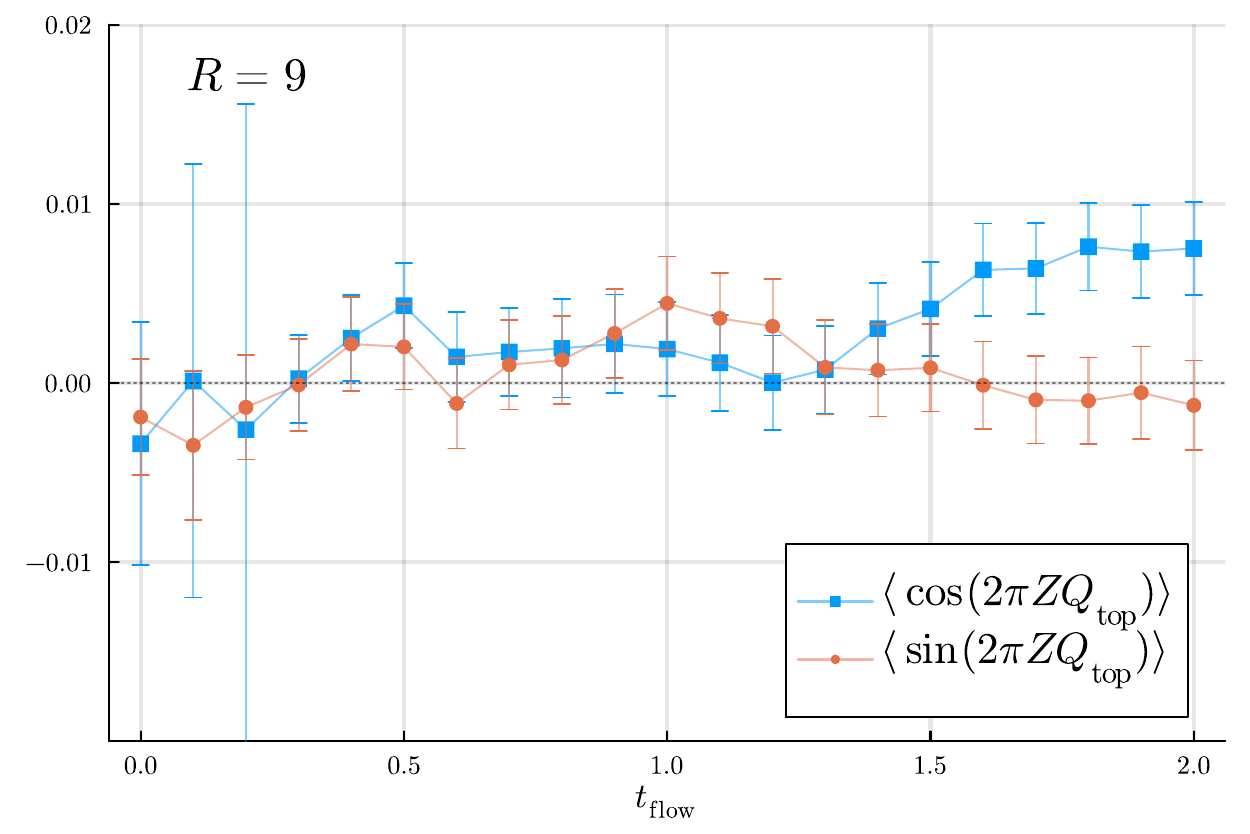}
    \includegraphics[width=0.32\linewidth]{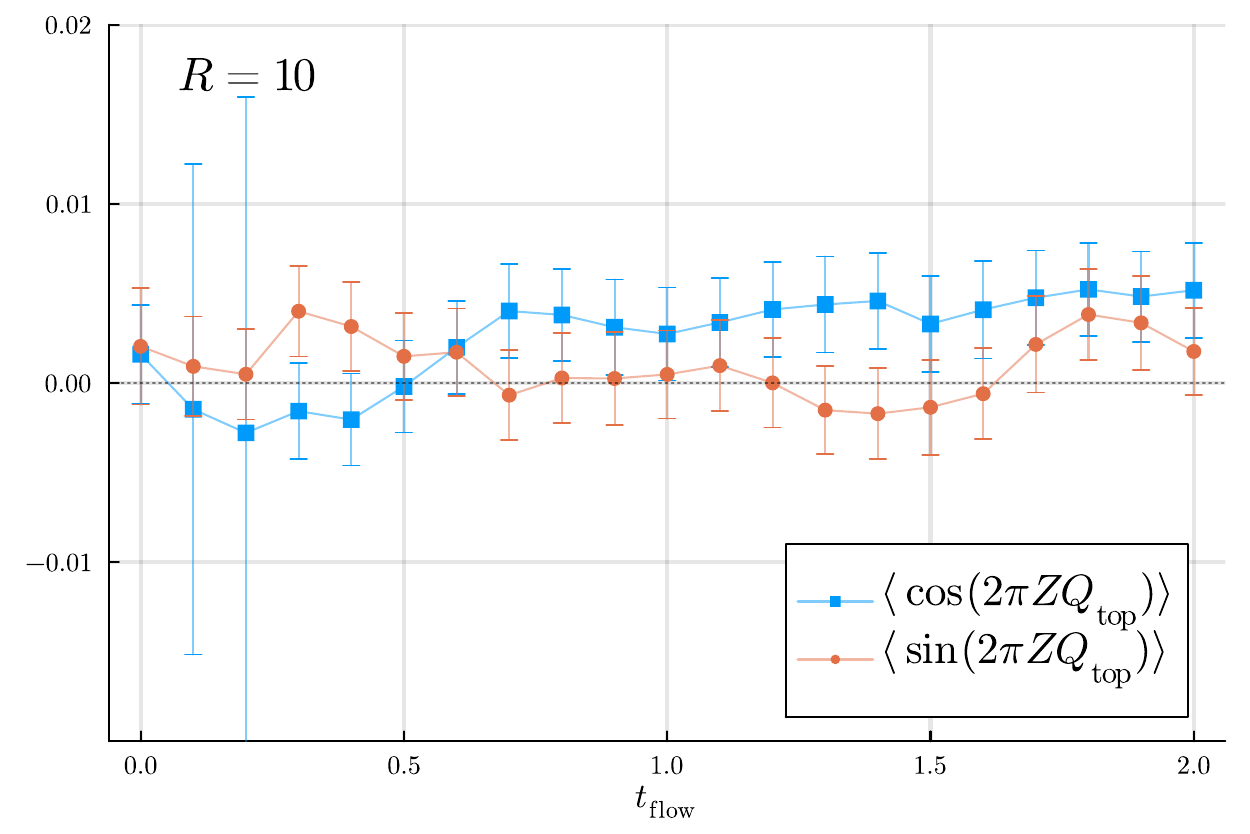}
    \caption{
    Flow-time dependence of the real and imaginary parts of \eqref{eq:H_MC} with $R=6$, \dots, $10$ on the $16^3\times 10$ lattice.
    }
    \label{fig:flowtime-dep_16x10_H_theta=2pi}
\end{figure}

Figure~\ref{fig:H_theta=2pi} plots the $R$ dependence of the real part of \eqref{eq:H_MC} for the $12^3 \times 8$ and $16^3\times 10$ lattices, with the log-scaled vertical axis. 
We also plot the flow-time dependence of the real and imaginary parts (with the linear scale) in Figs.~\ref{fig:flowtime-dep_12x8_H_theta=2pi} and \ref{fig:flowtime-dep_16x10_H_theta=2pi}, for the $12^3\times 8$ and $16^3\times 10$ lattices, respectively. 
The imaginary parts shown by orange data fluctuate around zero within roughly $2$ standard deviation, which reasonably passes the above consistency check. 
Note that the error bars are estimated by the bootstrap method, containing the statistical error of $Z$.

If the 't~Hooft loop at $\theta=2\pi$ shows the area law, its $R$ dependence for $2d \ll R\ll \Ns-2d$ (here $d\approx \sqrt{8 \tflow}$) is expected to be described by 
\begin{equation}
    A\left[\exp(-\tilde{\sigma} R)+\exp(-\tilde{\sigma}(\Ns-R))\right]=2A \exp\left(-\tilde{\sigma} \frac{\Ns}{2}\right)\cosh\left[\tilde{\sigma}\left(R-\frac{\Ns}{2}\right)\right].
    \label{eq:EyeGuide_AreaLaw}
\end{equation}
The discussion based on the Witten effect suggests that $\tilde{\sigma}$ is identical to the mass gap for the fundamental Polyakov loop. 
In Fig.~\ref{fig:H_theta=2pi}, we draw it with the dashed curves as the eye guide, and $\tilde{\sigma}$ for each figure is computed from the Polyakov-loop correlators. 
The overall factor $A$ is chosen so that the dashed curve passes through the central value at $R=\Ns/2$.


Let us then discuss if we can judge whether the 't~Hooft loop obeys the area or perimeter law from these results. 
For the $12^3 \times 8$ lattice with the flow time $\tflow = 0.8$, the smearing width is $d=\sqrt{8\tflow}\approx 2.53$ and the separation of test charges requires $2d\approx 5 \ll R\ll 7 \approx \Ns-2d$, which is satisfied only for $R=5,6,7$ even if we take a generous estimate. The effect that line operators overlap each other may not be negligible, and it is difficult to conclude if the 't~Hooft loop obeys the area law while its nontrivial $R$ dependence looks to be consistent with the area law of the theoretically expected string tension.

For the $16^3\times 10$ lattice with $\tflow=1.0$, the requirement of the test-charge separation gives $2d\approx 5.7 \ll R \ll 10.3 \approx \Ns-2d$, which is satisfied for our choice of $R = 6$, \dots, $10$. 
The signal values of $\langle \cos(Z \Qtop)\rangle$ for these $R$'s are roughly $1$--$3\times 10^{-3}$, and their magnitudes are roughly consistent with the predicted values from \eqref{eq:EyeGuide_AreaLaw} with $A=O(1)$ and $\tilde{\sigma}=0.785$, where $\tilde{\sigma}$ is determined from the Polyakov-loop correlators in the same lattice setup. 
However, these signals are contaminated by the same size of error bars. 
Due to non-quantization of $\Qtop$, $\cos(Z \Qtop)$ fluctuates from $-1$ to $1$ for Monte Carlo ensembles. 
Then, its statistical error can be estimated as $O(1)/\sqrt{\#{\rm samples}}$,\footnote{
The variance of the topological charge does not substantially alter the prefactor of this scaling:
Approximating its distribution as the Gaussian one $\exp(-\Qtop^2/v^2)$, then 
$\mathrm{Var}(\cos(2\pi \Qtop)) 
    =
    \langle \cos^2(2\pi \Qtop)\rangle - \langle \cos(2\pi \Qtop) \rangle^2
    = \frac{1}{2} + O(\rme^{-2\pi^2v^2})$,
and the correction term is negligible even when $v=O(1)$.
}
which is about $3.5\times 10^{-3}$ for $80,000$ configurations and consistent with the bootstrap estimate from our data. 
We would need, at least, 10 times more samples to get clear signals in this lattice setup for each $R$.\footnote{This is another difficulty peculiar to defect operators, like 't~Hooft loops. For Wilson loops, we can take the expectation values of different $R$'s from the single configuration, and moreover, the translational average helps a lot with giving a small prefactor for the error estimate. For defect observables, we need to prepare configurations for different $R$'s separately, and it is also not possible to take the translational average as the defect location is fixed at the stage of the configuration generation.} 

One might wonder why the obtained expectation value at $R=\Ns=12$ (shown as the rightmost point in the left panel of Fig.~\ref{fig:H_theta=2pi} and as the blue points in the lower-right panel of Fig.~\ref{fig:flowtime-dep_12x8_H_theta=2pi}) is rather small compared to values with what would be expected if the topological charges were integer-valued. We comment on this issue in Appendix~\ref{app:Qtop_R=12_12x8}.

\subsection{Dyonic Loop at \texorpdfstring{$\theta=2\pi$}{theta=2pi}}


\begin{figure}[t]
    \centering
    \includegraphics[width=0.49\linewidth]{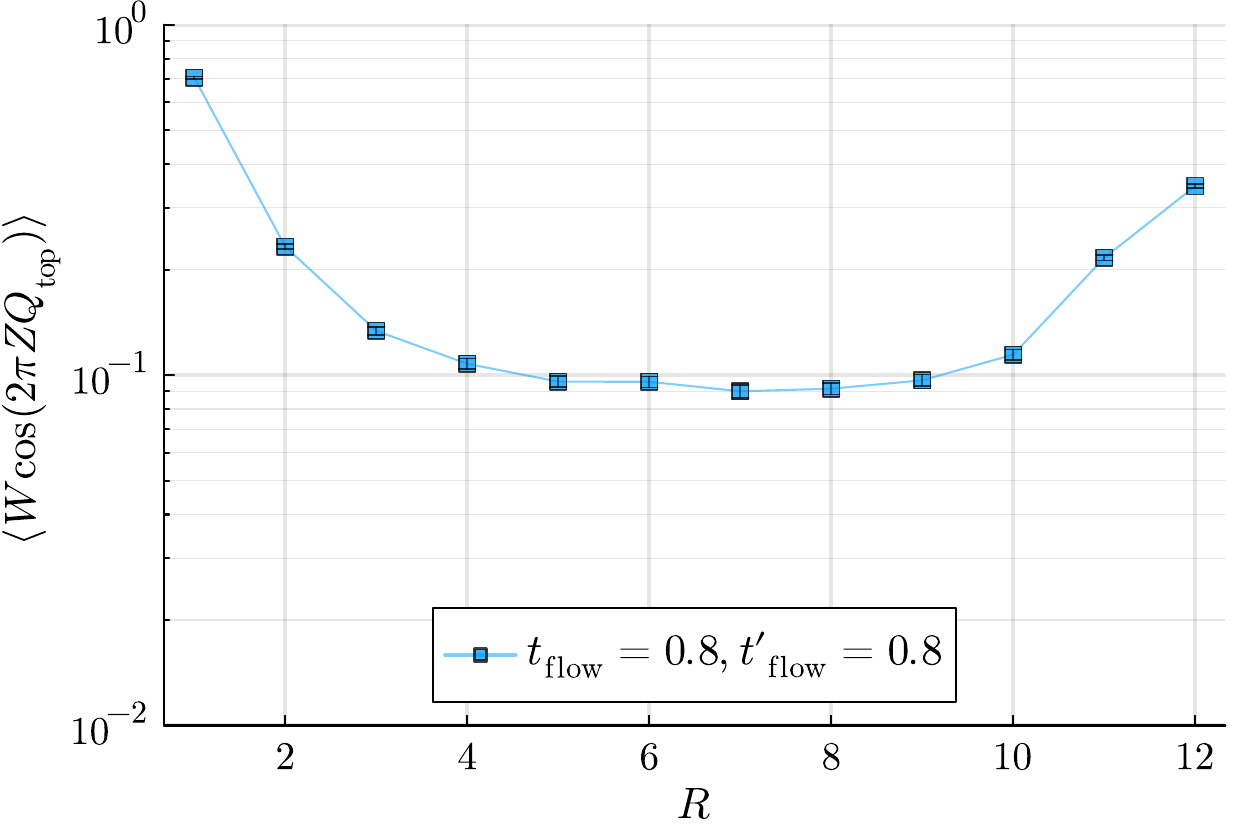}
    \includegraphics[width=0.49\linewidth]{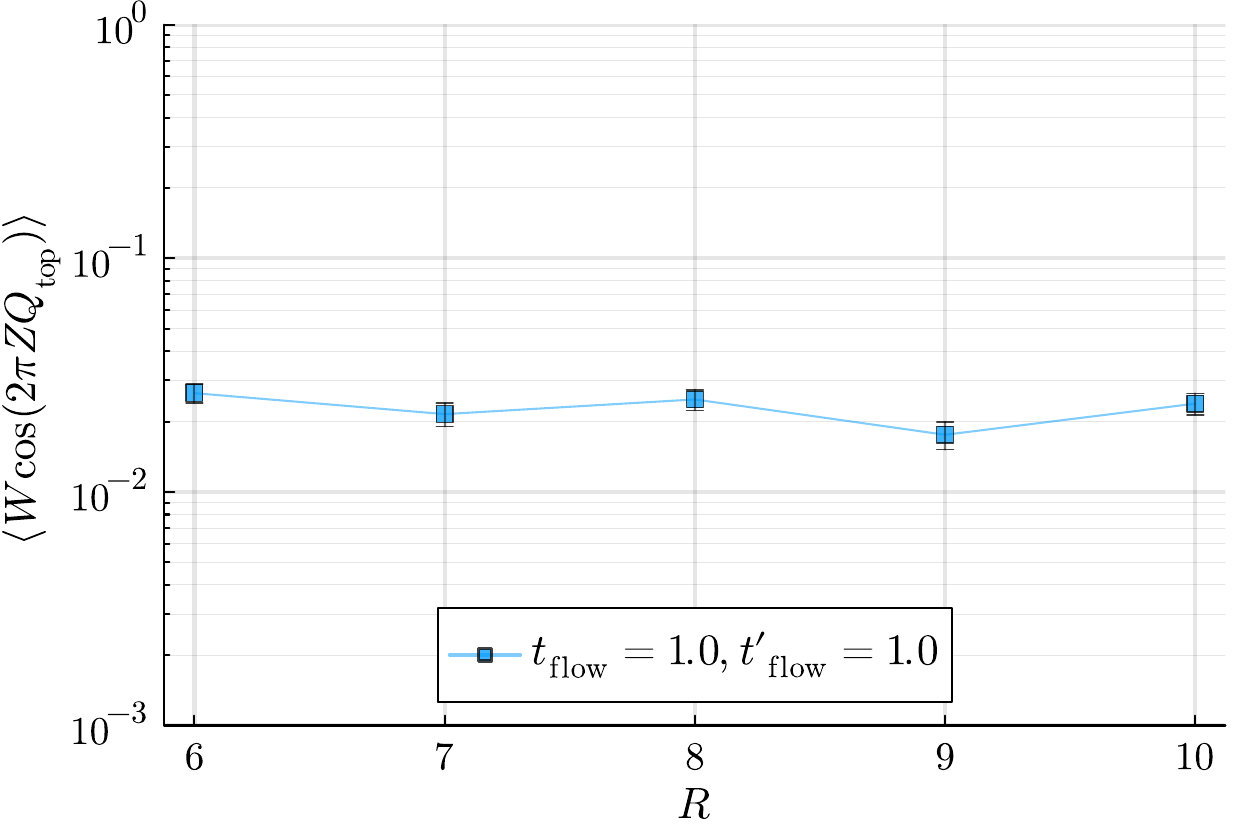}
    \caption{
    The real part of \eqref{eq:HW_MC}, $\ev{W^{(\tflow')}\cos(2\pi Z \Qtop^{(\tflow)})}$, for the $12^3\times 8$ lattice (left) and the $16^3\times 10$ lattice (right). 
    The smeared Wilson loop operators at $\tflow'=\tflow$ are used for these plots.
    }
    \label{fig:HW_theta=2pi}
\end{figure}

We next consider the dyon loop expectation value at $\theta = 2\pi$.
Here, we concentrate on evaluating the factor \eqref{eq:HW_MC}, $\vsup{\left\langle W^{(\tflow')}(C;\vsub{x}{base})\,
    \exp(2\pi\im Z\Qtop^{(\tflow)}[U_\ell, \delta[\tilde{\Sigma}]])\right\rangle}{MC}_{\SW[U_\ell, \delta[\tilde{\Sigma}]]}$.
Figure~\ref{fig:HW_theta=2pi} represents the $R$ dependence of the real part of \eqref{eq:HW_MC} for the $12^3\times 8$ and $16^3\times10$ lattices with choosing $\tflow=\tflow'=0.8$ and $1.0$, respectively.
In contrast to the 't~Hooft loop, the results are insensitive to the spatial extent $R$ for $4\lesssim R \lesssim 10$ in the left panel and the entire analyzed region, $6\le R\le 10$ in the right panel.\footnote{The asymmetry at $R=6$ for the $12^3\times 8$ lattice may be attributed to the fact that our dyon loop is defined by introducing the rectangular Wilson loop instead of using two Polyakov loops. See footnote~\ref{ftnt:def_dyonline}. In Appendix~\ref{app:more_numerical_results}, we show the results of the dyonic lines defined with Polyakov loops for the $12^3\times 8$ lattice, and we observe the consistent behavior for $R\gg 1$. } 
These results give a clear signal for the perimeter law of the dyonic lines at $\theta=2\pi$. 

\begin{figure}[t]
    \centering
    \includegraphics[width=0.32\linewidth]{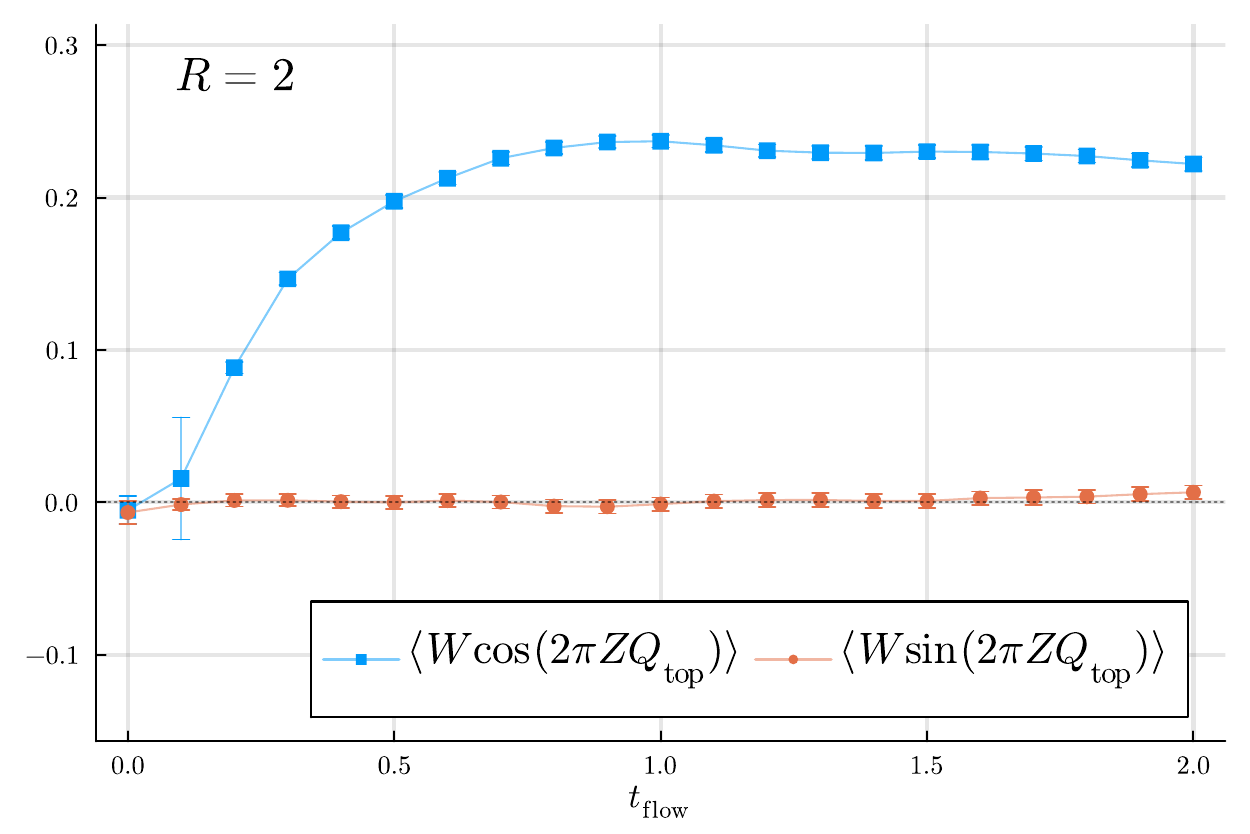}
    \includegraphics[width=0.32\linewidth]{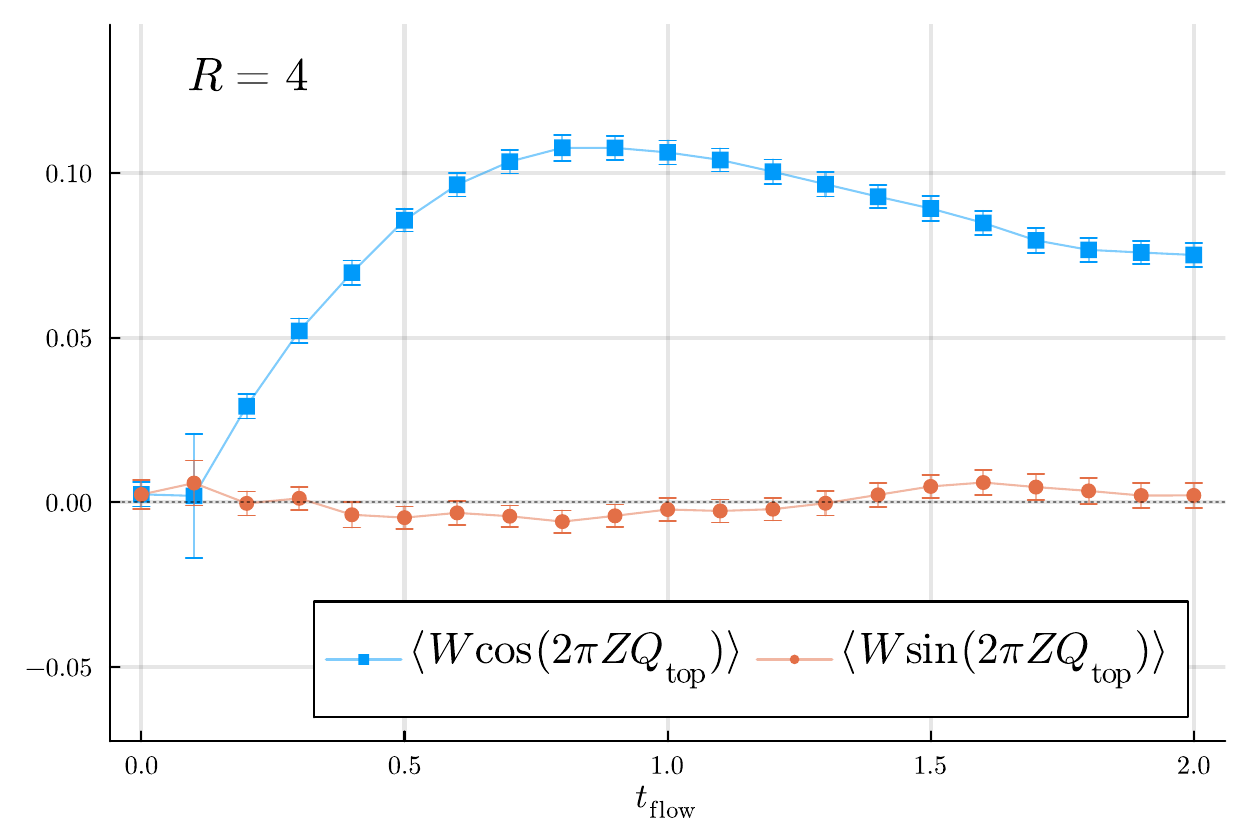}
    \includegraphics[width=0.32\linewidth]{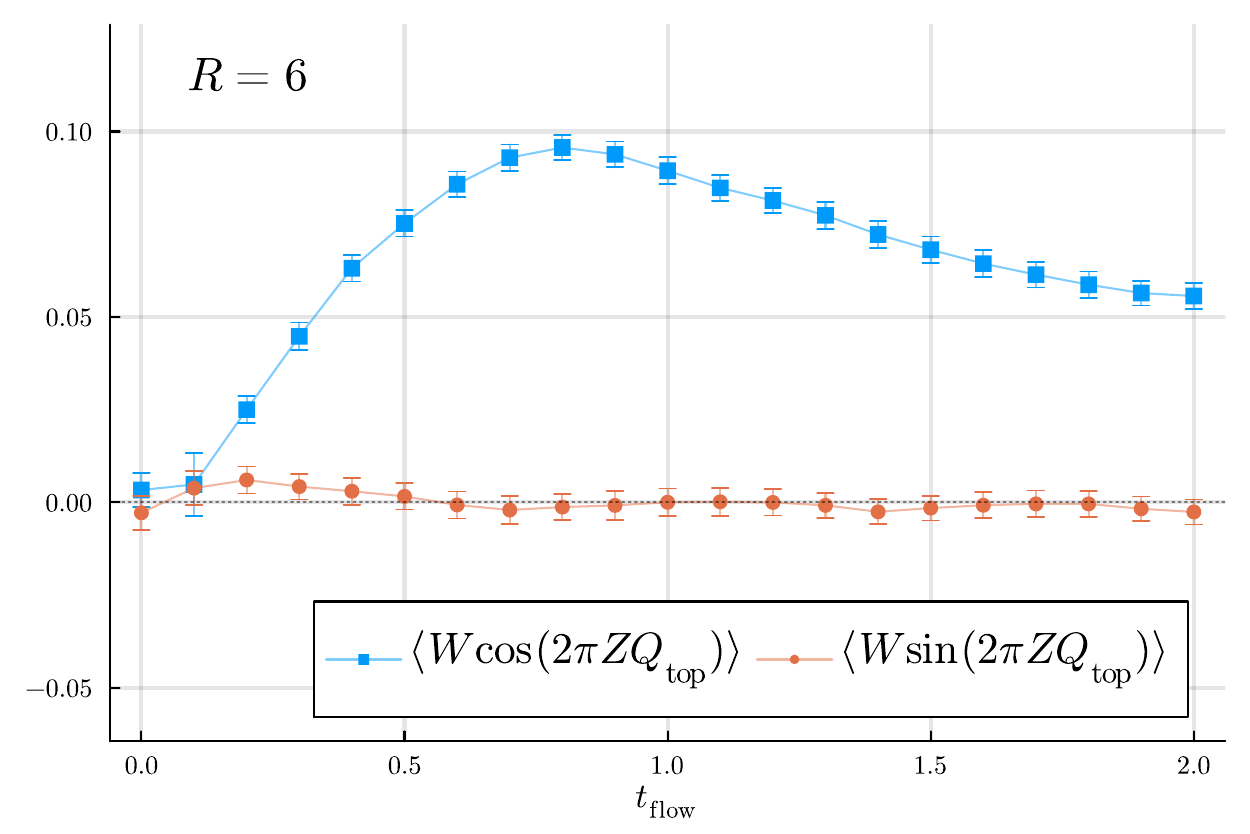}
    \\
    \includegraphics[width=0.32\linewidth]{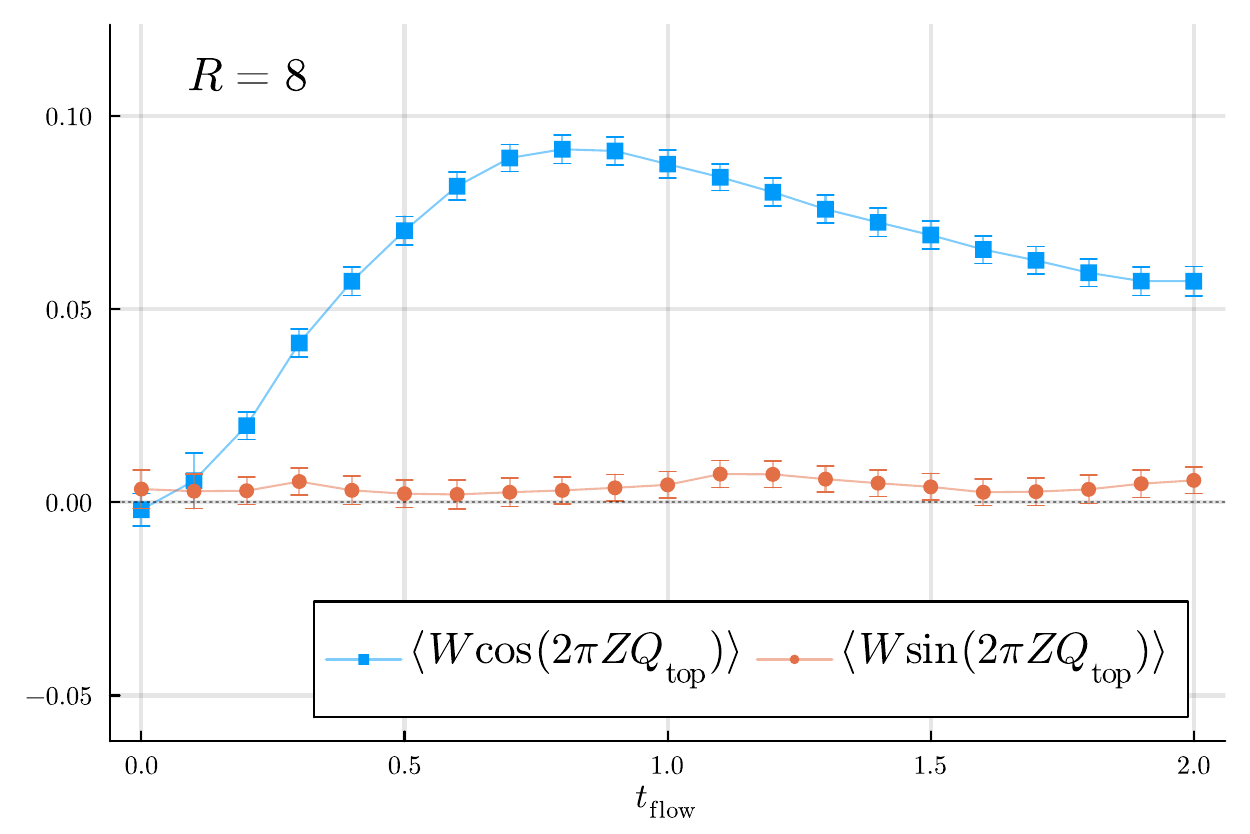}
    \includegraphics[width=0.32\linewidth]{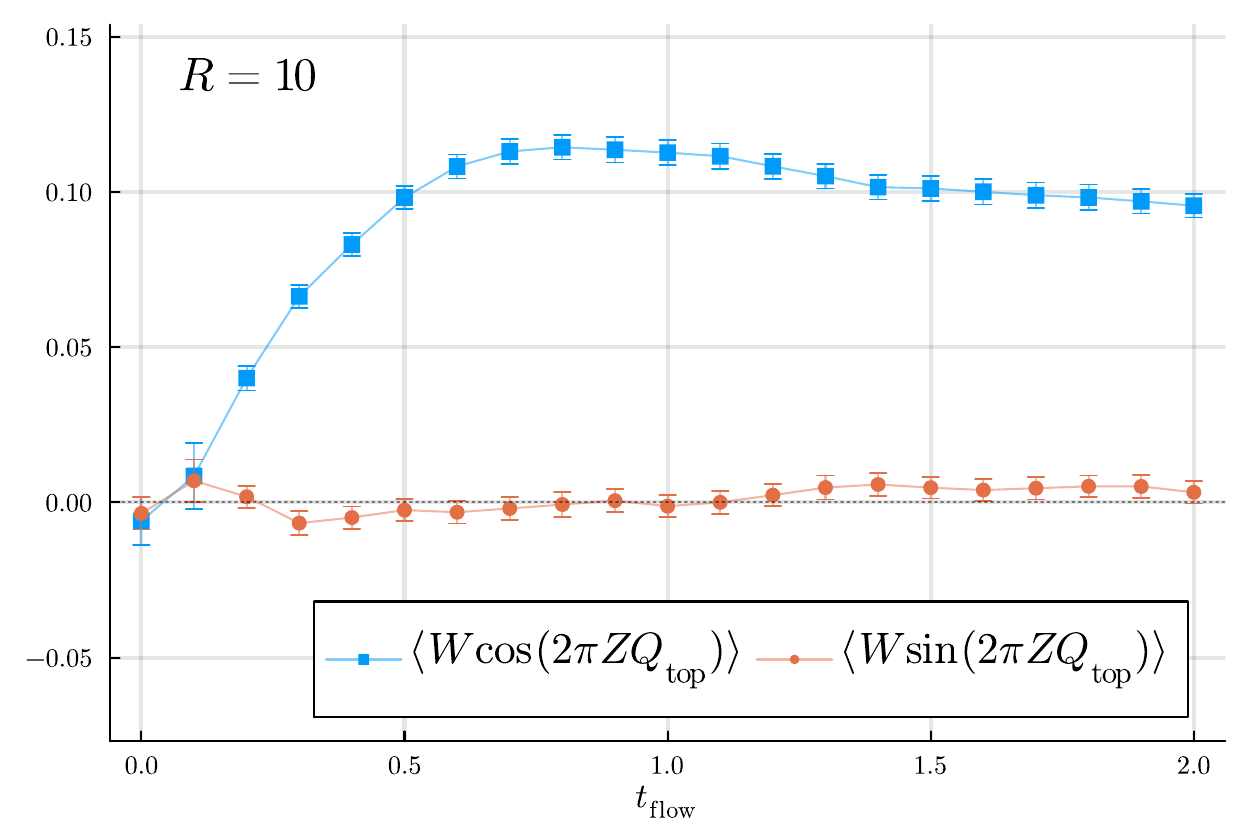}
    \includegraphics[width=0.32\linewidth]{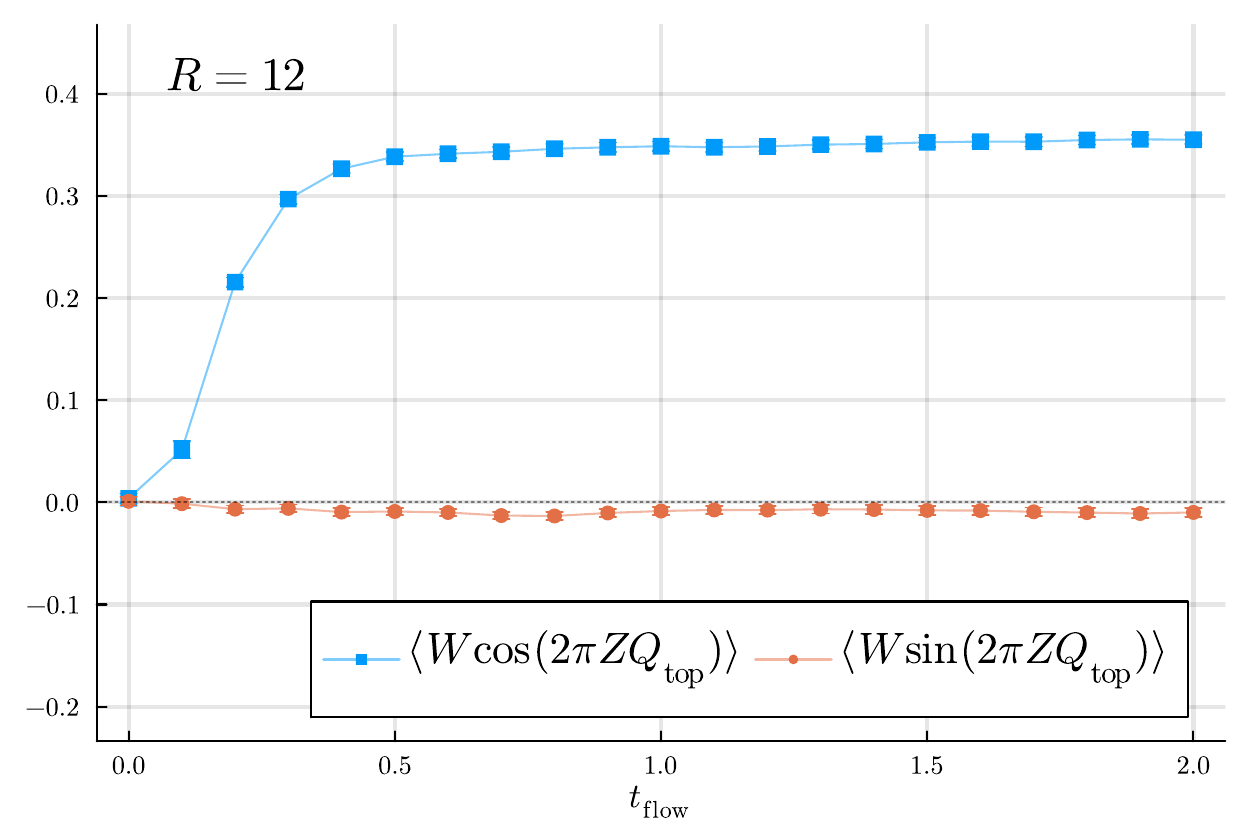}
    \caption{
    Flow-time dependence of the real and imaginary parts of \eqref{eq:HW_MC} for $R=2$, $4$, \dots, $12$ on the $12^3\times 8$ lattice.
    The flow time of the Wilson loop operators is chosen to $\tflow'=0.8$. 
    }
    \label{fig:flowtime-dep_12x8_HW_theta=2pi}
\end{figure}

Let us here discuss why we could obtain the clear signals for the dyon loops at $\theta=2\pi$ to conclude the perimeter law while it was not successful for the 't~Hooft loop with the same number of statistics. 
Assuming the area law for the 't~Hooft loops, its signal values should be exponentially small as we approach $R\to \Ns/2$, which become comparable with the statistical errors $O(1)/\sqrt{\# \mathrm{samples}}\approx 3\times 10^{-3}$. 
On the other hand, since the dyon loop is expected to obey the perimeter law, its expectation values do not drop when $1\ll R\ll \Ns$ and thus its expectation value can be much larger than the above statistical error. 
Thus, the large signal values of the dyon loop themselves already suggest the consistency with the dyon condensation. 
Let us emphasize that our conclusion is not affected by the precise choice of the flow time:
We plot the $\tflow$ dependence of the real and imaginary parts of \eqref{eq:HW_MC} in Figs.~\ref{fig:flowtime-dep_12x8_HW_theta=2pi} 
for the $12^3\times 8$ 
(Note that $\tflow'$ for the Wilson loop is not changed in these figures), and the $\tflow$ dependence is mild around $\tflow\approx 0.8$. We have confirmed the similar results for the $16^3\times 10$ lattice. 

\section{Summary and Discussion}
\label{sec:discussion}

In this paper, we have numerically studied the long-range behaviors of the 't~Hooft and dyon loop operators at $\theta=2\pi$ to
provide numerical evidence
that the quantum YM vacua at $\theta=0$ and $2\pi$ belong to different symmetry-protected topological (SPT) states. 
For this purpose, we need to confirm the area law of the 't~Hooft loop and the perimeter law of the dyon loop at $\theta=2\pi$, and we developed the reweighting formula that does not suffer from the sign problem at $\theta=2\pi$ when combined with the DBW2 flow. 
\rev{
For the 't Hooft loops at $\theta=2\pi$, 
their observed $R$ dependence of the $12^3\times 8$ lattice is consistent with the expected area law behavior, and the corresponding string-tension scale also agrees with that independently determined from the Polyakov-loop correlators as theoretically expected. However, the scale separation is not achieved due to the smearing via the DBW2 flow. 
On the $16^3\times 10$ lattice, their signal values are too small to make a conclusive statement about the asymptotic area law within our present statistics, while their magnitudes are comparable with the statistical errors expected from the area-law behavior.  
For the dyonic loops at $\theta=2\pi$, on the other hand, our results show a clear perimeter-law behavior.
The results can be summarized as the following Table~\ref{tab:summary_loop_results}. 
Due to the perimeter law of the dyonic Wilson--'t~Hooft loop, our observation provides the direct numerical evidence for the dyon condensation pattern as theoretically expected at $\theta=2\pi$.
\begin{table}[ht]
  \centering
\renewcommand{\arraystretch}{1.3}
\begin{tabular}{c|ccc}
    \hline 
    Loop & Criterion & $12^3\times8$ & $16^3\times10$ \\
    \hline 
    Dyonic & Long-range behavior & Perimeter & Perimeter 
    \\
    $\langle HW\rangle$ & Scale separation & Limited by smearing & Marginally achieved 
    \\
    & Statistical precision & Achieved & Achieved 
    \\
    \hline
    't Hooft & Long-range behavior & \multicolumn{2}{c}{Inconclusive for asymptotic behaviors, but} \\[-0.7ex]
    $\langle H \rangle$ & & \multicolumn{2}{c}{consistent with the area law with the string tension $\tilde{\sigma}$} 
    \\[0.7ex]
    & Scale separation & Limited by smearing & Marginally achieved 
    \\
    & Statistical precision & Achieved & Limited
    \\
    \hline
  \end{tabular}
    \caption{Summary of the numerical results for the loop operators at $\theta=2\pi$.}
  \label{tab:summary_loop_results}
\end{table}
}

About the 't~Hooft loop at $\theta=2\pi$, we have concluded that our numerical result itself does not clearly tell if it obeys the area law, and we would like to discuss this point from several perspectives here. 
Let us recall that the Wilson--'t~Hooft classification tells only one of the $W$, $H$, and $HW$ lines obeys the perimeter law in a gapped Lorentz-invariant vacuum of $SU(2)$ gauge theories (with adjoint matters), and the other two lines should obey the area law as a consequence of the locality, unitarity, and Lorentz invariance~\cite{tHooft:1977nqb, tHooft:1979rtg}. 
\rev{
Since we obtain clear numerical evidence for the perimeter law of the dyonic ($HW$) loop at $\theta=2\pi$, we emphasize that the Wilson--'t Hooft classification therefore provides an independent theoretical consistency argument for the 't Hooft-loop area law under the assumptions of a gapped, Lorentz-invariant vacuum.
This implication is consistent with, and strengthens, our direct but statistically inconclusive 't~Hooft-loop results about the area law. 
}

Still, one may ask if there is a way to obtain the clear direct signal for the area law of the 't~Hooft loop at $\theta=2\pi$. 
For this purpose, it would have been useful to study the 't~Hooft loops for the $16\times 12^2\times 8$ lattice, which are inserted as $R\times 8\subset 16 \times 8$. 
We prepared the larger lattice volume $16^3\times 10$ in this paper to find sufficient numbers of allowed $R$'s satisfying $2d\lesssim R\lesssim \Ns-2d$, where $2d\approx 5$ for $\tflow\approx 0.8$. 
For this criterion, the sizes of the temporal and $y,z$ directions are irrelevant as long as the system is in a confined phase, and we may take $\Nt=8$ instead of $10$ as we know it also belongs to the confinement phase from the result of the $12^3\times 8$ lattice. Then, the area of the cylindrical region surrounded by the magnetic loops becomes $8/10$ times smaller, which makes the exponential fall-off of the 't~Hooft loop at $\theta=2\pi$ milder, and we expect larger signals for the 't~Hooft loop. 
Moreover, the smaller volumes decrease computational costs of the gradient flow, which is the bottleneck in our numerical simulation, so it should also be useful to shrink the $y,z$ directions from $16$ to $12$. 
\rev{
Moreover, by considering the $N_s\times 12^2\times 8$ lattice, we can systematically achieve the scale separation $2d\ll R\ll N_s-2d$ with taking larger $N_s$ within the mild increase of the computational cost, and this would provide more ideal setups to study the asymptotic behaviors of Wilson--'t~Hooft operators. 
}

One of the interesting future directions is to study the interface tension of the spatial 't~Hooft and dyonic loops in the high-temperature deconfined phase at $\theta=2\pi$. 
At high temperatures, the $0$-form center symmetry acting on the Polyakov loop is spontaneously broken, and there exist domain walls connecting those $N$ symmetry-broken vacua. 
Since the spatial 't~Hooft loop is spanned by this broken $0$-form symmetry generator, its ``dual string tension'' is deeply related to the domain-wall tension, which acquired some attention (see, \textit{e.g.}, Refs.~\cite{Korthals-Altes:1999cqo, Bhattacharya:1992qb, Armoni:2008yp, Hidaka:2009hs}), and its lattice computation at $\theta=0$ has also been done in Refs.~\cite{deForcrand:2000fi, deForcrand:2005pb}. 
In a recent study~\cite{Hayashi:2026ijm}, it is uncovered that the $\mathbb{Z}_N$ domain-wall state encounters the phase transition at $\theta=\pi$ due to the generalized 't~Hooft anomaly~\eqref{eq:MixedAnomaly}: The domain-wall tension matches with the string tension of the spatial 't~Hooft loop for $-\pi<\theta<\pi$, but it becomes the dyonic loop that gives the domain-wall tension for $\pi<\theta<3\pi$. 
In Ref.~\cite{Hayashi:2026ijm}, this is explicitly demonstrated for the softly-broken $\mathcal{N}=1$ supersymmetric YM theory on $\mathbb{R}^3\times S^1$, which offers the semiclassically calculable setup for the confinement-deconfinement transition~\cite{Poppitz:2012sw, Anber:2013doa, Anber:2014lba, Chen:2020syd}. The direct numerical demonstration of the thermal pure YM theory should also be interesting.

\acknowledgments
This work was triggered by the conversation during the workshop, ``Nonperturbative Approaches to QFTs'' at the Komaba campus of the University of Tokyo in March 2024, and the authors thank the organizers for providing the opportunities. 
The authors also appreciate useful discussions with Yui Hayashi, Zohar Komargodski, Akira Matsumoto, Jun Nishimura, and Ryutaro Tsuji.
This work was partially supported by Japan Society for the Promotion of Science (JSPS)
Grant-in-Aid for Scientific Research Grant Numbers
JP25K17402 (O.M.), 20K14479, 22H05111, 22K03539, 22H05112 (Y.N.\ and A.T.), 23K22489, 26K07106 (Y.T.), 24K00630 (H.W.), 
by the Multidisciplinary Cooperative Research Program in CCS, University of Tsukuba, 
by MEXT as “Program for Promoting Researches on the Supercomputer Fugaku” (Grant Number
JPMXP1020230411, JPMXP1020230409), 
by JST BOOST, Japan Grant Number JPMJBY24F1 (A.T.), 
and by Center for Gravitational Physics and
Quantum Information (CGPQI) at Yukawa Institute for Theoretical Physics. 
This work also used computational resources provided by the Yukawa-21 at Yukawa Institute for Theoretical Physics, Kyoto University, and Ajiro in Center for Computational Sciences at the University of Tsukuba. 
O.M.\ acknowledges the RIKEN Special Postdoctoral Researcher Program
and RIKEN FY2025 Incentive Research Projects.

\appendix

\section{Topological charge with 't~Hooft flux}
\label{app:Qtop_R=12_12x8}

When the 't~Hooft loop of size $R=\Ns$ is inserted on the periodic lattice, the monopole singularity disappears, and the setup becomes equivalent to the insertion of 't~Hooft flux: $\diff B=0$, $\int_{(T^2)_{yz}}B=1$ mod $2$, and the fluxes for other two-cycles are zero.
In this case, the topological charge is integer quantized in the continuum limit (see \eqref{eq:GeometricTopologicalCharge}), and thus the integer-valued nature of the lattice topological charge is expected to be recovered after sufficient smearing. 
Let us check if this expectation holds for the gauge configurations of the $12^3\times 8$ lattice with $R=\Ns$. 

\begin{figure}[t]
    \centering
    \includegraphics[width=0.32\linewidth]{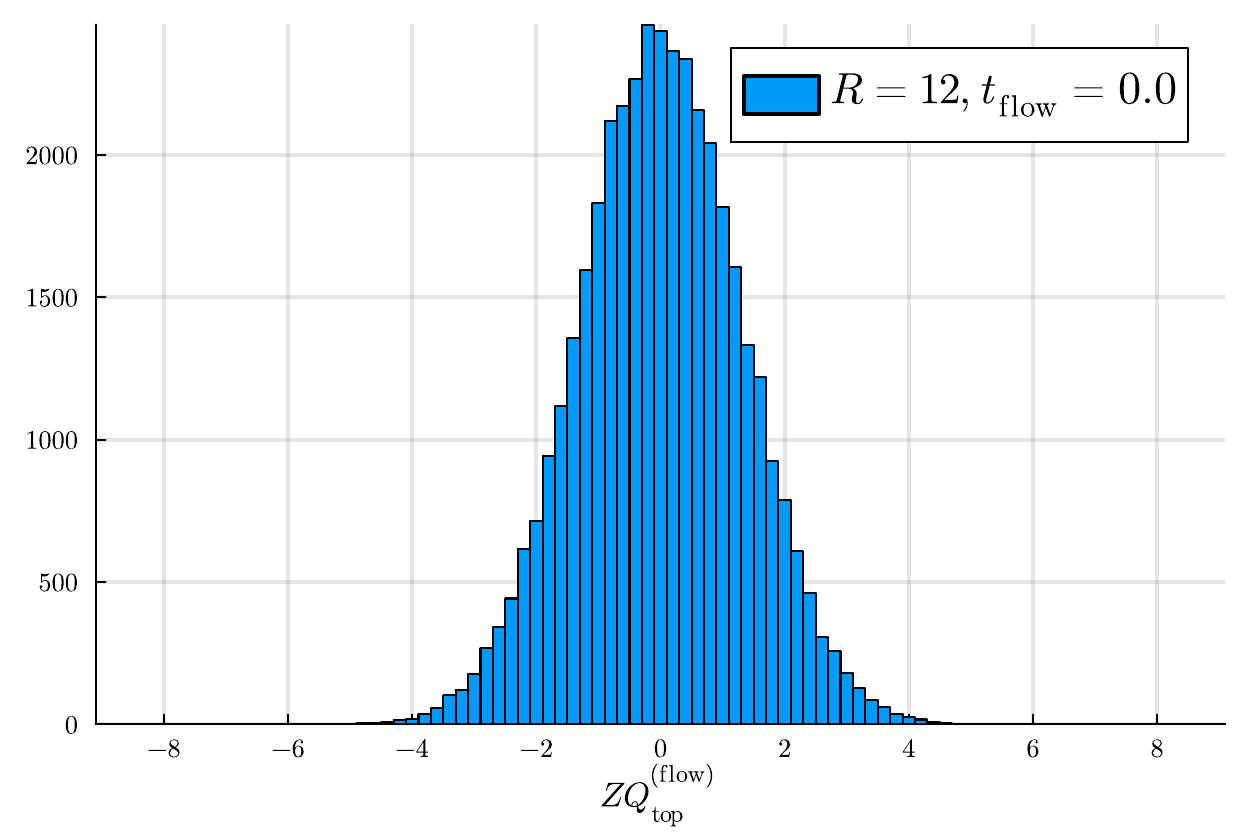}
    \includegraphics[width=0.32\linewidth]{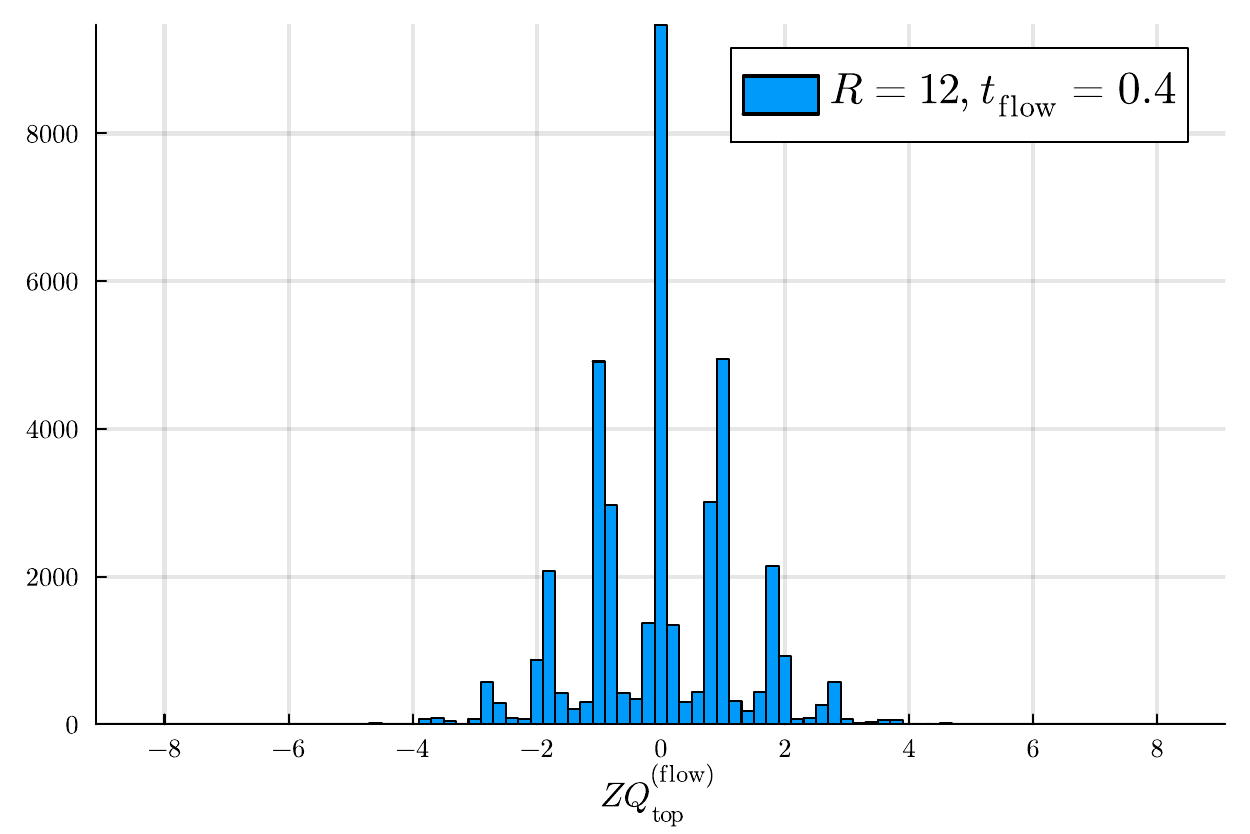}
    \includegraphics[width=0.32\linewidth]{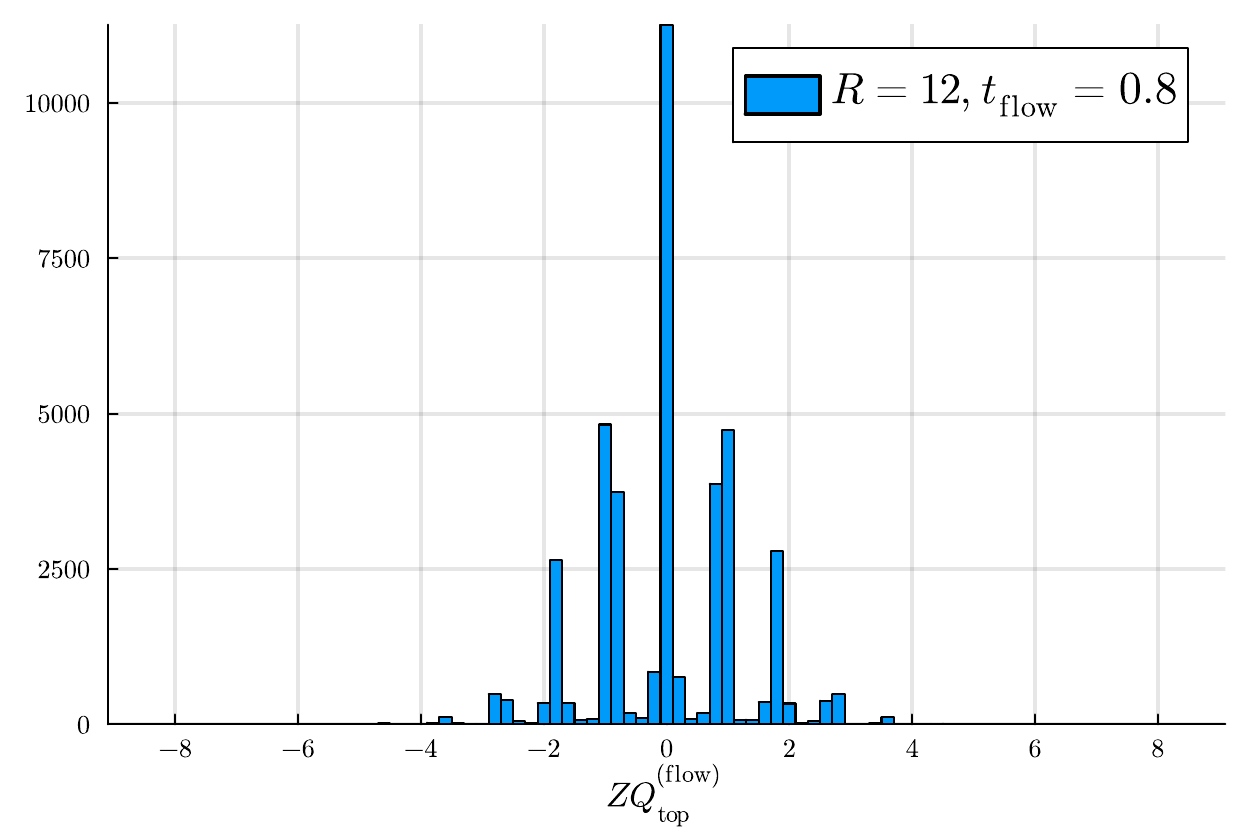}
    \caption{Histograms of $Z\Qtop^{(\tflow)}$ of the $12^3\times 8$ lattice at flow times $\tflow = 0$, $0.4$, and $0.8$.
    The spatial size of the inserted 't~Hooft loop is $R=\Ns=12$, which is equivalent to an insertion of 't~Hooft flux in the $x$ direction.
    }
    \label{fig:hist_ZQ_R=12_12x8}
\end{figure}

As shown in Fig.~\ref{fig:hist_ZQ_R=12_12x8}, the histograms of topological charge at $R=\Ns=12$, rescaled by $Z$ after gradient flow, exhibit peak structures. 
However, their peak locations are slightly off from integers.
\rev{
If one determines the rescaling factor for this ensemble of $R=12$, it is estimated as $\tilde Z = 1.1474(3)$, which shows a systematic discrepancy from $Z = 1.03818(8)$ computed by the ``$R = 0$'' ensemble. 
Here, let us emphasize that the rescaling factor $Z = 1.03818(8)$ determined from the $R=0$ ensemble is coherently used for the ensembles of any $R$, and we never used the above larger value of $\tilde{Z}$ determined at $R=N_s$ in the main text. 
}

\rev{
Still, it is a curious question to consider what causes the discrepancy between $Z$ and $\tilde{Z}$. 
This mismatch would be caused by the finite lattice spacings and come from the dimension-$8$ operator contribution:
}
Tree-level Symanzik improvement eliminates the dimension-$6$ term from the topological charge but $\Qtop[U_\ell, B_p]$ still has the $O(a^4)$ deviation from the continuum result via the dimension-$8$ operators~\cite{Luscher:1984xn, GarciaPerez:1993lic, Alexandrou:2017hqw}. 
\rev{
Since the $R=0$ and $R=12$ setups correspond to the different boundary conditions for the gauge fields, it would be natural to expect that the typical magnitudes of their gradients are different between those setups, which ends up with the difference of the dimension-$8$ contributions. 
Therefore, the discrepancy between $Z$ and $\tilde{Z}$ should quantify how large the gradients of smeared gauge fields are on average, and thus the difference should be suppressed by working on larger lattice volumes with weaker couplings and/or also with more improvements of the topological charge densities.
}

\section{More on numerical results}
\label{app:more_numerical_results}

In this appendix, we examine how the choice of loop operators on the lattice numerically affects the perimeter-law behavior. 
Note that, for the following analyses using the smeared Polyakov loops, we employ the same $12^3\times 8$ lattice ensembles but the last $30,000$ gauge configurations.

\subsection{Dyonic Loop at \texorpdfstring{$\theta=2\pi$}{theta=2pi}}

\begin{figure}[t]
    \centering
    \includegraphics[width=0.49\linewidth]{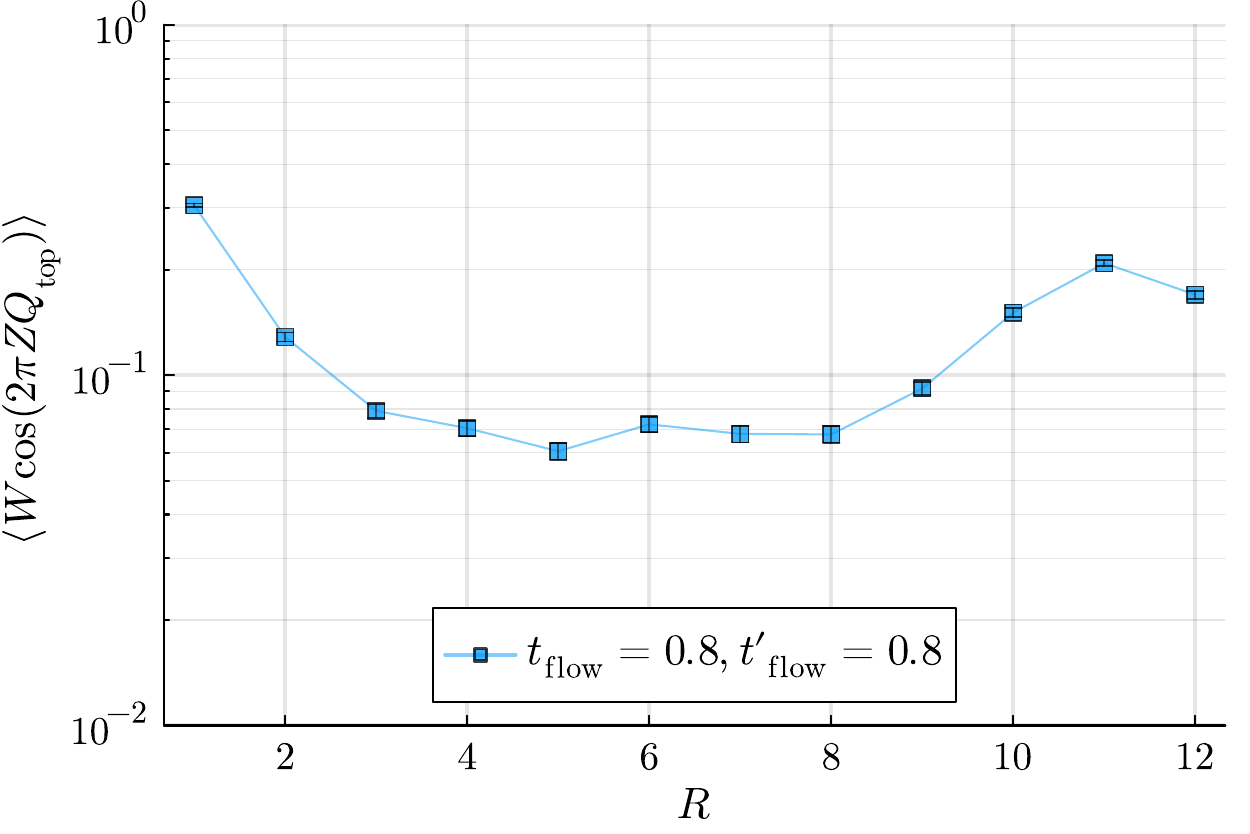}
    \includegraphics[width=0.49\linewidth]{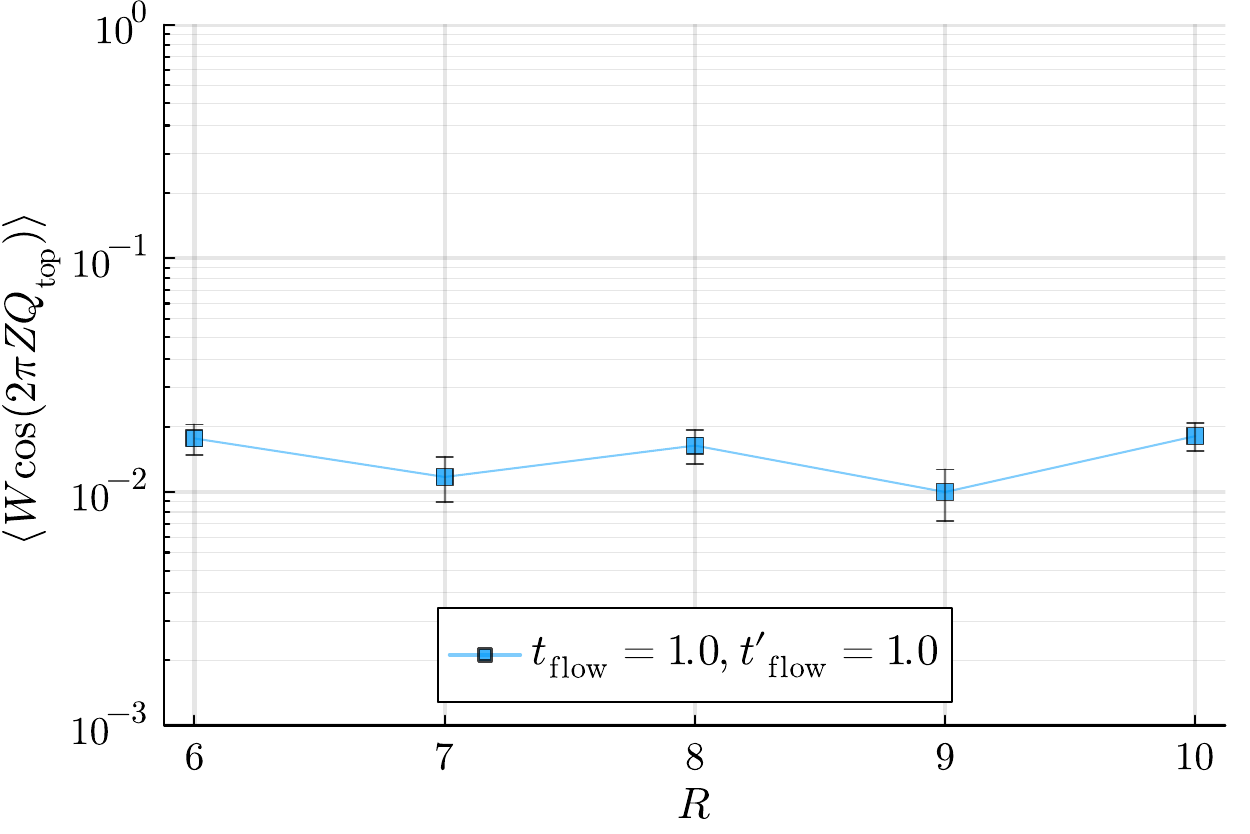}
    \caption{
    Same format as Fig.~\ref{fig:HW_theta=2pi}, but with the modified Wilson-loop contour.
    }
    \label{fig:HW_large_theta=2pi}
\end{figure}

\begin{figure}[t]
    \centering
    \includegraphics[width=0.49\linewidth]{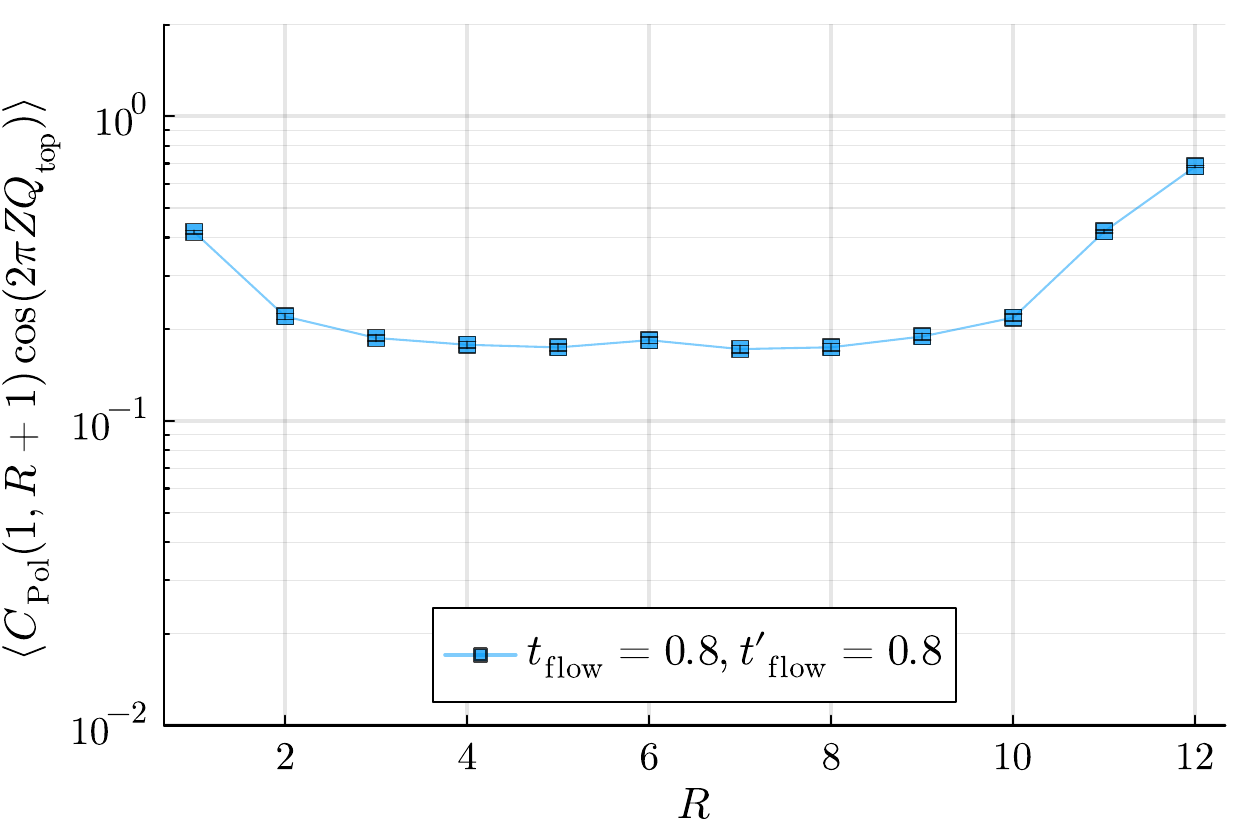}
    \includegraphics[width=0.49\linewidth]{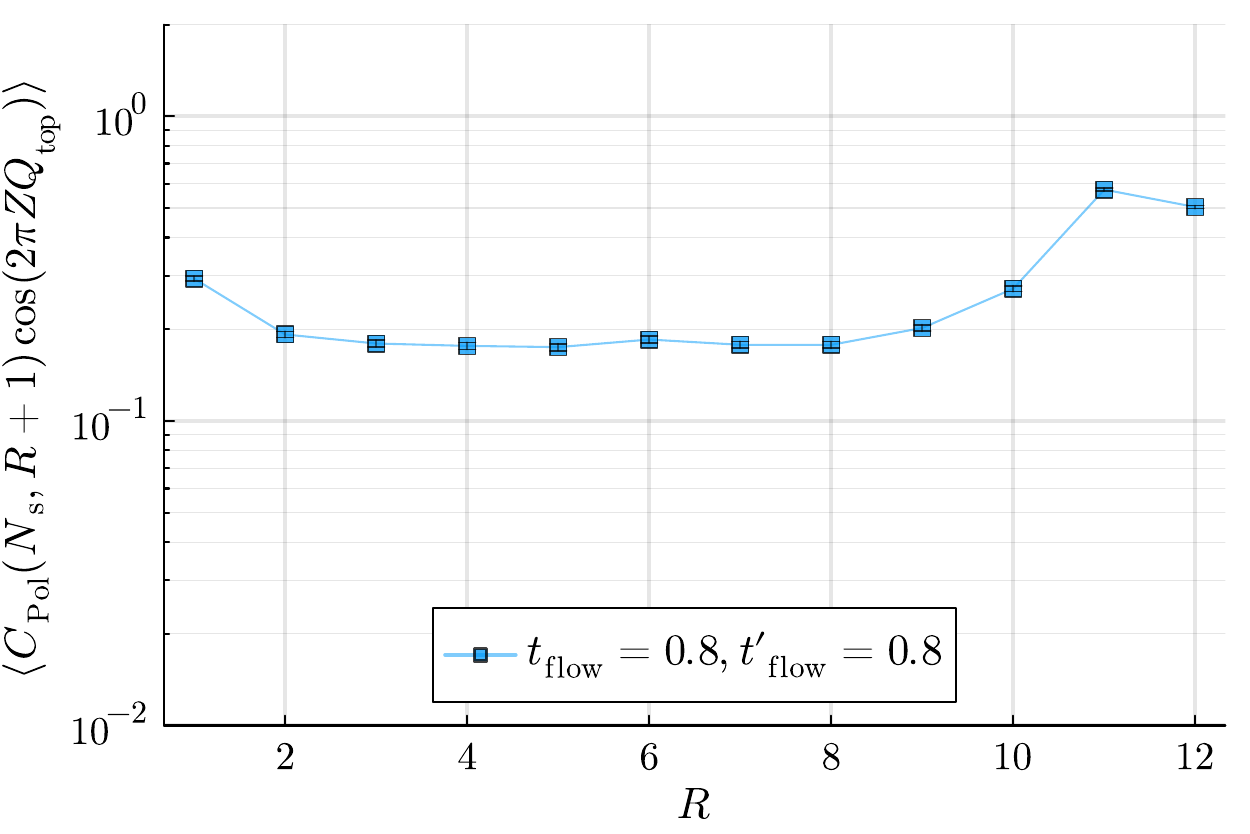}
    \caption{
    Same format as the left panel of Fig.~\ref{fig:HW_theta=2pi}, but with the dyonic loop defined using fundamental Polyakov loops as in \eqref{eq:HW_MC_PP}. The locations of the Polyakov loops are set to $x_0=1$, $R'=R+1$ (left), and to $x_0=\Ns\equiv-1$, $R'=R+1$ (right). We set the flow time of the smeared Polyakov-loop operators to $\tflow'=\tflow=0.8$.
    }
    \label{fig:HW_PP_theta=2pi}
\end{figure}

We first consider the modification of the dyonic loop at $\theta=2\pi$.
For the definition~\eqref{eq:HW_MC} used in the main text, the Wilson loop contour $C$ is chosen as the rectangle that is shifted from the 't~Hooft loop insertion by a half-lattice unit. 
Here, let us instead introduce the probe electric charges so that the dyonic line becomes reflection symmetric, and the electric charges are located at $(x,y,z)=(\Ns,1,1)$ and $(R+1,1,1)$.
In other words, the contour $C$ forms a rectangle of size $(R+1)\times\Nt$ that includes the site $\vsub{x}{base}-\hat{e}_x= (-1,1,1,1)\equiv (\Ns,1,1,1)$, where $x_0=\Ns$ is identified with $x_0=-1$ by periodicity.
The result is shown in Fig.~\ref{fig:HW_large_theta=2pi}, and we again find the perimeter-law behavior for $4\lesssim R\lesssim 8$ for the $12^3\times 8$ lattice and all the analyzed region $6\le R\le 10$ for the $16^3\times 10$ lattice. 
When the location of charges becomes closer for the $12^3\times 8$ lattice, the large difference between Figs.~\ref{fig:HW_theta=2pi} and \ref{fig:HW_large_theta=2pi} appears. 

Next, let us discuss the temporal dyonic line operators defined by the Polyakov-loop correlators instead of the rectangular Wilson loops.
That is, we replace \eqref{eq:HW_MC} by
\begin{equation}
    \vsup{\left\langle P^{(\tflow')}(x_0,1,1)P^{(\tflow')}(R',1,1)\,
    \exp(2\pi\im Z\Qtop^{(\tflow)}[U_\ell, \delta[\tilde{\Sigma}]])\right\rangle}{MC}_{\SW[U_\ell, \delta[\tilde{\Sigma}]]}, 
    \label{eq:HW_MC_PP}
\end{equation} 
where $P^{(\tflow')}(x,y,z)$ is the fundamental Polyakov loop at $(x,y,z)$ with the flowed configuration by the flow time $\tflow'$. 
Here, we again choose the same flow time $\tflow'=\tflow$.
Unlike the definition with the rectangular Wilson loop, the back-and-forth Wilson line along the $x$-direction is absent, and thus the $R'$-dependent renormalization factor does not appear from the beginning. 
When $R'\gg 1$, those two definitions should give the identical result up to an overall numerical factor as discussed in footnote~\ref{ftnt:def_dyonline}. 

Figure~\ref{fig:HW_PP_theta=2pi} represents the $R$ dependence of the real part of \eqref{eq:HW_MC_PP} for the $12^3\times 8$ lattice.
The left panel corresponds to the choice of $x_0 = 1$ and $R' = R+1$, which is the analog of the Wilson loop of size $R\times\Nt$ examined around Fig.~\ref{fig:HW_theta=2pi}.
On the other hand, the right panel corresponds to the choice of $x_0 = \Ns \equiv -1$ and $R' = R+1$, which is the analog of the Wilson loop of size $(R+1)\times \Nt$ discussed in this Appendix.
In both cases, a clear perimeter-law behavior is observed in the region $4\lesssim R \lesssim 8$, and the plateau appears around $0.2$.

\subsection{'t Hooft Loop at \texorpdfstring{$\theta=4\pi$}{theta=4pi}}

\begin{figure}[t]
    \centering
    \includegraphics[width=0.49\linewidth]{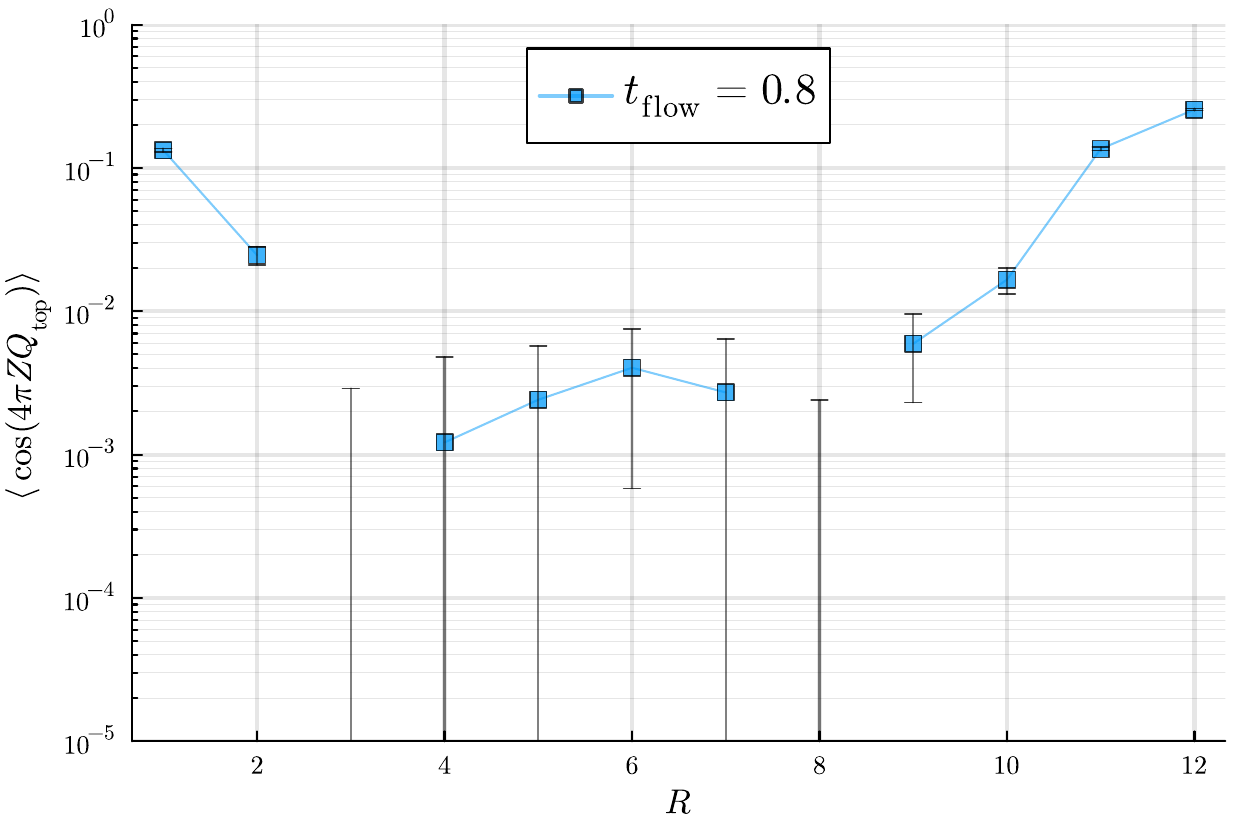}
    \includegraphics[width=0.49\linewidth]{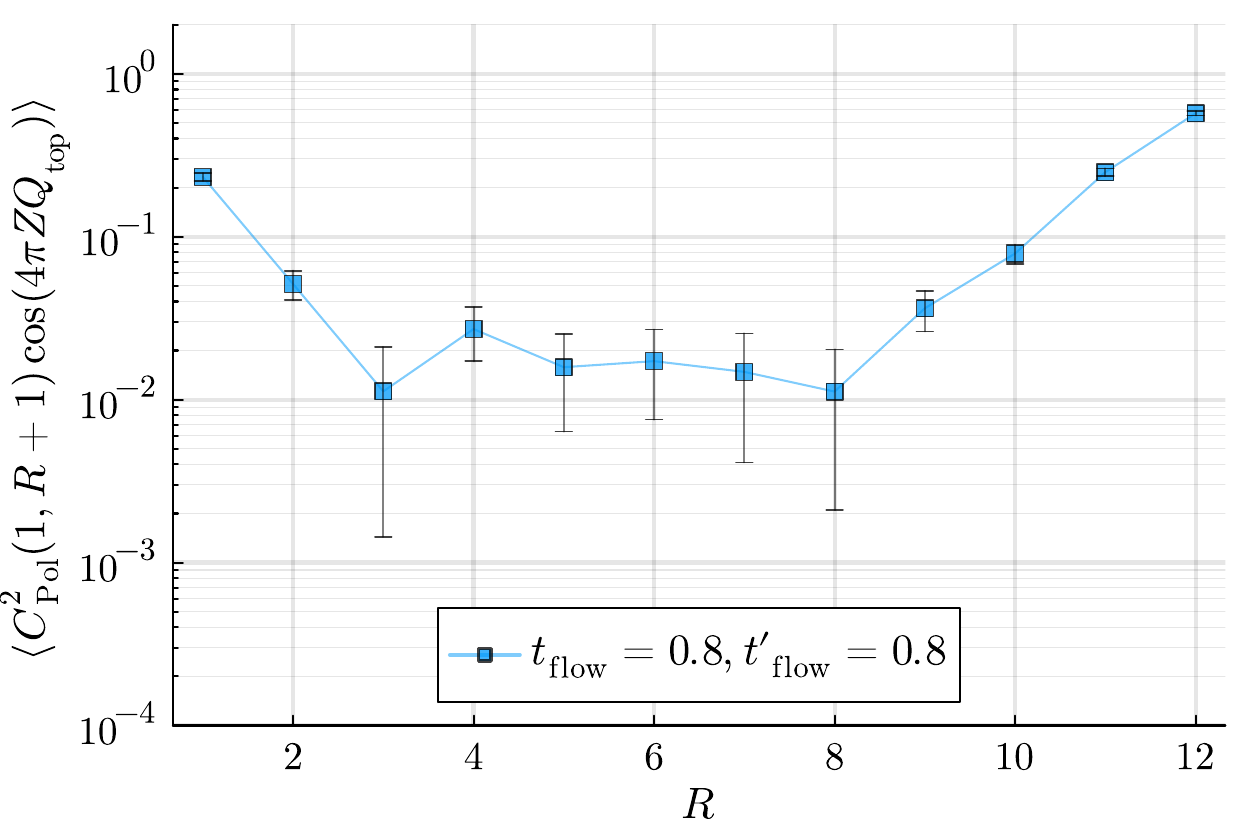}
    \caption{
    The real part of \eqref{eq:H_MC_4pi}, $\left\langle \cos(4\pi Z\Qtop^{(\tflow)})\right\rangle$ (left) and 
    the real part of \eqref{eq:H_MC_PP_4pi}, $\left\langle \left(P^{(\tflow')}(1,1,1)P^{(\tflow')}(R+1,1,1)\right)^2\cos(4\pi Z\Qtop^{(\tflow)})\right\rangle$ (right), for the $12^3 \times 8$ lattice.
    For the smeared Polyakov loops, we set $\tflow'=\tflow=0.8$.
    }
    \label{fig:H_theta=4pi}
\end{figure}

In the main text, we have studied the expectation values of loop operators at $\theta = 2\pi$. 
The 't~Hooft anomaly relation~\eqref{eq:MixedAnomaly} predicts that the confinement vacua at $\theta=0$ and $2\pi N$ belong to the same SPT phase. 
Therefore, in the present $SU(2)$ setup, it is natural for the loop operators at $\theta=4\pi$ to obey the same area/perimeter law as those at $\theta=0$.
To confirm this, we examine if the 't~Hooft loop at $\theta=4\pi$ obeys the perimeter law in this subsection.

By substituting $\theta=4\pi$ in \eqref{eq:H_MC}, the long-range behavior of the 't~Hooft loop at $\theta=4\pi$ is determined by
\begin{equation}
    \vsup{\left\langle \exp(4\pi\im Z\Qtop^{(\tflow)}[U_\ell, \delta[\tilde{\Sigma}]])\right\rangle}{MC}_{\SW[U_\ell, \delta[\tilde{\Sigma}]]}. 
    \label{eq:H_MC_4pi}
\end{equation} 
The result is shown in the left panel of Fig.~\ref{fig:H_theta=4pi}. 
As we can imagine from the Monte Carlo result for $\theta=2\pi$, its expectation value becomes so small that we cannot obtain clear signals within current statistics. 
Physical interpretation is that the Witten effect for changing $\theta:0\to 4\pi$ induces the adjoint electric charge, so the perimeter law would appear only after the adjoint string breaking. 

To confirm it, let us dress the adjoint Polyakov loops at each location of the monopoles:
\begin{equation}
    \vsup{\left\langle \left(P^{(\tflow')}(1,1,1)P^{(\tflow')}(R+1,1,1))\right)^2\,
    \exp(4\pi\im Z\Qtop^{(\tflow)}[U_\ell, \delta[\tilde{\Sigma}]])\right\rangle}{MC}_{\SW[U_\ell, \delta[\tilde{\Sigma}]]}. 
    \label{eq:H_MC_PP_4pi}
\end{equation} 
The result is shown in the right panel of Figure~\ref{fig:H_theta=4pi}.
The data points in the middle $R$ region tend to align the value $1.5\times 10^{-2}$ up to the numerical precision, and these findings provide a positive indication of the expected perimeter-law behavior for the 't~Hooft loop at $\theta = 4\pi$ up to the dressing of adjoint Wilson loops. 
This encourages further lattice Monte Carlo studies in the future.


\bibliographystyle{utphys}
\bibliography{QFT,ref,coderef}

@article{DElia:2003zne,
    author = "D'Elia, Massimo",
    title = "{Field theoretical approach to the study of theta dependence in Yang-Mills theories on the lattice}",
    eprint = "hep-lat/0302007",
    archivePrefix = "arXiv",
    reportNumber = "GEF-TH-2003-02",
    doi = "10.1016/S0550-3213(03)00311-0",
    journal = "Nucl. Phys. B",
    volume = "661",
    pages = "139--152",
    year = "2003"
}

@article{Luscher:1981zq,
    author = "L{\"u}scher, M.",
    title = "{Topology of Lattice Gauge Fields}",
    reportNumber = "BUTP-10/1981-BERN",
    doi = "10.1007/BF02029132",
    journal = "Commun. Math. Phys.",
    volume = "85",
    pages = "39",
    year = "1982"
}

@article{Abe:2023ncy,
    author = "Abe, Motokazu and Morikawa, Okuto and Onoda, Soma and Suzuki, Hiroshi and Tanizaki, Yuya",
    title = "{Topology of SU(N) lattice gauge theories coupled with {\ensuremath{\mathbb{Z}}}$_{N}$ 2-form gauge fields}",
    eprint = "2303.10977",
    archivePrefix = "arXiv",
    primaryClass = "hep-lat",
    reportNumber = "KYUSHU-HET-258, OU-HET-1177, YITP-23-37, OU-HET-1177,",
    doi = "10.1007/JHEP08(2023)118",
    journal = "JHEP",
    volume = "08",
    pages = "118",
    year = "2023"
}

@article{Maskawa:1974vs,
    author = "Maskawa, Toshihide and Nakajima, Hideo",
    title = "{Spontaneous Symmetry Breaking in Vector-Gluon Model}",
    reportNumber = "KUNS-272",
    doi = "10.1143/PTP.52.1326",
    journal = "Prog. Theor. Phys.",
    volume = "52",
    pages = "1326--1354",
    year = "1974"
}

@article{Kobayashi:1971qz,
    author = "Kobayashi, M. and Kondo, H. and Maskawa, T.",
    title = "{Symmetry breaking of the chiral u(3) x u(3) and the quark model}",
    doi = "10.1143/PTP.45.1955",
    journal = "Prog. Theor. Phys.",
    volume = "45",
    pages = "1955--1959",
    year = "1971"
}

@article{Bonati:2015sqt,
    author = "Bonati, Claudio and D'Elia, Massimo and Scapellato, Aurora",
    title = "{$\theta$ dependence in $SU(3)$ Yang-Mills theory from analytic continuation}",
    eprint = "1512.01544",
    archivePrefix = "arXiv",
    primaryClass = "hep-lat",
    reportNumber = "IFUP-TH-2015-14",
    doi = "10.1103/PhysRevD.93.025028",
    journal = "Phys. Rev. D",
    volume = "93",
    number = "2",
    pages = "025028",
    year = "2016"
}

@article{Bonati:2019kmf,
    author = "Bonati, Claudio and Cardinali, Marco and D'Elia, Massimo and Mazziotti, Fabrizio",
    title = "{$\theta$-dependence and center symmetry in Yang-Mills theories}",
    eprint = "1912.02662",
    archivePrefix = "arXiv",
    primaryClass = "hep-lat",
    doi = "10.1103/PhysRevD.101.034508",
    journal = "Phys. Rev. D",
    volume = "101",
    number = "3",
    pages = "034508",
    year = "2020"
}

@article{Bonanno:2023hhp,
    author = "Bonanno, Claudio and D'Elia, Massimo and Verzichelli, Lorenzo",
    title = "{The $\theta$-dependence of the $SU(N)$ critical temperature at large $N$}",
    eprint = "2312.12202",
    archivePrefix = "arXiv",
    primaryClass = "hep-lat",
    doi = "10.1007/JHEP02(2024)156",
    journal = "JHEP",
    volume = "02",
    pages = "156",
    year = "2024"
}

@article{Hirasawa:2024fjt,
    author = "Hirasawa, Mitsuaki and Honda, Masazumi and Matsumoto, Akira and Nishimura, Jun and Yosprakob, Atis",
    title = "{Evidence of a CP broken deconfined phase in 4D SU(2) Yang-Mills theory at {\ensuremath{\theta}} = {\ensuremath{\pi}} from imaginary {\ensuremath{\theta}} simulations}",
    eprint = "2412.03683",
    archivePrefix = "arXiv",
    primaryClass = "hep-th",
    reportNumber = "YITP-24-166, KEK-TH-2669, RIKEN-iTHEMS-Report-24, STUPP-24-274",
    doi = "10.1007/JHEP05(2025)009",
    journal = "JHEP",
    volume = "05",
    pages = "009",
    year = "2025"
}

@article{Yamada:2024pjy,
    author = "Yamada, Norikazu and Yamazaki, Masahito and Kitano, Ryuichiro",
    title = "{{\ensuremath{\theta}} dependence of T$_{c}$ in SU(2) Yang-Mills theory}",
    eprint = "2411.00375",
    archivePrefix = "arXiv",
    primaryClass = "hep-lat",
    doi = "10.1007/JHEP02(2025)211",
    journal = "JHEP",
    volume = "02",
    pages = "211",
    year = "2025"
}

@article{Yamada:2024vsk,
    author = "Yamada, Norikazu and Yamazaki, Masahito and Kitano, Ryuichiro",
    title = "{Subvolume method for SU(2) Yang-Mills theory at finite temperature: topological charge distributions}",
    eprint = "2403.10767",
    archivePrefix = "arXiv",
    primaryClass = "hep-lat",
    doi = "10.1007/JHEP07(2024)198",
    journal = "JHEP",
    volume = "07",
    pages = "198",
    year = "2024"
}

@article{Kitano:2021jho,
    author = "Kitano, Ryuichiro and Matsudo, Ryutaro and Yamada, Norikazu and Yamazaki, Masahito",
    title = "{Peeking into the {\ensuremath{\theta}} vacuum}",
    eprint = "2102.08784",
    archivePrefix = "arXiv",
    primaryClass = "hep-lat",
    reportNumber = "KEK-CP-0379, KEK-TH-2301",
    doi = "10.1016/j.physletb.2021.136657",
    journal = "Phys. Lett. B",
    volume = "822",
    pages = "136657",
    year = "2021"
}

@article{Kitano:2020mfk,
    author = "Kitano, Ryuichiro and Yamada, Norikazu and Yamazaki, Masahito",
    title = "{Is $N = 2$ Large?}",
    eprint = "2010.08810",
    archivePrefix = "arXiv",
    primaryClass = "hep-lat",
    reportNumber = "KEK-TH-2264",
    doi = "10.1007/JHEP02(2021)073",
    journal = "JHEP",
    volume = "02",
    pages = "073",
    year = "2021"
}

@article{deForcrand:2000fi,
    author = "de Forcrand, Philippe and D'Elia, Massimo and Pepe, Michele",
    title = "{A Study of the 't Hooft loop in SU(2) Yang-Mills theory}",
    eprint = "hep-lat/0007034",
    archivePrefix = "arXiv",
    reportNumber = "IFUP-TH-25-2000",
    doi = "10.1103/PhysRevLett.86.1438",
    journal = "Phys. Rev. Lett.",
    volume = "86",
    pages = "1438",
    year = "2001"
}

@article{Nguyen:2023fun,
    author = {Nguyen, Mendel and Tanizaki, Yuya and {\"U}nsal, Mithat},
    title = "{Study of gapped phases of 4d gauge theories using temporal gauging of the~$\mathbb{Z}_N$ 1-form symmetry}",
    eprint = "2306.02485",
    archivePrefix = "arXiv",
    primaryClass = "hep-th",
    reportNumber = "YITP-23-72",
    doi = "10.1007/JHEP08(2023)013",
    journal = "JHEP",
    volume = "08",
    pages = "013",
    year = "2023"
}

@article{Maeda:2025ycr,
    author = "Maeda, Jun and Tanizaki, Yuya",
    title = "{Twisted partition functions as order parameters}",
    eprint = "2505.16546",
    archivePrefix = "arXiv",
    primaryClass = "hep-th",
    reportNumber = "YITP-25-42, KUNS-3051",
    doi = "10.1007/JHEP08(2025)128",
    journal = "JHEP",
    volume = "08",
    pages = "128",
    year = "2025"
}

@article{Hayashi:2026ijm,
    author = "Hayashi, Yui and Tanizaki, Yuya",
    title = "{Wilson-{\textquoteright}t Hooft classification and the perimeter law for dyonic loops in 3d monopole semiclassics}",
    eprint = "2601.02058",
    archivePrefix = "arXiv",
    primaryClass = "hep-th",
    reportNumber = "YITP-25-197",
    doi = "10.1007/JHEP05(2026)303",
    journal = "JHEP",
    volume = "05",
    pages = "303",
    year = "2026"
}

@article{Tanizaki:2022plm,
    author = {Tanizaki, Yuya and {\"U}nsal, Mithat},
    title = "{Semiclassics with {\textquoteright}t Hooft flux background for QCD with 2-index quarks}",
    eprint = "2205.11339",
    archivePrefix = "arXiv",
    primaryClass = "hep-th",
    reportNumber = "YITP-22-45",
    doi = "10.1007/JHEP08(2022)038",
    journal = "JHEP",
    volume = "08",
    pages = "038",
    year = "2022"
}

@article{Hayashi:2023wwi,
    author = "Hayashi, Yui and Tanizaki, Yuya and Watanabe, Hiromasa",
    title = "{Semiclassical analysis of the bifundamental QCD on~$\mathbb{R}^2\times T^2$ with {\textquoteright}t Hooft flux}",
    eprint = "2307.13954",
    archivePrefix = "arXiv",
    primaryClass = "hep-th",
    reportNumber = "YITP-23-96",
    doi = "10.1007/JHEP10(2023)146",
    journal = "JHEP",
    volume = "10",
    pages = "146",
    year = "2023"
}

@article{Hayashi:2024gxv,
    author = "Hayashi, Yui and Tanizaki, Yuya and Watanabe, Hiromasa",
    title = "{Non-supersymmetric duality cascade of QCD(BF) via semiclassics on {\ensuremath{\mathbb{R}}}$^{2}${\texttimes} T$^{2}$ with the baryon-{\textquoteright}t Hooft flux}",
    eprint = "2404.16803",
    archivePrefix = "arXiv",
    primaryClass = "hep-th",
    reportNumber = "YITP-24-41",
    doi = "10.1007/JHEP07(2024)033",
    journal = "JHEP",
    volume = "07",
    pages = "033",
    year = "2024"
}

@article{Hayashi:2024qkm,
    author = "Hayashi, Yui and Tanizaki, Yuya",
    title = "{Semiclassics for the QCD vacuum structure through T$^{2}$-compactification with the baryon-{\textquoteright}t Hooft flux}",
    eprint = "2402.04320",
    archivePrefix = "arXiv",
    primaryClass = "hep-th",
    reportNumber = "YITP-24-15",
    doi = "10.1007/JHEP08(2024)001",
    journal = "JHEP",
    volume = "08",
    pages = "001",
    year = "2024"
}

@article{Komargodski:2018odf,
    author = "Komargodski, Zohar",
    title = "{Baryons as Quantum Hall Droplets}",
    eprint = "1812.09253",
    archivePrefix = "arXiv",
    primaryClass = "hep-th",
    month = "12",
    year = "2018"
}

@Article{Aharony:2013hda,
  Title                    = {{Reading between the lines of four-dimensional gauge theories}},
  Author                   = {Aharony, Ofer and Seiberg, Nathan and Tachikawa, Yuji},
  Journal                  = {JHEP},
  Year                     = {2013},
  Pages                    = {115},
  Volume                   = {08},

  Archiveprefix            = {arXiv},
  Doi                      = {10.1007/JHEP08(2013)115},
  Eprint                   = {1305.0318},
  Primaryclass             = {hep-th},
  Reportnumber             = {UT-13-15, IPMU13-0081, WIS-03-13-APR-DPPA},
  Slaccitation             = {%%CITATION = ARXIV:1305.0318;%%}
}

@Article{Anber:2013doa,
  Title                    = {{Deconfinement in $\mathcal{N}=1$ super Yang-Mills theory on $\mathbb{R}^3 \times \mathbb{S}^1$ via dual-Coulomb gas and "affine" XY-model}},
  Author                   = {Anber, Mohamed M. and Collier, Scott and Poppitz, Erich and Strimas-Mackey, Seth and Teeple, Brett},
  Journal                  = {JHEP},
  Year                     = {2013},
  Pages                    = {142},
  Volume                   = {11},

  Archiveprefix            = {arXiv},
  Doi                      = {10.1007/JHEP11(2013)142},
  Eprint                   = {1310.3522},
  Primaryclass             = {hep-th},
  Slaccitation             = {%%CITATION = ARXIV:1310.3522;%%}
}

@Article{Anber:2014lba,
  Title                    = {{Deconfinement and continuity between thermal and (super) Yang-Mills theory for all gauge groups}},
  Author                   = {Anber, Mohamed M. and Poppitz, Erich and Teeple, Brett},
  Journal                  = {JHEP},
  Year                     = {2014},
  Pages                    = {040},
  Volume                   = {09},

  Archiveprefix            = {arXiv},
  Doi                      = {10.1007/JHEP09(2014)040},
  Eprint                   = {1406.1199},
  Primaryclass             = {hep-th},
  Slaccitation             = {%%CITATION = ARXIV:1406.1199;%%}
}

@Article{vanBaal:1982ag,
  Title                    = {{Some Results for SU(N) Gauge Fields on the Hypertorus}},
  Author                   = {van Baal, Pierre},
  Journal                  = {Commun. Math. Phys.},
  Year                     = {1982},
  Pages                    = {529},
  Volume                   = {85},

  Doi                      = {10.1007/BF01403503},
  Reportnumber             = {Print-82-0072 (UTRECHT)},
  Slaccitation             = {%%CITATION = CMPHA,85,529;%%}
}

@Article{Bhattacharya:1992qb,
  Title                    = {{Z(N) interface tension in a hot SU(N) gauge theory}},
  Author                   = {Bhattacharya, Tanmoy and Gocksch, Andreas and Korthals Altes, Chris and Pisarski, Robert D.},
  Journal                  = {Nucl. Phys.},
  Year                     = {1992},
  Pages                    = {497-524},
  Volume                   = {B383},

  Archiveprefix            = {arXiv},
  Doi                      = {10.1016/0550-3213(92)90086-Q},
  Eprint                   = {hep-ph/9205231},
  Primaryclass             = {hep-ph},
  Reportnumber             = {BNL-47625},
  Slaccitation             = {%%CITATION = HEP-PH/9205231;%%}
}

@Article{Cardy:1981fd,
  Title                    = {{Duality and the Theta Parameter in Abelian Lattice Models}},
  Author                   = {Cardy, John L.},
  Journal                  = {Nucl. Phys.},
  Year                     = {1982},
  Pages                    = {17-26},
  Volume                   = {B205},

  Doi                      = {10.1016/0550-3213(82)90464-3},
  Reportnumber             = {UCSB-TH-41-1981},
  Slaccitation             = {%%CITATION = NUPHA,B205,17;%%}
}

@Article{Cardy:1981qy,
  Title                    = {{Phase Structure of Z(p) Models in the Presence of a Theta Parameter}},
  Author                   = {Cardy, John L. and Rabinovici, Eliezer},
  Journal                  = {Nucl. Phys.},
  Year                     = {1982},
  Pages                    = {1-16},
  Volume                   = {B205},

  Doi                      = {10.1016/0550-3213(82)90463-1},
  Reportnumber             = {PRINT-81-0453 (HEBREW)},
  Slaccitation             = {%%CITATION = NUPHA,B205,1;%%}
}

@Article{DElia:2013uaf,
  Title                    = {{Phase diagram of Yang-Mills theories in the presence of a $\theta$ term}},
  Author                   = {D'Elia, Massimo and Negro, Francesco},
  Journal                  = {Phys. Rev.},
  Year                     = {2013},
  Number                   = {3},
  Pages                    = {034503},
  Volume                   = {D88},

  Archiveprefix            = {arXiv},
  Doi                      = {10.1103/PhysRevD.88.034503},
  Eprint                   = {1306.2919},
  Primaryclass             = {hep-lat},
  Reportnumber             = {IFUP-TH-2013-13},
  Slaccitation             = {%%CITATION = ARXIV:1306.2919;%%}
}

@Article{DElia:2012pvq,
  Title                    = {{$\theta$ dependence of the deconfinement temperature in Yang-Mills theories}},
  Author                   = {D'Elia, Massimo and Negro, Francesco},
  Journal                  = {Phys. Rev. Lett.},
  Year                     = {2012},
  Pages                    = {072001},
  Volume                   = {109},

  Archiveprefix            = {arXiv},
  Doi                      = {10.1103/PhysRevLett.109.072001},
  Eprint                   = {1205.0538},
  Primaryclass             = {hep-lat},
  Reportnumber             = {IFUP-TH-2012-07},
  Slaccitation             = {%%CITATION = ARXIV:1205.0538;%%}
}

@Article{Dirac:1931kp,
  author       = {Dirac, Paul A. M.},
  title        = {{Quantized Singularities in the Electromagnetic Field}},
  doi          = {10.1098/rspa.1931.0130},
  pages        = {60-72},
  volume       = {A133},
  journal      = {Proc. Roy. Soc. Lond.},
  reportnumber = {RX-722},
  slaccitation = {%%CITATION = PRSLA,A133,60;%%},
  year         = {1931},
}

@Article{Gaiotto:2017yup,
  Title                    = {{Theta, Time Reversal, and Temperature}},
  Author                   = {Gaiotto, Davide and Kapustin, Anton and Komargodski, Zohar and Seiberg, Nathan},
  Journal                  = {JHEP},
  Year                     = {2017},
  Pages                    = {091},
  Volume                   = {05},

  Archiveprefix            = {arXiv},
  Doi                      = {10.1007/JHEP05(2017)091},
  Eprint                   = {1703.00501},
  Primaryclass             = {hep-th},
  Slaccitation             = {%%CITATION = ARXIV:1703.00501;%%}
}

@Article{Gaiotto:2014kfa,
  Title                    = {{Generalized Global Symmetries}},
  Author                   = {Gaiotto, Davide and Kapustin, Anton and Seiberg, Nathan and Willett, Brian},
  Journal                  = {JHEP},
  Year                     = {2015},
  Pages                    = {172},
  Volume                   = {02},

  Archiveprefix            = {arXiv},
  Doi                      = {10.1007/JHEP02(2015)172},
  Eprint                   = {1412.5148},
  Primaryclass             = {hep-th},
  Slaccitation             = {%%CITATION = ARXIV:1412.5148;%%}
}

@Article{Gaiotto:2017tne,
  Title                    = {{Time-reversal breaking in QCD$_{4}$, walls, and dualities in 2 + 1 dimensions}},
  Author                   = {Gaiotto, Davide and Komargodski, Zohar and Seiberg, Nathan},
  Journal                  = {JHEP},
  Year                     = {2018},
  Pages                    = {110},
  Volume                   = {01},

  Archiveprefix            = {arXiv},
  Doi                      = {10.1007/JHEP01(2018)110},
  Eprint                   = {1708.06806},
  Primaryclass             = {hep-th},
  Slaccitation             = {%%CITATION = ARXIV:1708.06806;%%}
}

@Article{Goddard:1976qe,
  Title                    = {{Gauge Theories and Magnetic Charge}},
  Author                   = {Goddard, P. and Nuyts, J. and Olive, David I.},
  Journal                  = {Nucl. Phys.},
  Year                     = {1977},
  Pages                    = {1-28},
  Volume                   = {B125},

  Doi                      = {10.1016/0550-3213(77)90221-8},
  Reportnumber             = {CERN-TH-2255},
  Slaccitation             = {%%CITATION = NUPHA,B125,1;%%}
}

@Article{tHooft:1981bkw,
  Title                    = {{Topology of the Gauge Condition and New Confinement Phases in Nonabelian Gauge Theories}},
  Author                   = {'t Hooft, Gerard},
  Journal                  = {Nucl. Phys.},
  Year                     = {1981},
  Pages                    = {455-478},
  Volume                   = {B190},

  Doi                      = {10.1016/0550-3213(81)90442-9},
  Reportnumber             = {CALT-68-819},
  Slaccitation             = {%%CITATION = NUPHA,B190,455;%%}
}

@Article{tHooft:1979rtg,
  Title                    = {{A Property of Electric and Magnetic Flux in Nonabelian Gauge Theories}},
  Author                   = {'t Hooft, Gerard},
  Journal                  = {Nucl. Phys.},
  Year                     = {1979},
  Pages                    = {141-160},
  Volume                   = {B153},

  Doi                      = {10.1016/0550-3213(79)90595-9},
  Reportnumber             = {PRINT-79-0117 (UTRECHT)},
  Slaccitation             = {%%CITATION = NUPHA,B153,141;%%}
}

@Article{tHooft:1976rip,
  Title                    = {{Symmetry Breaking Through Bell-Jackiw Anomalies}},
  Author                   = {'t Hooft, Gerard},
  Journal                  = {Phys. Rev. Lett.},
  Year                     = {1976},
  Pages                    = {8-11},
  Volume                   = {37},

  Doi                      = {10.1103/PhysRevLett.37.8},
  Reportnumber             = {PRINT-76-0254 (HARVARD)},
  Slaccitation             = {%%CITATION = PRLTA,37,8;%%}
}

@Article{Kapustin:2014gua,
  Title                    = {{Coupling a QFT to a TQFT and Duality}},
  Author                   = {Kapustin, Anton and Seiberg, Nathan},
  Journal                  = {JHEP},
  Year                     = {2014},
  Pages                    = {001},
  Volume                   = {04},

  Archiveprefix            = {arXiv},
  Doi                      = {10.1007/JHEP04(2014)001},
  Eprint                   = {1401.0740},
  Primaryclass             = {hep-th},
  Slaccitation             = {%%CITATION = ARXIV:1401.0740;%%}
}

@Article{Kikuchi:2017pcp,
  Title                    = {{Global inconsistency, 't~Hooft anomaly, and level crossing in quantum mechanics}},
  Author                   = {Kikuchi, Yuta and Tanizaki, Yuya},
  Journal                  = {Prog. Theor. Exp. Phys.},
  Year                     = {2017},
  Pages                    = {113B05},
  Volume                   = {2017},

  Archiveprefix            = {arXiv},
  Doi                      = {10.1093/ptep/ptx148},
  Eprint                   = {1708.01962},
  Primaryclass             = {hep-th},
  Reportnumber             = {RBRC-1247},
  Slaccitation             = {%%CITATION = ARXIV:1708.01962;%%}
}

@Article{Komargodski:2017dmc,
  author        = {Komargodski, Zohar and Sharon, Adar and Thorngren, Ryan and Zhou, Xinan},
  title         = {{Comments on Abelian Higgs Models and Persistent Order}},
  journal       = {SciPost Phys.},
  year          = {2019},
  volume        = {6},
  number        = {1},
  pages         = {003},
  doi           = {10.21468/SciPostPhys.6.1.003},
  eprint        = {1705.04786},
  archiveprefix = {arXiv},
  primaryclass  = {hep-th},
  slaccitation  = {%%CITATION = ARXIV:1705.04786;%%},
}

@InCollection{Mandelstam:1974pi,
  Title                    = {{Vortices and Quark Confinement in Nonabelian Gauge Theories}},
  Author                   = {Mandelstam, S.},
  Booktitle                = {{Phys. Rep. 23 (1976) 245-249, In *Gervais, J.L. (Ed.), Jacob, M. (Ed.): Non-linear and Collective Phenomena In Quantum Physics*, 12-16}},
  Year                     = {1976},
  Pages                    = {245-249},
  Volume                   = {23},

  Doi                      = {10.1016/0370-1573(76)90043-0},
  Journal                  = {Phys. Rept.},
  Reportnumber             = {PRINT-74-1623 (UC,BERKELEY)},
  Slaccitation             = {%%CITATION = PRPLC,23,245;%%}
}

@Article{Nambu:1974zg,
  Title                    = {{Strings, Monopoles and Gauge Fields}},
  Author                   = {Nambu, Yoichiro},
  Journal                  = {Phys. Rev.},
  Year                     = {1974},
  Pages                    = {4262},
  Volume                   = {D10},

  Doi                      = {10.1103/PhysRevD.10.4262},
  Reportnumber             = {EFI 74/40},
  Slaccitation             = {%%CITATION = PHRVA,D10,4262;%%}
}

@Article{Polyakov:1975rs,
  Title                    = {{Compact Gauge Fields and the Infrared Catastrophe}},
  Author                   = {Polyakov, Alexander M.},
  Journal                  = {Phys. Lett.},
  Year                     = {1975},
  Pages                    = {82-84},
  Volume                   = {B59},

  Doi                      = {10.1016/0370-2693(75)90162-8},
  Slaccitation             = {%%CITATION = PHLTA,B59,82;%%}
}

@Article{Schwinger:1966nj,
  author       = {Schwinger, Julian S.},
  title        = {{Magnetic charge and quantum field theory}},
  doi          = {10.1103/PhysRev.144.1087},
  pages        = {1087-1093},
  volume       = {144},
  journal      = {Phys. Rev.},
  slaccitation = {%%CITATION = PHRVA,144,1087;%%},
  year         = {1966},
}

@Article{Seiberg:1994aj,
  Title                    = {{Monopoles, duality and chiral symmetry breaking in N=2 supersymmetric QCD}},
  Author                   = {Seiberg, N. and Witten, Edward},
  Journal                  = {Nucl. Phys.},
  Year                     = {1994},
  Pages                    = {484-550},
  Volume                   = {B431},

  Archiveprefix            = {arXiv},
  Doi                      = {10.1016/0550-3213(94)90214-3},
  Eprint                   = {hep-th/9408099},
  Primaryclass             = {hep-th},
  Reportnumber             = {RU-94-60, IASSNS-HEP-94-55},
  Slaccitation             = {%%CITATION = HEP-TH/9408099;%%}
}

@Article{Tanizaki:2018wtg,
  Title                    = {{Anomaly constraint on massless QCD and the role of Skyrmions in chiral symmetry breaking}},
  Author                   = {Tanizaki, Yuya},
  Journal                  = {JHEP},
  Year                     = {2018},
  Pages                    = {171},
  Volume                   = {08},

  Archiveprefix            = {arXiv},
  Doi                      = {10.1007/JHEP08(2018)171},
  Eprint                   = {1807.07666},
  Primaryclass             = {hep-th},
  Reportnumber             = {RBRC-1287},
  Slaccitation             = {%%CITATION = ARXIV:1807.07666;%%}
}

@Article{Tanizaki:2017bam,
  Title                    = {{Vacuum structure of bifundamental gauge theories at finite topological angles}},
  Author                   = {Tanizaki, Yuya and Kikuchi, Yuta},
  Journal                  = {JHEP},
  Year                     = {2017},
  Pages                    = {102},
  Volume                   = {06},

  Archiveprefix            = {arXiv},
  Doi                      = {10.1007/JHEP06(2017)102},
  Eprint                   = {1705.01949},
  Primaryclass             = {hep-th},
  Reportnumber             = {RBRC-1239},
  Slaccitation             = {%%CITATION = ARXIV:1705.01949;%%}
}

@Article{Tanizaki:2017mtm,
  Title                    = {{Anomaly matching for phase diagram of massless $\mathbb{Z}_N$-QCD}},
  Author                   = {Tanizaki, Yuya and Kikuchi, Yuta and Misumi, Tatsuhiro and Sakai, Norisuke},
  Journal                  = {Phys. Rev.},
  Year                     = {2018},
  Pages                    = {054012},
  Volume                   = {D97},

  Archiveprefix            = {arXiv},
  Doi                      = {10.1103/PhysRevD.97.054012},
  Eprint                   = {1711.10487},
  Primaryclass             = {hep-th},
  Reportnumber             = {RBRC-1264},
  Slaccitation             = {%%CITATION = ARXIV:1711.10487;%%}
}

@Article{Tanizaki:2018xto,
  Title                    = {{Anomaly and global inconsistency matching: $\theta$-angles, $SU(3)/U(1)^2$ nonlinear sigma model, $SU(3)$ chains and its generalizations}},
  Author                   = {Tanizaki, Yuya and Sulejmanpasic, Tin},
  Journal                  = {Phys. Rev.},
  Year                     = {2018},
  Number                   = {11},
  Pages                    = {115126},
  Volume                   = {B98},

  Archiveprefix            = {arXiv},
  Doi                      = {10.1103/PhysRevB.98.115126},
  Eprint                   = {1805.11423},
  Primaryclass             = {cond-mat.str-el},
  Reportnumber             = {RBRC-1285},
  Slaccitation             = {%%CITATION = ARXIV:1805.11423;%%}
}

@Article{Wilson:1974sk,
  Title                    = {{Confinement of Quarks}},
  Author                   = {Wilson, Kenneth G.},
  Journal                  = {Phys. Rev.},
  Year                     = {1974},
  Pages                    = {2445-2459},
  Volume                   = {D10},

  Doi                      = {10.1103/PhysRevD.10.2445},
  Reportnumber             = {CLNS-262},
  Slaccitation             = {%%CITATION = PHRVA,D10,2445;%%}
}

@Article{Witten:1998uka,
  Title                    = {{Theta dependence in the large N limit of four-dimensional gauge theories}},
  Author                   = {Witten, Edward},
  Journal                  = {Phys. Rev. Lett.},
  Year                     = {1998},
  Pages                    = {2862-2865},
  Volume                   = {81},

  Archiveprefix            = {arXiv},
  Doi                      = {10.1103/PhysRevLett.81.2862},
  Eprint                   = {hep-th/9807109},
  Primaryclass             = {hep-th},
  Reportnumber             = {IASSNS-HEP-98-65},
  Slaccitation             = {%%CITATION = HEP-TH/9807109;%%}
}

@Article{Witten:1980sp,
  Title                    = {{Large N Chiral Dynamics}},
  Author                   = {Witten, Edward},
  Journal                  = {Annals Phys.},
  Year                     = {1980},
  Pages                    = {363},
  Volume                   = {128},

  Doi                      = {10.1016/0003-4916(80)90325-5},
  Reportnumber             = {HUTP-80/A005},
  Slaccitation             = {%%CITATION = APNYA,128,363;%%}
}

@Article{Witten:1979ey,
  Title                    = {{Dyons of Charge $e\theta/2\pi$}},
  Author                   = {Witten, Edward},
  Journal                  = {Phys. Lett.},
  Year                     = {1979},
  Pages                    = {283-287},
  Volume                   = {B86},

  Doi                      = {10.1016/0370-2693(79)90838-4},
  Reportnumber             = {CERN-TH-2724},
  Slaccitation             = {%%CITATION = PHLTA,B86,283;%%}
}

@Article{Zwanziger:1968rs,
  Title                    = {{Quantum field theory of particles with both electric and magnetic charges}},
  Author                   = {Zwanziger, Daniel},
  Journal                  = {Phys. Rev.},
  Year                     = {1968},
  Pages                    = {1489-1495},
  Volume                   = {176},

  Doi                      = {10.1103/PhysRev.176.1489},
  Slaccitation             = {%%CITATION = PHRVA,176,1489;%%}
}

@Article{Yonekura:2019vyz,
  author        = {Yonekura, Kazuya},
  title         = {{Anomaly matching in QCD thermal phase transition}},
  journal       = {JHEP},
  year          = {2019},
  volume        = {05},
  pages         = {062},
  doi           = {10.1007/JHEP05(2019)062},
  eprint        = {1901.08188},
  archiveprefix = {arXiv},
  primaryclass  = {hep-th},
  slaccitation  = {%%CITATION = ARXIV:1901.08188;%%},
}

@Article{Kobayashi:1970ji,
  author       = {Kobayashi, M. and Maskawa, T.},
  title        = {{Chiral symmetry and eta-x mixing}},
  journal      = {Prog. Theor. Phys.},
  year         = {1970},
  volume       = {44},
  pages        = {1422-1424},
  doi          = {10.1143/PTP.44.1422},
  slaccitation = {%%CITATION = PTPKA,44,1422;%%},
}

@Article{Seiberg:1994rs,
  author        = {Seiberg, N. and Witten, Edward},
  title         = {{Electric - magnetic duality, monopole condensation, and confinement in N=2 supersymmetric Yang-Mills theory}},
  journal       = {Nucl. Phys.},
  year          = {1994},
  volume        = {B426},
  pages         = {19-52},
  note          = {[Erratum: Nucl. Phys.B430,485(1994)]},
  doi           = {10.1016/0550-3213(94)90124-4},
  eprint        = {hep-th/9407087},
  archiveprefix = {arXiv},
  primaryclass  = {hep-th},
  reportnumber  = {RU-94-52, IASSNS-HEP-94-43},
  slaccitation  = {%%CITATION = HEP-TH/9407087;%%},
}

@Article{Sulejmanpasic:2019ytl,
  author        = {Sulejmanpasic, Tin and Gattringer, Christof},
  title         = {{Abelian gauge theories on the lattice: $\theta$-terms and compact gauge theory with(out) monopoles}},
  journal       = {Nucl. Phys.},
  year          = {2019},
  volume        = {B943},
  pages         = {114616},
  doi           = {10.1016/j.nuclphysb.2019.114616},
  eprint        = {1901.02637},
  archiveprefix = {arXiv},
  primaryclass  = {hep-lat},
  slaccitation  = {%%CITATION = ARXIV:1901.02637;%%},
}

@Article{Cordova:2019jnf,
  author        = {C\'ordova, Clay and Freed, Daniel S. and Lam, Ho Tat and Seiberg, Nathan},
  title         = {{Anomalies in the Space of Coupling Constants and Their Dynamical Applications I}},
  doi           = {10.21468/SciPostPhys.8.1.001},
  eprint        = {1905.09315},
  number        = {1},
  pages         = {001},
  volume        = {8},
  archiveprefix = {arXiv},
  journal       = {SciPost Phys.},
  primaryclass  = {hep-th},
  year          = {2020},
}

@Article{Cordova:2019uob,
  author        = {C\'ordova, Clay and Freed, Daniel S. and Lam, Ho Tat and Seiberg, Nathan},
  title         = {{Anomalies in the Space of Coupling Constants and Their Dynamical Applications II}},
  doi           = {10.21468/SciPostPhys.8.1.002},
  eprint        = {1905.13361},
  number        = {1},
  pages         = {002},
  volume        = {8},
  archiveprefix = {arXiv},
  journal       = {SciPost Phys.},
  primaryclass  = {hep-th},
  year          = {2020},
}

@Article{tHooft:1977nqb,
  author       = {'t Hooft, Gerard},
  title        = {On the Phase Transition Towards Permanent Quark Confinement},
  journal      = {Nucl.Phys.B},
  year         = {1978},
  volume       = {138},
  pages        = {1--25},
  doi          = {10.1016/0550-3213(78)90153-0},
  reportnumber = {Print-78-0099 (UTRECHT)},
}

@Article{Bonati:2016tvi,
  author        = {Bonati, Claudio and D'Elia, Massimo and Rossi, Paolo and Vicari, Ettore},
  title         = {{$\theta$ dependence of 4D $SU(N)$ gauge theories in the large-$N$ limit}},
  journal       = {Phys. Rev.},
  year          = {2016},
  volume        = {D94},
  number        = {8},
  pages         = {085017},
  doi           = {10.1103/PhysRevD.94.085017},
  eprint        = {1607.06360},
  archiveprefix = {arXiv},
  primaryclass  = {hep-lat},
  slaccitation  = {%%CITATION = ARXIV:1607.06360;%%},
}

@Article{Poppitz:2012sw,
  author        = {Poppitz, Erich and Sch\"{a}fer, Thomas and \"{U}nsal, Mithat},
  title         = {{Continuity, Deconfinement, and (Super) Yang-Mills Theory}},
  journal       = {JHEP},
  year          = {2012},
  volume        = {10},
  pages         = {115},
  doi           = {10.1007/JHEP10(2012)115},
  eprint        = {1205.0290},
  archiveprefix = {arXiv},
  primaryclass  = {hep-th},
  slaccitation  = {%%CITATION = ARXIV:1205.0290;%%},
}

@Article{Chen:2020syd,
  author        = {Chen, Shi and Fukushima, Kenji and Nishimura, Hiromichi and Tanizaki, Yuya},
  title         = {{Deconfinement and $\mathcal {CP}$ breaking at $\theta=\pi$ in Yang-Mills theories and a novel phase for SU(2)}},
  doi           = {10.1103/PhysRevD.102.034020},
  eprint        = {2006.01487},
  number        = {3},
  pages         = {034020},
  volume        = {102},
  archiveprefix = {arXiv},
  journal       = {Phys. Rev. D},
  primaryclass  = {hep-th},
  reportnumber  = {YITP-20-78},
  year          = {2020},
}

@Article{DiVecchia:1980yfw,
  author       = {Di Vecchia, P. and Veneziano, G.},
  title        = {{Chiral Dynamics in the Large n Limit}},
  doi          = {10.1016/0550-3213(80)90370-3},
  pages        = {253--272},
  volume       = {171},
  journal      = {Nucl. Phys. B},
  reportnumber = {CERN-TH-2814},
  year         = {1980},
}

@Article{Anber:2019nze,
  author        = {Anber, Mohamed M. and Poppitz, Erich},
  title         = {{On the baryon-color-flavor (BCF) anomaly in vector-like theories}},
  doi           = {10.1007/JHEP11(2019)063},
  eprint        = {1909.09027},
  pages         = {063},
  volume        = {11},
  archiveprefix = {arXiv},
  journal       = {JHEP},
  primaryclass  = {hep-th},
  year          = {2019},
}

@Article{Honda:2020txe,
  author        = {Honda, Masazumi and Tanizaki, Yuya},
  title         = {{Topological aspects of $4$D Abelian lattice gauge theories with the $\theta$ parameter}},
  doi           = {10.1007/JHEP12(2020)154},
  eprint        = {2009.10183},
  pages         = {154},
  volume        = {12},
  archiveprefix = {arXiv},
  journal       = {JHEP},
  primaryclass  = {hep-th},
  reportnumber  = {YITP-20-99},
  year          = {2020},
}

@Article{Kapustin:2013uxa,
  author        = {Kapustin, Anton and Thorngren, Ryan},
  title         = {{Higher symmetry and gapped phases of gauge theories}},
  eprint        = {1309.4721},
  archiveprefix = {arXiv},
  month         = {9},
  primaryclass  = {hep-th},
  year          = {2013},
}

@Article{Gukov:2013zka,
  author        = {Gukov, Sergei and Kapustin, Anton},
  title         = {{Topological Quantum Field Theory, Nonlocal Operators, and Gapped Phases of Gauge Theories}},
  eprint        = {1307.4793},
  archiveprefix = {arXiv},
  month         = {7},
  primaryclass  = {hep-th},
  year          = {2013},
}

@Article{Kapustin:2013qsa,
  author        = {Kapustin, Anton and Thorngren, Ryan},
  title         = {{Topological Field Theory on a Lattice, Discrete Theta-Angles and Confinement}},
  doi           = {10.4310/ATMP.2014.v18.n5.a4},
  eprint        = {1308.2926},
  number        = {5},
  pages         = {1233--1247},
  volume        = {18},
  archiveprefix = {arXiv},
  journal       = {Adv. Theor. Math. Phys.},
  primaryclass  = {hep-th},
  year          = {2014},
}

@Article{Tanizaki:2022ngt,
  author        = {Tanizaki, Yuya and \"Unsal, Mithat},
  title         = {{Center vortex and confinement in Yang-Mills theory and QCD with anomaly-preserving compactifications}},
  doi           = {10.1093/ptep/ptac042},
  eprint        = {2201.06166},
  pages         = {04A108},
  volume        = {2022},
  archiveprefix = {arXiv},
  journal       = {PTEP},
  primaryclass  = {hep-th},
  reportnumber  = {YITP-22-04},
  year          = {2022},
}

@Article{Hayashi:2022fkw,
  author        = {Hayashi, Yui and Tanizaki, Yuya},
  title         = {{Non-invertible self-duality defects of Cardy-Rabinovici model and mixed gravitational anomaly}},
  doi           = {10.1007/JHEP08(2022)036},
  eprint        = {2204.07440},
  pages         = {036},
  volume        = {08},
  archiveprefix = {arXiv},
  journal       = {JHEP},
  primaryclass  = {hep-th},
  reportnumber  = {YITP-22-12},
  year          = {2022},
}

@article{Nagai:2024yaf,
    author = "Nagai, Yuki and Tomiya, Akio",
    title = "{JuliaQCD: Portable lattice QCD package in Julia language}",
    eprint = "2409.03030",
    archivePrefix = "arXiv",
    primaryClass = "hep-lat",
    month = "9",
    year = "2024"
}

@misc{bezanson2012juliafastdynamiclanguage,
      title={Julia: A Fast Dynamic Language for Technical Computing}, 
      author={Jeff Bezanson and Stefan Karpinski and Viral B. Shah and Alan Edelman},
      year={2012},
      eprint={1209.5145},
      archivePrefix={arXiv},
      primaryClass={cs.PL},
      url={https://arxiv.org/abs/1209.5145}, 
}

@article{Halliday:1981tm,
    author = "Halliday, I. G. and Schwimmer, A.",
    title = "{$Z$(2) Monopoles in Lattice Gauge Theories}",
    reportNumber = "ICTP/80/81-20",
    doi = "10.1016/0370-2693(81)90630-4",
    journal = "Phys. Lett. B",
    volume = "102",
    pages = "337--340",
    year = "1981"
}

@article{Halliday:1981te,
    author = "Halliday, I. G. and Schwimmer, A.",
    editor = "Julve, J. and Ram{\'o}n-Medrano, M.",
    title = "{The Phase Structure of SU(N)/Z(N) Lattice Gauge Theories}",
    reportNumber = "ICTP/80-81/15",
    doi = "10.1016/0370-2693(81)90055-1",
    journal = "Phys. Lett. B",
    volume = "101",
    pages = "327",
    year = "1981"
}

@article{deForcrand:2004jt,
    author = "de Forcrand, Philippe and Lucini, Biagio and Vettorazzo, Michele",
    editor = "Bodwin, Geoffrey T. and Sinclair, D. K. and Eichten, E. and Holmgren, D. and Kronfeld, Andreas S. and Mackenzie, P. and Okamoto, M. and Simone, J. and El-Khadra, Aida X.",
    title = "{Measuring interface tensions in 4d SU(N) lattice gauge theories}",
    eprint = "hep-lat/0409148",
    archivePrefix = "arXiv",
    doi = "10.1016/j.nuclphysbps.2004.11.260",
    journal = "Nucl. Phys. B Proc. Suppl.",
    volume = "140",
    pages = "647--649",
    year = "2005"
}

@article{Anosova:2022cjm,
    author = "Anosova, Mariia and Gattringer, Christof and Sulejmanpasic, Tin",
    title = "{Self-dual U(1) lattice field theory with a {\ensuremath{\theta}}-term}",
    eprint = "2201.09468",
    archivePrefix = "arXiv",
    primaryClass = "hep-lat",
    doi = "10.1007/JHEP04(2022)120",
    journal = "JHEP",
    volume = "04",
    pages = "120",
    year = "2022"
}

@article{Chen:2024ddr,
    author = "Chen, Jing-Yuan",
    title = "{Instanton density operator in lattice QCD from higher category theory}",
    eprint = "2406.06673",
    archivePrefix = "arXiv",
    primaryClass = "hep-lat",
    doi = "10.21468/SciPostPhys.19.6.158",
    journal = "SciPost Phys.",
    volume = "19",
    number = "6",
    pages = "158",
    year = "2025"
}

@article{Zhang:2024sgm,
    author = "Zhang, Peng and Chen, Jing-Yuan",
    title = "{An explicit categorical construction of instanton density in lattice Yang-Mills theory}",
    eprint = "2411.07195",
    archivePrefix = "arXiv",
    primaryClass = "hep-lat",
    doi = "10.1007/JHEP06(2025)085",
    journal = "JHEP",
    volume = "06",
    pages = "085",
    year = "2025"
}

@article{Katayama:2025pmz,
    author = "Katayama, Nagare and Tanizaki, Yuya",
    title = "{2d Cardy-Rabinovici model with the modified Villain lattice: exact dualities and symmetries}",
    eprint = "2505.19412",
    archivePrefix = "arXiv",
    primaryClass = "hep-th",
    reportNumber = "YITP-25-78",
    doi = "10.1007/JHEP11(2025)004",
    journal = "JHEP",
    volume = "11",
    pages = "004",
    year = "2025"
}

@article{Kovacs:2000sy,
    author = "Kov\'acs, Tamas G. and Tomboulis, E. T.",
    title = "{Computation of the vortex free energy in SU(2) gauge theory}",
    eprint = "hep-lat/0002004",
    archivePrefix = "arXiv",
    reportNumber = "UCLA-00-TEP-06, INLO-PUB-02-00",
    doi = "10.1103/PhysRevLett.85.704",
    journal = "Phys. Rev. Lett.",
    volume = "85",
    pages = "704--707",
    year = "2000"
}

@article{Hoelbling:2000su,
    author = "Hoelbling, Christian and Rebbi, C. and Rubakov, V. A.",
    title = "{Free energy of an SU(2) monopole - anti-monopole pair}",
    eprint = "hep-lat/0003010",
    archivePrefix = "arXiv",
    reportNumber = "BUHEP-00-5",
    doi = "10.1103/PhysRevD.63.034506",
    journal = "Phys. Rev. D",
    volume = "63",
    pages = "034506",
    year = "2001"
}

@article{Korthals-Altes:1999cqo,
    author = "Korthals-Altes, C. and Kovner, A. and Stephanov, Misha A.",
    title = "{Spatial 't Hooft loop, hot QCD and Z(N) domain walls}",
    eprint = "hep-ph/9909516",
    archivePrefix = "arXiv",
    reportNumber = "OUTP-99-52P, CPT-P-3891",
    doi = "10.1016/S0370-2693(99)01242-3",
    journal = "Phys. Lett. B",
    volume = "469",
    pages = "205--212",
    year = "1999"
}

@article{Hidaka:2009hs,
    author = "Hidaka, Yoshimasa and Pisarski, Robert D.",
    title = "{Hard thermal loops, to quadratic order, in the background of a spatial 't Hooft loop}",
    eprint = "0906.1751",
    archivePrefix = "arXiv",
    primaryClass = "hep-ph",
    reportNumber = "KUNS-2216",
    doi = "10.1103/PhysRevD.80.036004",
    journal = "Phys. Rev. D",
    volume = "80",
    number = "3",
    pages = "036004",
    year = "2009",
    note = "[Erratum: Phys.Rev.D 102, 059902 (2020)]"
}

@article{Armoni:2008yp,
    author = "Armoni, Adi and Kumar, S. Prem and Ridgway, Jefferson M.",
    title = "{Z(N) Domain walls in hot N=4 SYM at weak and strong coupling}",
    eprint = "0812.0773",
    archivePrefix = "arXiv",
    primaryClass = "hep-th",
    doi = "10.1088/1126-6708/2009/01/076",
    journal = "JHEP",
    volume = "01",
    pages = "076",
    year = "2009"
}

@article{deForcrand:2005pb,
    author = "de Forcrand, Philippe and Noth, David",
    title = "{Precision lattice calculation of SU(2) 't Hooft loops}",
    eprint = "hep-lat/0506005",
    archivePrefix = "arXiv",
    doi = "10.1103/PhysRevD.72.114501",
    journal = "Phys. Rev. D",
    volume = "72",
    pages = "114501",
    year = "2005"
}

@article{Luscher:2010iy,
    author = {L{\"u}scher, Martin},
    title = "{Properties and uses of the Wilson flow in lattice QCD}",
    eprint = "1006.4518",
    archivePrefix = "arXiv",
    primaryClass = "hep-lat",
    reportNumber = "CERN-PH-TH-2010-143",
    doi = "10.1007/JHEP08(2010)071",
    journal = "JHEP",
    volume = "08",
    pages = "071",
    year = "2010",
    note = "[Erratum: JHEP 03, 092 (2014)]"
}

@article{Luscher:1984xn,
    author = "L{\"u}scher, M. and Weisz, P.",
    title = "{On-shell improved lattice gauge theories}",
    reportNumber = "DESY-84-030",
    doi = "10.1007/BF01205792",
    journal = "Commun. Math. Phys.",
    volume = "98",
    number = "3",
    pages = "433",
    year = "1985",
    note = "[Erratum: Commun.Math.Phys. 98, 433 (1985)]"
}

@article{Iwasaki:1983bv,
    author = "Iwasaki, Y. and Yoshie, T.",
    title = "{Instantons and Topological Charge in Lattice Gauge Theory}",
    reportNumber = "UTHEP-109",
    doi = "10.1016/0370-2693(83)91111-5",
    journal = "Phys. Lett. B",
    volume = "131",
    pages = "159--164",
    year = "1983"
}

@article{Iwasaki:1983iya,
    author = "Iwasaki, Y.",
    title = "{Renormalization Group Analysis of Lattice Theories and Improved Lattice Action. II. Four-dimensional non-Abelian SU(N) gauge model}",
    eprint = "1111.7054",
    archivePrefix = "arXiv",
    primaryClass = "hep-lat",
    reportNumber = "UTHEP-118",
    month = "12",
    year = "1983"
}

@article{GarciaPerez:1993lic,
    author = "Garcia Perez, Margarita and Gonzalez-Arroyo, Antonio and Snippe, Jeroen R. and van Baal, Pierre",
    title = "{Instantons from over - improved cooling}",
    eprint = "hep-lat/9309009",
    archivePrefix = "arXiv",
    reportNumber = "INLO-PUB-11-93, FTUAM-93-31",
    doi = "10.1016/0550-3213(94)90631-9",
    journal = "Nucl. Phys. B",
    volume = "413",
    pages = "535--552",
    year = "1994"
}

@article{deForcrand:1995bq,
    author = "de Forcrand, Philippe and Kim, Seyong",
    editor = "Kieu, T. D. and McKellar, B. H. J. and Guttmann, A. J.",
    title = "{Topological susceptibility and instanton size distribution from over improved cooling}",
    eprint = "hep-lat/9509081",
    archivePrefix = "arXiv",
    reportNumber = "IPS-95-22",
    doi = "10.1016/0920-5632(96)00056-4",
    journal = "Nucl. Phys. B Proc. Suppl.",
    volume = "47",
    pages = "278--281",
    year = "1996"
}

@article{Tanizaki:2024zsu,
    author = "Tanizaki, Yuya and Tomiya, Akio and Watanabe, Hiromasa",
    title = "{Lattice gradient flows (de-)stabilizing topological sectors}",
    eprint = "2411.14812",
    archivePrefix = "arXiv",
    primaryClass = "hep-lat",
    reportNumber = "YITP-24-159",
    doi = "10.1007/JHEP04(2025)123",
    journal = "JHEP",
    volume = "04",
    pages = "123",
    year = "2025"
}

@article{QCD-TARO:1999mox,
    author = "de Forcrand, P. and Garcia Perez, M. and Hashimoto, T. and Hioki, S. and Matsufuru, H. and Miyamura, O. and Nakamura, A. and Stamatescu, I. O. and Takaishi, T. and Umeda, T.",
    collaboration = "QCD-TARO",
    title = "{Renormalization group flow of SU(3) lattice gauge theory: Numerical studies in a two coupling space}",
    eprint = "hep-lat/9911033",
    archivePrefix = "arXiv",
    doi = "10.1016/S0550-3213(00)00145-0",
    journal = "Nucl. Phys. B",
    volume = "577",
    pages = "263--278",
    year = "2000"
}

@article{Butti:2025rnu,
    author = {Butti, Pietro and Della Morte, Michele and J{\"a}ger, Benjamin and Martins, Sofie and Tsang, J. Tobias},
    title = "{Comparison of smoothening flows for the topological charge in QCD-like theories}",
    eprint = "2504.10197",
    archivePrefix = "arXiv",
    primaryClass = "hep-lat",
    reportNumber = "CERN-TH-2025-079",
    doi = "10.1103/53vh-wm6v",
    journal = "Phys. Rev. D",
    volume = "112",
    number = "1",
    pages = "014504",
    year = "2025"
}

@article{Bonati:2014tqa,
    author = "Bonati, Claudio and D'Elia, Massimo",
    title = "{Comparison of the gradient flow with cooling in $SU(3)$ pure gauge theory}",
    eprint = "1401.2441",
    archivePrefix = "arXiv",
    primaryClass = "hep-lat",
    doi = "10.1103/PhysRevD.89.105005",
    journal = "Phys. Rev. D",
    volume = "89",
    number = "10",
    pages = "105005",
    year = "2014"
}

@article{Alexandrou:2017hqw,
    author = "Alexandrou, Constantia and Athenodorou, Andreas and Cichy, Krzysztof and Dromard, Arthur and Garcia-Ramos, Elena and Jansen, Karl and Wenger, Urs and Zimmermann, Falk",
    title = "{Comparison of topological charge definitions in Lattice QCD}",
    eprint = "1708.00696",
    archivePrefix = "arXiv",
    primaryClass = "hep-lat",
    reportNumber = "DESY 17-115, DESY-17-115",
    doi = "10.1140/epjc/s10052-020-7984-9",
    journal = "Eur. Phys. J. C",
    volume = "80",
    number = "5",
    pages = "424",
    year = "2020"
}

@article{Abe:2023uan,
    author = "Abe, Motokazu and Morikawa, Okuto and Onoda, Soma and Suzuki, Hiroshi and Tanizaki, Yuya",
    title = "{Magnetic operators in 2D compact scalar field theories on the lattice}",
    eprint = "2304.14815",
    archivePrefix = "arXiv",
    primaryClass = "hep-lat",
    reportNumber = "KYUSHU-HET-260, OU-HET-1185, YITP-23-58",
    doi = "10.1093/ptep/ptad078",
    journal = "PTEP",
    volume = "2023",
    number = "7",
    pages = "073B01",
    year = "2023"
}
\end{document}